\documentclass[reprint,amsmath,amssymb, aps, prx,superscriptaddress]{revtex4-2}

\usepackage{graphicx}
\usepackage{epsfig}
\usepackage{color}
\usepackage{amsmath}
\usepackage{amsfonts}
\usepackage{mathrsfs}
\usepackage{txfonts}
\usepackage{color}
\usepackage{subfig}
\usepackage{ulem}

\usepackage[breaklinks=true]{hyperref} 
\hypersetup{colorlinks=true, linkcolor=blue, citecolor=blue, filecolor=blue, urlcolor=blue} 

\newcommand{\bst}{\mathcal{T}}
\newcommand{\bea}{\begin{eqnarray}}
\newcommand{\eea}{\end{eqnarray}}
\newcommand{\dket}[1]{|#1\rangle}
\newcommand{\dbra}[1]{\langle#1|}
\newcommand{\trace}[1]{\text{Tr}~#1}
\newcommand{\expval}[1]{\langle#1\rangle}
\newcommand{\imth}{\hspace{1pt}\mathrm{i}\hspace{1pt}}
\newcommand{\bpm}{\begin{pmatrix}}
\newcommand{\epm}{\end{pmatrix}}

\begin{document}
	\title{Bilayer construction for mixed state phenomena with strong, weak symmetries and symmetry breakings}
	\author{Shuangyuan Lu}
 \author{Penghao Zhu}
	\author{Yuan-Ming Lu}
	\affiliation{Department of Physics, The Ohio State University, Columbus OH 43210, USA}
	\date{\today}
	
	\begin{abstract}
        We introduce the bilayer construction, as a specific purification scheme for a general mixed state, where each mixed state has a one-to-one correspondence with a bilayer pure state with two constraints: non-negativity of the bilayer wavefunction; and the presence of an anti-unitary layer-exchange symmetry $\bst$. Different from the Choi-Jamiołkowski isomorphism, any mixed state can be realized as the monolayer reduced density matrix of a bilayer pure state, and its physical properties can be experimentally realized and detected in non-magnetic bilayer $2D$ materials with a layer-exchange mirror symmetry. We study a variety of mixed state phenomena in the bilayer construction: (1) strong and weak symmetries, their explicit and spontaneous breakings in mixed states can be understood as usual Landau-type symmetry breakings in the bilayer pure state, and their criteria can be derived accordingly; (2) decoherence of a pure state by local errors can be realized by quantum quench dynamics of the bilayer pure states; (3) mixed symmetry protected topological (SPT) states and mixed state topological orders can be classified, characterized and realized as pure state SPTs and topological orders in the bilayer. We further study examples of strong-to-weak spontaneous symmetry breaking and their critical scalings at the symmetry-breaking transition in the bilayer construction. 
    \end{abstract}
	\maketitle

\tableofcontents

\section{Introduction} 

In the past few decades, tremendous progress has been made in the study of zero-temperature topological phases\cite{Prange1990,Wen2007B,Nayak2008,Hasan2010,Qi2011,Balents2010,Senthil2015,Ando2015,Chiu2016,Wen2017,Armitage2018,Zeng2019}, which are ground states of local Hamiltonians in closed quantum systems. The huge success of classifying and characterizing pure quantum states and phase transitions between them has motivated recent investigations into the topological properties of mixed states in open quantum systems\cite{deGroot2022,Lee2022,Ma2023,Wang2023,Ma2024a,Fan2024,Bao2023,Lee2023,Sala2024,You2024,Lessa2024,Zhang2024,Soha2024,Ellison2024,Guo2024}, which has become a flourishing research area in the past few years. In zero-temperature topological phases, the interplay of symmetry and topology in the space of gapped local Hamiltonians gives rise to a rich structure of symmetry protected topological phases\cite{Chen2013,Levin2012,Lu2012,Senthil2015} and symmetry enriched topological orders\cite{Essin2013,Mesaros2013,Lu2016,Tarantino2016,Barkeshli2019}. In the case of open quantum systems, it turns out that symmetries in mixed states have a more complicated structure than in pure states, as they can be grouped into weak and strong symmetries\cite{deGroot2022,Ma2023,Lessa2024,Sala2024}. While strong symmetries in mixed states analogous to symmetries in pure states, implying the conservation of symmetry quantum numbers in the whole ensemble of the mixed state, weak symmetries in mixed states has no counterparts in pure states. It has been demonstrated that the presence of both strong and weak symmetries can lead to new mixed state phenomena, which are believed to not exist in pure states\cite{Ma2023,Ma2024a,Lessa2024,Sala2024,You2024,Zhang2024,Soha2024,Ellison2024}. Many recent efforts have been devoted to the theoretical classification of mixed states with symmetries and associated phase transitions\cite{deGroot2022,Ma2023,Ma2023a,Ma2024a,Fan2024,Bao2023,Lee2023,Sala2024,You2024,Lessa2024,Zhang2024,Soha2024,Ellison2024,Guo2024}. Since a large amount of knowledge has been accumulated in the past few decades on many-body pure state phenomena, on how to realize them and to detect them in condensed matters, one natural question is, is it possible to realize various mixed state phenomena with strong and weak symmetries in a pure state of condensed matters? 

In this work, we provide a positive answer to this question, by introducing a bilayer construction as a specific purification of a mixed state. The bilayer construction makes use of a one-to-one correspondence between a mixed state (in its diagonal basis)
\bea\label{general mixed state}
\rho=\sum_{j,k}\rho_{j,k}\dket{e_j}\dbra{e_k},~~~\rho=\rho^\dagger\geq0,~\trace{\rho}=1.
\eea
where $\{\dket{e_k}\}$ is a complete orthonormal basis of the Hilbert space $\mathcal{H}$, and a ``non-negative''\cite{Ma2024a} pure state in a bilayer construction consisting of ``system layer'' $A$ and ``environment layer'' $B$ with a doubled Hilbert space $\mathcal{H}_A\otimes\mathcal{H}_B$:
\bea\label{bilayer purification}
\dket\psi=\sum_{j,k}\big(\sqrt{\rho}\big)_{j,k}\dket{e_j,A}\otimes\dket{e_k,B}
\eea
constrained by an anti-unitary layer-exchange symmetry $\bst$:
\bea
\bst\dket{e_k,A}=\dket{e_k,B},~~\bst\dket{e_k,B}=\dket{e_k,A},~~\forall~k.
\eea
Note that the wavefunction $\psi_{j,k}$ in the orthonormal basis $\dket{e_j,A}\otimes\dket{e_k,B}$ must be non-negative ($\psi=\sqrt\rho\geq0$ ) in order for the one-to-one correspondence between pure state $\dket\psi$ and mixed state $\rho$. In the literature, the above bilayer pure state $\dket\psi$ is also known as the canonical purification\cite{Weinstein2024} or thermofield double\cite{Liu2024} of a mixed state $\rho$.

As a specific $\bst$-preserving purification scheme for mixed state $\rho$, the above bilayer state $\dket{\psi}$ is different from the Choi state 
\bea\label{choi state}
\dket{\psi_\text{Choi}}=\sum_{j,k}\rho_{j,k}\dket{e_j,A}\otimes\dket{e_k,B}\notag
\eea
from Choi-Jamiołkowski isomorphism between operator $\rho$ and state $\dket{\psi_\text{Choi}}$, which was used a lot in previous studies of mixed state phenomena\cite{Lee2022,Sala2024,Ma2024a}. Although the bilayer state $\dket{\psi}$ and Choi state $\dket{\psi_\text{Choi}}$ are similar in terms of symmetry (including strong, weak and $\bst$ symmetries as we will clarify in section \ref{sec:bilayer symmetry}) and non-negativity, the bilayer purification $\dket\psi$ have two unique advantages over the Choi state:

(1) Since mixed state $\rho$ is nothing but the reduced density matrix of layer $A$ in bilayer pure state $\dket\psi$
\bea
\rho_A=\trace{_B\big(\dket\psi\dbra\psi\big)}
\eea
any physical property of mixed state $\rho$ can be realized and experimentally detected as a property of layer $A$ in the bilayer pure state $\dket\psi$.

(2) The anti-unitary layer-exchange symmetry $\bst$, crucial for the one-to-one correspondence between mixed state $\rho$ and bilayer pure state $\dket\psi$ (as we will clarify in section \ref{sec:bilayer construction}), can be realized as the combination of time-reversal and layer-exchange mirror reflection operations. Experimentally, such a symmetry can be generally implemented in two-dimensional (2D) bilayer materials with a layer-exchange mirror symmetry, 
even in the presence of an in-plane magnetic field. 
This provides a route to realize and to observe 2D mixed state phenomena as pure state phenomena in 2D quantum materials. 

In this work, we lay out the framework of bilayer construction for mixed states with strong and weak symmetries, and apply the framework to understand mixed state phenomena such as strong-to-weak spontaneous symmetry breaking\cite{Lessa2024,Sala2024,Ding2024} and decoherence of topological orders by local errors\cite{Dennis2002,Fan2024,Bao2023,Lee2023,Chen2024a,Sala2024b,Chirame2024}. 

The rest of the paper is organized as follows. After introducing the bilayer construction for mixed state in section \ref{sec:bilayer construction}, we discuss how to describe strong and weak symmetries in the bilayer construction in section \ref{sec:bilayer symmetry}, and derive the criteria (see Table \ref{tab:swssb}) for spontaneous and explicit symmetry breakings in mixed states using the correspondence to symmetry breakings in bilayer pure state in section \ref{sec:swssb}. Our criteria summarized in Table \ref{tab:swssb} is equivalent to those proposed by Ref.\cite{Lessa2024} in terms of fidelity correlator\cite{Liu2024,Weinstein2024}. We also use the bilayer framework to discuss the failure of Renyi-2 correlator\cite{Lessa2024,Sala2024} in describing strong-to-weak spontaneous symmetry breaking (SWSSB) in section \ref{sec:renyi-2}. Next, we investigate two examples of SWSSB in mixed states, which are realized as usual spontaneous symmetry breakings (SSBs) in the bilayer construction: section \ref{sec:example:1d Ising} studies SWSSB for discrete $Z_2$ symmetry in a two-leg Ising ladder, whose critical scalings of vairous correlators are studied in section \ref{sec:Ising:criticality}; section \ref{sec:example:U(1)} studies SWSSB for continuous $U(1)$ symmetry as the formation of interlayer superconductivity in the bilayer construction. In section \ref{sec:realizations} we further discuss various mixed state phenomena can be realized in the bilayer construction: in section \ref{sec:decoherence} we show the decoherence of topological orders by local quantum channels\cite{Dennis2002,Fan2024,Bao2023,Lee2023,Chen2024a,Sala2024b,Chirame2024} can be realized as quantum quench dynamics\cite{Mitra2018} of the bialyer pure state, and in section \ref{sec:mSPT+mTO} we show how symmetry protected topological mixed states and mixed state topologial orders can be constructed from their pure state counterparts in the bilayer construction. Finally we conclude in section \ref{sec:summary} with a summary and outlook for future directions.

\section{The bilayer construction}\label{sec:bilayer construction}

We consider a bilayer Hamiltonian where the layer $A$ and $B$ correspond to the system and environment respectively. For any given symmetry group $G$, we consider a symmetric Hamiltonian $\hat H_{G}$ preserving both symmetry $G_A$ for the system, and symmetry $G_B$ for the environment. If the full symmetry $G_A\times G_B$ of the Hamiltonian is spontaneously broken down to a diagonal subgroup $G_{AB}=\{g_Ag_B|g\in G\}$ in the ground state $|\psi\rangle$, the long-range order in $|\psi\rangle$ can be characterized by a local order parameter $\sum_{\alpha}\langle O^\alpha_{A,i}\otimes (O^\alpha_{B,i})^\ast\rangle\neq0$, which transforms nontrivially under $g_A$ and $g_B$, but remains invariant under the combined operation $g_A\cdot g_B$. In other words, $O_{A}$ forms a nontrivial unitary $G_A$ group representation $\mathcal{R}$ of group $G_A$: 
\bea\label{rep:A}
g_AO_{A,i}^{\alpha}g_A^{-1}=(\mathcal{R}_g)_{\alpha,\beta}O_{A,i}^\beta,~~~\mathcal{R}_g^{-1}=\mathcal{R}_g^\dagger,~~1\leq\alpha,\beta\leq \text{dim}\mathcal{R}
\eea
and $O^{\ast}_{B}$ forms its complex conjugate representation $\bar{\mathcal{R}}$ of $G_B$ with the same dimension:
\bea\label{rep:B}
g_B(O^{\alpha}_{B,i})^\ast g_B^{-1}=(\mathcal{R}_g)^\ast_{\alpha,\beta}(O_{B,i}^\beta)^\ast,~~~1\leq\alpha,\beta\leq \text{dim}\mathcal{R}
\eea
so that their tensor product must contain the trivial scalar representation $1$:
\bea
\mathcal{R}\otimes\bar{\mathcal{R}}=1\oplus\cdots
\eea
The above consideration of spontaneous symmetry breaking (SSB) motivates the following bilayer Hamiltonian\cite{Moudgalya2023}
{\bea\notag
&\hat H_G=h_A+h_B^\ast-\sum_{i,j}V_{i,j}O_{A,i}^\alpha(O_{B,i}^\alpha)^\ast (O_{B,j}^\beta)^T(O_{A,j}^\beta)^\dagger \\
&+\sum_{i,j}M_{i,j}\big[(O_{A,i}^\alpha)^\dagger O_{A,j}^\alpha+(O_{B,i}^\alpha)^T(O_{B,j}^\alpha)^\ast+h.c.\big]+\cdots\label{bilayer ham}
\eea}
where $V_{i,j}=V_{j,i}>0$ and $M_{i,j}=M_{j,i}>0$ are positive parameters, and $\cdots$ represents other symmetry-allowed small terms that explicitly break any extra accidental symmetry of the Hamiltonian. Hereafter we shall drop indices $\alpha,\beta$ and follow the dot product notation for simplicity:
\bea
O_{A,i}\cdot O^\ast_{B,i}\equiv\sum_\alpha O_{A,i}^\alpha(O^\alpha_{B,i})^\ast
\eea

We have implemented an anti-unitary layer exchange symmetry $\bst$ in bilayer Hamiltonian (\ref{bilayer ham}):
\bea\label{T:order parameter}
\bst O_A\bst^{-1}=O_B^\ast
\eea
where $O_A$ is any operator defined in layer $A$. To be specific, anti-unitary symmetry $\bst$ maps a real complete orthonormal basis $\{\dket{e_j,A}\}$ of layer $A$ to that of layer $B$:
\bea\label{def:T}
\bst\dket{e_j,A}=\dket{e_j,B},~~~1\leq j\leq \text{dim}\mathcal{H}_A
\eea
Since symmetry $\bst$ permutes $G_A$ and $G_B$:
\bea\label{T exchange}
\bst g_A\bst^{-1}=g_B,~~~\forall~g\in G.
\eea
the matrix representations of symmetry $g_A$ and $g_B$ in the above orthonormal basis are given by $U_g$ and $U_g^\ast$ respectively. It is straightforward to verify that symmetry transformations (\ref{rep:A}) for $O_{A,i}$ and (\ref{rep:B}) for $O_{B,i}$ are consistent in this basis. 

The above anti-unitary layer-exchange symmetry $\bst$ can be realized in non-magnetic two-dimensional solid state materials with a mirror-symmetric bilayer structure, where the mirror symmetry flips the two layers. After applying a uniform in-plane magnetic field, the bilayer still preserves an anti-unitary symmetry $\bst$, which is the combination of time reversal and the mirror reflection operations. As long as symmetry $\bst$ is preserved by the ground state $\dket{\psi}$ of Hamiltonian (\ref{bilayer ham}), the pure state $\dket{\psi}$ has a one-to-one correspondence with the mixed state $\rho_A$ in system layer $A$, as we will show below.

In bilayer Hamiltonian (\ref{bilayer ham}), as we increase the amplitude of inter-layer couplings $V_{i,j}=V_{j,i}>0$, the ground state $\dket\psi$ will go through a quantum phase transition, where the full symmetry $G_A\times G_B$ of the Hamiltonian spontaneously breaks down to the diagonal subgroup $G_{AB}=\{g_Ag_B|g\in G\}$ in ground state $\dket{\psi}$. Instead of directly examining the properties of bilayer pure state $\dket{\psi}$, below we examine this problem from the viewpoint of the system only, in terms of the density matrix of system layer $A$ by tracing out the environment layer:
\bea\label{rho_A}
\rho_A=\text{Tr}_B\big(|\psi\rangle\langle\psi|\big)
\eea
We can always construct a bilayer system where we can Schmidt decompose the (normalized) ground state $\dket{\psi}$ as follows, treating system layer $A$ as a subsystem of the bilayer:
\bea\label{schmidt:psi}
\dket{\psi}=\sum_k\sqrt{p_k}~\dket{k,A}\otimes\dket{k,B}^\ast,~~~\sum_kp_k=1.
\eea
where $\{\dket{k,A}\}$ and $\{\dket{k,B}^\ast\}$ are the Schmidt vectors, and $\{\sqrt{p_k}\}$ are the Schmidt coefficients. According to the transformation rule (\ref{def:T}), the anti-unitary layer-exchange symmetry $\mathcal{T}$ in ground state $\dket{\psi}$ implies:
\bea\label{T:state}
\mathcal{T}\dket{k,A}=\dket{k,B}^\ast,~~~\mathcal{T}\dket{k,B}=\dket{k,A}^\ast
\eea
and from (\ref{T:order parameter}) we have
\bea\label{T:matrix element}
\expval{k,A|O_{A,i}^\dagger|j,A}=\expval{j,B|^\ast O^\ast_{B,i}|k,B}^\ast
\eea
The reduced density matrix (\ref{rho_A}) for system layer $A$ is also diagonalized by the Schmidt vectors:
\bea\label{schmidt:rho}
\rho_A=\sum_kp_k\dket{k,A}\dbra{k,A},~~~\sum_kp_k=1. 
\eea
This specific bilayer purification scheme maps a mixed state $\rho_A$ to a pure state $\dket\psi$, and vice versa. 
Notice that the wavefucntion of pure state $\dket\psi$ in the $\bst$-symmetric Schmidt basis (\ref{schmidt:psi}) are all non-negative ($\sqrt p_k\geq0$), a property known as \emph{non-negativity} for pure state $\dket\psi$\cite{Ma2024a}. Most generally, the eigenstates of a bilayer system with $\mathcal{T}$ symmetry can be decomposed as $\dket{\psi}=\sum_{k}\sigma_{k}\dket{k,A}\otimes\dket{k,B}^{\ast}$, where $\sigma_{k}$ is real but not necessarily positive. This is because in a general bilayer pure state
\bea\label{bilayer pure state}
\dket{\psi}=\sum_{i,j}\psi_{i,j}\dket{e_{i},A}\otimes\dket{e_{j},B}
\eea
$\mathcal{T}$ symmetry enforces the coefficient matrix $\psi$ to be Hermitian 
\bea
\psi_{i,j}=\psi_{j,i}^\ast\notag
\eea
but not necessarily non-negative. As a result, $\psi$ can be diagonalized with real eigenvalues $\{\sigma_{k}=\sigma_k^\ast\}$. The non-negativity of $\dket{\psi}$ is defined as the non-negativity of Hermitian matrix $\psi_{i.j}$, and it imposes a strong constraint on the bilayer pure state $\dket\psi$. Note that the above definition of non-negativity is indpendent of the choice of the complete orthonormal basis $\{\dket{e_j,A}\}$, and therefore is an intrinsic property of any $\bst$-symmetric bilayer state $\dket\psi$.

Note that both the presence of symmetry $\bst$ and non-negativity are crucial to uniquely determine a bilayer purification $\dket{\psi}$ for a generic density matrix $\rho_A$, and to establish a {\bf one-to-one correspondence between a non-negative bilayer pure state $\dket{\psi}$ with anti-unitary layer-exchange symmetry $\bst$, and a mixed state $\rho_A$ for system layer $A$}. Moreover, the mixed state $\rho_A$ is nothing but the reduced density matrix of pure state $\dket\psi$ in system layer $A$. Therefore, any physical property of mixed state $\rho_A$ can be experimentally observed by physical measurements on pure state $\dket\psi$, which are performed solely in the system layer $A$. 

Following the routine of Landau paradigm for broken symmetries, to describe the low-energy physics of the weak symmetric ($G_{AB}$-symmetric) phase where strong symmetry ($G_A\times G_B$) is spontaneously broken, instead of looking into the original Hamiltonian (\ref{bilayer ham}), we can examine the following mean-field Hamiltonian:
\bea\label{mean field ham}
H^\text{MF}_G=h_A+h_B^\ast-\sum_{i,j}(\Delta_{i,j}O_{A,i}\cdot O_{B,j}^\ast+h.c.)+\cdots
\eea
where anti-unitary symmetry $\bst$ implies the following property of mean-field parameters $\{\Delta_{i,j}\}$:
\bea
\Delta_{i,j}=\Delta_{j,i}^\ast
\eea
In mean-field Hamiltonian (\ref{mean field ham}), the full symmetry $G_A\times G_B$ of bilayer Hamiltonian (\ref{bilayer ham}) is explicitly broken, while the diagonal subgroup $G_{AB}$ is still preserved.

\section{Strong versus weak symmetries and symmetry breakings in the bilayer construction}\label{sec:symmetry}

\subsection{Strong and weak symmetries in the bilayer construction}\label{sec:bilayer symmetry}

Now that we have established the one-to-one correspondence between bilayer pure state $\dket{\psi}$ and system-layer density matrix $\rho_A$, we are ready to discuss strong symmetry, weak symmetry, and strong-to-weak spontaneous symmetry breaking (SWSSB) in the bilayer pure state $\dket{\psi}$. 

First, strong symmetry of the system layer $A$ is defined as\cite{deGroot2022}
\bea\label{strong sym:left}
g_A\rho_A=e^{\imth\theta_g}\rho_A,~~~\forall~g\in G.
\eea
which also implies that
\bea\label{strong sym:right}
\rho_Ag_A^{-1}=e^{-\imth\theta_g}\rho_A,~~~\forall~g\in G.
\eea
In other words, all Schmidt vectors $\dket{k,A}$ of $\dket{\psi}$ have the same symmetry eigenvalues. This means the bilayer pure state $\dket{\psi}$ being an eigenstate of symmetry $G_A$
\bea\label{strong sym:A}
g_A\dket{\psi}=e^{\imth\theta_g}\dket{\psi},~~~\forall~g\in G. 
\eea
and of symmetry $G_B$ as well 
\bea\label{strong sym:B}
g_B\dket{\psi}=e^{-\imth\theta_g}\dket{\psi},~~~\forall~g\in G. 
\eea
due to property (\ref{T exchange}) of the layer-exchange symmetry $\bst$. This is equivalent to the fact that the bilayer Hamiltonian (\ref{bilayer ham}) preserves the full symmetry $G_A\times G_B$, and its ground state $\dket{\psi}$ is a symmetric eigenstate in both the symmetry-preserved and spontaneously broken phases. It is important to note that in the spontaneously symmetry-broken phase, the ground state $\dket{\psi}$ is a symmetric Greenberger-Horne-Zeilinger (GHZ) type state with long-range entanglement~\cite{zeng2015}.

Formally, in our bilayer purification, due to the one-to-one correspondence between the density matrix $\rho_A$ in (\ref{schmidt:rho}) and pure state $\dket\psi$ in (\ref{schmidt:psi}), and the anti-unitary symmetry $\bst$ satisfying (\ref{T:state}) and (\ref{T exchange}), it is straightforward to establish the correspondence between strong symmetry $G$ in density matrix $\rho_A$, and symmetry $G_A\times G_B$ in pure state $\dket\psi$. Specifically, (\ref{strong sym:left})-(\ref{strong sym:right}) for mixed state $\rho_A$ correspond to (\ref{strong sym:A})-(\ref{strong sym:B}) for pure state $\dket\psi$. 

Next we consider weak symmetry $G$ of system $A$, defined as\cite{deGroot2022}
\bea
g_A\rho_Ag_A^{-1}=\rho_A,~~~\forall~g\in G.
\eea
Due to the aforementioned correspondence between density matrix and pure state, this definition corresponds to nothing but the diagonal subgroup symmetry $G_{AB}=\{g_Ag_B|g\in G\}$ in pure state $\dket\psi$. As a result, the strong-to-weak spontaneous symmetry breaking in pure state $\dket\psi$ exactly corresponds to the spontaneous breaking of symmetry $G_A\times G_B$ down to diagonal subgroup $G_{AB}$ in pure state $\dket\psi$.  

\onecolumngrid

\begin{table}[h]
    \centering
    \begin{tabular}{|c|c|c|c|c|c|}
    \hline
         Strong symmetry $G_A\times G_B$ & Weak symmetry $G_{AB}$ & $\lim_{|i-j|\rightarrow\infty}C_0(i,j)$ in (\ref{wightman correlator})&$\sum_{\alpha}\trace{\big[\sqrt\rho O_i^\alpha\sqrt\rho (O_i^\alpha)^\dagger\big]}$ in (\ref{expval:AB})& $\lim_{|i-j|\rightarrow\infty}\expval{O_i\cdot O_j^\dagger}$&$\expval{O_i}$\\
    \hline
        Preserved&Preserved&0&0&0&0\\
        \hline
        Spontaneously broken&Preserved&$O(1)$&0&0&0\\
        \hline Explicitly broken&Preserved&$O(1)$&$O(1)$&0&0\\
        \hline Spontaneously broken&Spontaneously broken&$O(1)$&0&$O(1)$&0\\
                \hline Explicitly broken&Spontaneously broken&$O(1)$&$O(1)$&$O(1)$&0\\
        \hline Explicitly broken&Explicitly broken&$O(1)$&$O(1)$&$O(1)$&$O(1)$\\
    \hline
    \end{tabular}
    \caption{Criteria for explicit and spontaneous breakings of strong and weak symmetries in the bilayer construction.}
    \label{tab:swssb}
\end{table}

\twocolumngrid

\subsection{Criteria for spontaneous and explicit breaking of strong and weak symmetries}\label{sec:swssb}

Since the definition and characterization of spontaneous symmetry breaking in pure state $\dket\psi$ is well understood, we can use the correspondence between a bilayer pure state $\dket\psi$ and mixed state $\rho_A$ to define spontaneous breaking of strong and weak symmetries in mixed state $\rho_A$. The characterization of spontaneous and explicit symmetry breakings are summarized in Table \ref{tab:swssb}. Below, we show how the definition of symmetry breakings in mixed state $\rho_A$ is equivalent to those proposed in Ref.\cite{Lessa2024}.

We start by writing down a few useful equalities between correlators of pure state $\dket{\psi}$ and correlators of mixed state $\rho_A$, which will be used later to establish our results in Table \ref{tab:swssb}.

First, the correlation function of order parameter $O_{i,A}\cdot O^\ast_{i,B}$ in pure state $\dket{\psi}$
\begin{widetext}
{\bea
\notag&C_0(i,j)\equiv\langle\psi|(O_{A,i}\cdot O^\ast_{B,i})(O_{B,j}^T\cdot O^\dagger_{A,j})|\psi\rangle=\sum_{k,l,\alpha,\beta}\sqrt{p_kp_l}\expval{k,A|O_{A,i}^\alpha (O_{A,j}^\beta)^\dagger|l,A}\expval{k,B|^\ast (O^\alpha_{B,i})^\ast(O_{B,j}^\beta)^T|l,B}^\ast\\
&=\sum_{k,l,\alpha,\beta}\sqrt{p_kp_l}\expval{k,A|O_{A,i}^\alpha (O_{A,j}^\beta)^\dagger|l,A}\expval{l,A| O_{A,j}^\beta (O^\alpha_{A,i})^\dagger |k,A}=\sum_{\alpha,\beta}\trace{\big[\sqrt{\rho_A}O_{A,i}^\alpha (O_{A,j}^\beta)^\dagger\sqrt{\rho_A} O_{A,j}^\beta(O^\alpha_{A,i})^\dagger\big]}\label{wightman correlator} 
\eea}  
\end{widetext}
is equal to the Wightman correlator\cite{Liu2024} (also known as Renyi-1 correlator\cite{Weinstein2024}) $C_0(i,j)$ of mixed state $\rho_A$. We have used relation (\ref{T:matrix element}) in the above derivation. As shown in Ref.\cite{Liu2024,Weinstein2024}, the long-range order in the Wightman correlator is equivalent to long-range order in the fedility correlator (\ref{fidelity correlator}), which signals the spontaneous breaking of strong symmetry $G$ in the picture of mixed state $\rho_A$. In the picture of pure state $\dket{\psi}$, on the other hand, long-range correlation of bilayer order parameter $O_{A,i}\cdot O^\ast_{B,i}$ in two-point correlator $C_0(i,j)$ signals the spontaneous breaking of full symmetry $G_A\times G_B$ down to the diagonal subgroup $G_{AB}$. The equality (\ref{wightman correlator}) confirms the equivalence between these two pictures. 

Note that Ref.\cite{Lessa2024} defines the strong-to-weak spontaneous symmetry breaking (SWSSB) as the long-range correlation of order parameter operator $\hat {O}_i$ in the fidelity correlator, but the absence of long-range order in the usual two-point correlator:
\bea\label{swssb}
\lim_{|i-j|\rightarrow\infty}F_O(i,j)\neq0,~~\lim_{|i-j|\rightarrow\infty}\trace{\big[\rho O_i O_j^\dagger\big]}=0.
\eea
where the fidelity correlator is defined as:
\bea\label{fidelity correlator}
F_O(i,j)\equiv\trace{\sqrt{\sqrt{\rho}O_iO_j^\dagger\rho O_i^\dagger O_j\sqrt{\rho}}}
\eea
As shown in Ref.\cite{Liu2024,Weinstein2024}, for bounded operator $O_{i}$, the fidelity correlator is lower bounded by the Wightman correlator and upper bounded by its square root up to $O(1)$ prefactors. Therefore the long-range correlation in Wightman correlator $C_0(i,j)$ is \emph{equivalent}\cite{Liu2024,Weinstein2024} to the long-range correlator in the fidelity correlator (\ref{fidelity correlator})
\bea\label{lro:fidelity correlator}
\lim_{|i-j|\rightarrow\infty}C_0(i,j)\neq0\Longleftrightarrow \lim_{|i-j|\rightarrow\infty}F_O(i,j)\neq0
\eea
Meanwhile, the usual two-point correlation function for order parameter $O_{A}$ is given by
\begin{equation}\label{ldlro correlator}
\expval{O_{i}\cdot O_{j}^\dagger}\equiv\expval{\psi|O_{A,i}\cdot O^\dagger_{A,j}|\psi}=\trace{\big[{\rho_A}(O_{A,i}\cdot O^\dagger_{A,j})\big]}    
\end{equation}
Since the SWSSB for mixed state $\rho_A$ corresponds to spontaneous breaking of full symmetry $G_A\times G_B$ to diagonal subgroup $G_{AB}$, there should be no long-range correlation for the above two-point correlator
\bea\label{no lro:two-point correlator}
\lim_{|i-j|\rightarrow\infty}\expval{O_i\cdot O_j^\dagger}=0
\eea
As a result, the above criteria (\ref{lro:fidelity correlator}) and (\ref{no lro:two-point correlator}) for spontaneous symmetry breaking from $G_A\times G_B$ to $G_{AB}$ in the bilayer construction is equivalent to the definition (\ref{swssb}) of SWSSB proposed in Ref.\cite{Lessa2024}.

A strong symmetry can either be spontaneously or explicitly broken, corresponding to spontaneous breaking of $G_A\times G_B$ to $G_{AB}$ in bilayer Hamiltonian (\ref{bilayer ham}), or its explicit breaking in mean-field Hamiltonian (\ref{mean field ham}). The criteria for explicit symmetry breaking can be analyzed in parallel to the spontaneous symmetry breaking discussed above. First, explicit breaking of $G_{A}\times G_B$ to $G_{AB}$ can be directly reflected in the ground state $\dket{\psi}$, which is no longer symmetric under $G_A\times G_B$ but only symmetric under $G_{AB}$. This symmetry breaking is characterized by a nonvanishing expectation value of the order parameter $O_{A,i}\cdot O^\ast_{B,i}$:
\bea\notag
&\expval{\psi|O_{A,i}\cdot O^\ast_{B,i}|\psi}\\
\notag&=\sum_{\alpha,k,l}\sqrt{p_lp_k}\expval{k.A|O_{i,A}^\alpha|l.A}\expval{k,B|^\ast (O^\alpha)^\ast_{B,i}|l,B}^\ast\\
\notag&=\sum_{\alpha,k,l}\sqrt{p_lp_k}\expval{k.A|O_{i,A}^\alpha|l.A}\expval{l,A| (O^\alpha)^\dagger_{A,i}|k,A}\\
&=\sum_{\alpha}\trace{\big[\sqrt\rho_A O_i^\alpha\sqrt\rho_A (O_i^\alpha)^\dagger\big]}.\label{expval:AB}
\eea
The above equality defines explicit breaking of strong symmetry down to weak symmetry for mixed state $\rho_A$. Finally, explicit breaking of weak symmetry is characterized as usual by nonzero expectation value of the order parameter $O_{A,i}$:
\bea\label{expval:A}
\expval{O_{A,i}}\equiv\expval{\psi|O_{A,i}|\psi}=\trace{\big[{\rho_A}O_{A,i}\big]}
\eea
We summarize all criteria for the spontaneous and/or explicit breaking of strong and/or weak symmetries in Table \ref{tab:swssb}.\\

\subsection{Failure of Renyi-2 correlator in diagnosing SWSSB}\label{sec:renyi-2}

Finally, we comment on the Renyi-2 correlator\cite{Lessa2024,Sala2024,Ma2024a}
\bea\label{renyi-2 correlator}
C_1(i,j)\equiv\frac{\text{Tr}\big(\rho_AO_{A,i}O^\dagger_{A,j}\rho_A O_{A,j} O_{A,i}^\dagger\big)}{\text{Tr}(\rho_A^2)}
\eea
which has been proposed as an alternative to fidelity correlator (\ref{fidelity correlator}), whose long-range correlation can be used to diagnose SWSSB in the absence of long-range two-point correlation (\ref{no lro:two-point correlator}). In the bilayer construction, SWSSB is defined as long-range correlation in the Wightman correlator (\ref{wightman correlator}), which is equivalent to long-range correlation in fidelity operator (\ref{fidelity correlator}). However, it is unclear how long-range correlation in the Renyi-2 correlator is related to that of Wightman correlator. To make such a connection, we consider the following continuous family of correlators that interpolate between Wightman correlator $C_0(i,j)$ in (\ref{wightman correlator}) and Renyi-2 correlator $C_1(i,j)$ in (\ref{renyi-2 correlator}):
\bea\label{family of correlation}
C_t(i,j)\equiv\frac1{{{\trace{\rho_A^{t+1}}}}}\trace{\big[\rho_A^{(t+1)/2}O_{A,i}O^\dagger_{A,j}\rho_A^{(t+1)/2}O^\dagger_{A,i}O_{A,j}\big]}
\eea
in the parameter range $0\leq t\leq 1$. 

Below we study the long-range behavior of correlator family $\{C_t(i,j)|0\leq t\leq 1\}$ in the limit of $|i-j|\rightarrow\infty$ in two examples, introduced first in Ref.\cite{Lessa2024}. The goal is to demonstrate that these correlators are not equivalent, in the sense that in a given state, the presence of long-range correlation in one correlator does not necessarily imply that the others will exhibit long-range correlation as well. 

In the first example with $Z_2$ symmetry, the density matrix is 
\bea
\rho = \frac{1}{2} |++ \cdots \rangle \langle ++ \cdots| + \frac{1}{2^{L+1}} (\mathbb{I} + X)
\eea
with $X\equiv\prod_jX_j$ and the observable operator is $O_{A, i} = Z_i$. We calculate the correlator for any two different site $i\neq j$:
\bea
C_t(i, j) = \frac{2(2^{-1} + 2^{-L})^\frac{t+1}{2} 2^{-\frac{L(t+1)}{2}} + (2^{L-1} - 2) 2 ^{-L(t+1)}}{(2^{-1} + 2^{-L})^{t+1} + (2 ^{L - 1} -1) 2^{-L(t+1)}}
\eea
In the thermodynamic limit $L\rightarrow\infty$, the correlator is nonzero only for $t = 0$:
\bea
\lim_{L\rightarrow \infty} C_t(i, j) = 
\begin{cases}
\frac12 &\quad t=0\\
0 &\quad 0<t\leq 1
\end{cases}
\eea

In the second example, we use the model in Appendix A of Ref.\cite{Lessa2024}. 
\bea
\rho = \sum_{m=1} ^{2^N} p_m |\psi_m\rangle \langle \psi_m |
\eea
where $p_m \sim m^{-2/3}$ and $\lim_{|x-y|\rightarrow \infty} \langle \psi_m | Z_x Z_y |\psi_n\rangle = o_m^\prime \delta_{mn} \sim \delta_{mn}/ m^2$. The correlator is:
\bea
C_t(i, j) = \frac{\sum_m p_m^{t+1} |o_m^\prime|^2}{\sum_m p_m^{t+1}} \sim \frac{\sum_m m^{-(2t + 14) / 3}}{\sum_m m^{-2(t+1) / 3}}
\eea
The numerator is always finite since the exponent is smaller than $-1$, while the numerator is finite when $-2(t+1)/3 < -1$ and infinite otherwise. Thus,
\bea
C_t(i, j) = 
\begin{cases}
    O(1) &\quad  1/2 < t \leq 1\\
    0 &\quad  0 \leq t \leq 1/2
\end{cases}
\eea
the correlator is finite when $t$ is larger than $1/2$.

\section{Examples of strong-to-weak spontaneous symmetry breaking}\label{sec:example:swssb}

\subsection{Discrete SWSSB with $G=Z_2$ in a 2-leg Ising ladder}\label{sec:example:1d Ising}

We consider the following bilayer Hamiltonian of a 2-leg Ising ladder:
\bea\label{ising}
\hat H_{Z_2}=-J\sum_{i}Z^A_iZ^B_iZ^A_{i+1}Z^B_{i+1}-\sum_i(X_i^A+X_i^B)-g\sum_iX_i^AX_i^B.~~
\eea
The strong symmetry group $Z_2^A\times Z_2^B$ of the above Hamiltonian is generated by 
\bea
g_A=\prod_iX^A_i,~~~g_B=\prod_iX^B_i
\eea
In the large $J$ phase, the full $Z_2^A\times Z_2^B$ symmetry is spontaneously broken down to a $Z_2$ subgroup generated by
\bea
g_Ag_B=\prod_i\big(X^A_iX^B_i\big)
\eea
The associated long-range order is characterized by the correlation function, $C_{0}(i,j)$, of the local order parameter $O_{A,i}\cdot O_{B,i}$ with $O_{A,i}=Z_i^A,~~~O_{B,i}=Z_i^B$. Given that the 2-leg Ising ladder in Eq.~\eqref{ising} is exactly solvable, it can be analytically shown that $C_{0}(i,j)\neq0$, as demonstrated in the following.

\subsubsection{Spontaneous breaking of strong symmetry}

Since the last $g$ term of Hamiltonian (\ref{ising}) commute with all other terms, we can take the $g\gg1,J$ limit and work in its low-energy sector with 
\bea\label{low energy subspace}
X_i^AX_i^B=1,~~~\forall~i.
\eea
The effective Hamiltonian in this subspace is simply a transverse field Ising model
\bea\label{tfim}
H^{eff}_{Z_2}=-J\sum_{i}\sigma_i^z\sigma^z_{i+1}-2\sum_i\sigma_i^x
\eea
where the effective spin-$1/2$ operators in the low-energy subspace (\ref{low energy subspace}) are given by
\bea\label{eff spin}
\sigma_i^z\equiv Z_i^AZ_i^B,~\sigma_i^x\equiv X_i^A=X^B_i.
\eea
It can be solved exactly using the Jordan-Wigner transformation:
\bea
Z^A_iZ^B_i=\chi_{2i-1}\prod_{r<i}(\imth\chi_{2r-1}\chi_{2r}),~~~X_i^A=\imth\chi_{2i-1}\chi_{2i}
\eea
where $\chi_j$ are Majorana fermions satisfying $\{\chi_i,\chi_j\}=2\delta_{i,j}$. The effective Hamiltonian (\ref{tfim}) is a non-interacting mode of fermions:
\bea\notag
&H^{eff}_{Z_2}=-J\sum_i\imth\chi_{2i}\chi_{2i+1}-2\sum_i\imth\chi_{2i-1}\chi_{2i}\\
&=\sum_k\bpm f_k^\dagger,f_{-k}\epm\bpm J\cos k-2&\imth J\sin k\\-\imth J\sin k&2-J\cos k\epm\bpm f_k\\f^\dagger_{-k}\epm
\eea
where we have defined complex fermions $f_r$ as
\bea
f_r\equiv\frac{\chi_{2r-1}+\imth\chi_{2r}}2=\frac1{\sqrt L}\sum_ke^{\imth kr}f_k
\eea
Given the ground state of transverse field Ising model (\ref{tfim}) in the $\{\sigma_i^x\}$ basis:
\bea
\dket{\psi_\sigma}=\sum_{\{\sigma^x_i=\pm1\}}C_{\{\sigma_i^x\}}\otimes_i\dket{\sigma_i^x}
\eea
the bilayer ground state of full Hamiltonian (\ref{ising}) is simply
\bea
\dket{\psi}=\sum_{\{\sigma^x_i=\pm1\}}C_{\{\sigma_i^x\}}\otimes_i\big(\dket{X^A_i=\sigma_i^x}\otimes\dket{X^B_i=\sigma_i^x}\big)
\eea
and the reduced density matrix for system $A$ is diagonal in $X_i^A$ basis:
\bea
\rho_A=\trace_B{\dket{\psi}\dbra{\psi}}=\sum_{\{\sigma^x_i=\pm1\}}|C_{\{\sigma_i^x\}}|^2\otimes_i\dket{X^A_i=\sigma_i^x}\dbra{X^A_i=\sigma_i^x}
\eea
Note that according to Perron-Frobenius theorem, $C_{\{\sigma_i^x\}}\geq0$ since the transverse field Ising model is a negative matrix in the $\sigma^x_i$ basis. Therefore bilayer ground state $\dket\psi$ is a non-negative state which has a one-to-one correspondence with mixed state $\rho_A$. When $J\gg2$, the ground state $\dket{\psi_{\sigma}}$ of the transverse Ising model is the symmetric GHZ state that is a linear combination of the all-up state and the all-down state with the same probability amplitude $1/\sqrt{2}$. It is straightforward to verify that $\operatorname{Tr}(\sqrt{\rho_A}Z_{i}^A\sqrt{\rho_A}Z_{i}^A)=0$, $\langle Z_{i}^{A}Z_{j}^{A}\rangle=0$, $\langle Z_{i}^{A}\rangle=0$, and the Wightman correlator is nonvanishing:
\bea
C_{0}(i,j)=\sum_{\{\sigma^x_k=\pm1\}}|C_{\{\sigma_k^x\}}C_{\sigma_1^x,\cdots,-\sigma_i^x,\cdots,-\sigma_j^x,\cdots}|=1.
\eea
regardless of $i,j$. These results align with the second line in Table~\ref{tab:swssb}, confirming that $\rho_{A}$ is a mixed state with a spontaneously broken strong $Z_2$ symmetry. 

\subsubsection{Universal power-law scaling of $C_t(i,j)$ at the critical point}
\label{sec:Ising:criticality}

\begin{figure}
    \centering
    \includegraphics[width=\columnwidth]{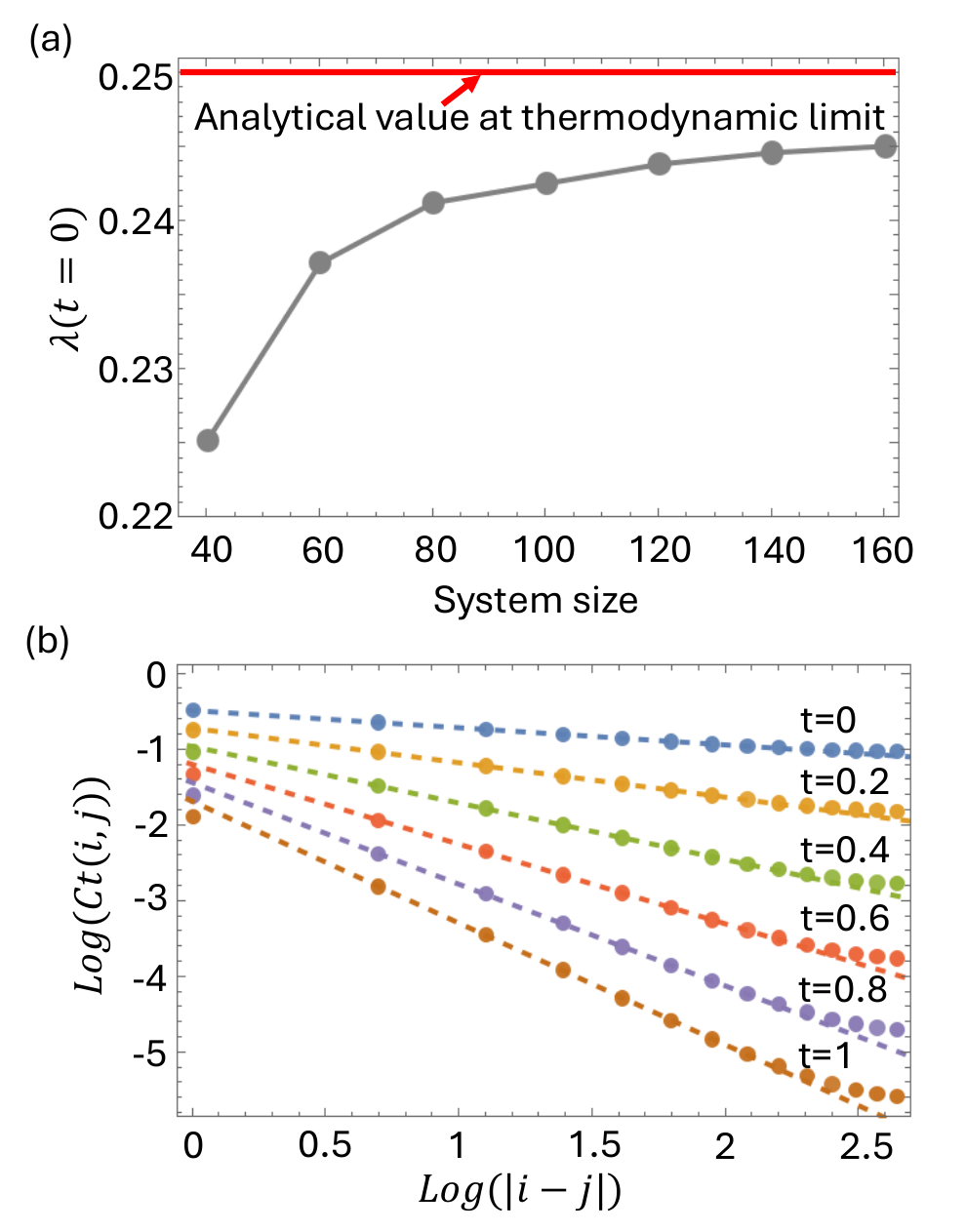}
    \caption{ (a) The plot of  $\lambda(t=0)$ versus system size demonstrates that as the system size increases, $\lambda(t=0)$ approaches the theoretical value of 0.25 in the thermodynamic limit.
    (b) Scaling of correlations $C_{t}(i,j)$ with $|i-j|$ for a 30-unit-cell periodic chain. By fitting the data points in the range $2\leqslant|i-j|\leqslant8$, we obtain the exponent
    $\lambda(t)$ being $0.23,0.46,0.75,1.05,1.34,1.61$ for $t=0,0.2,0.4,0.6,0.8,1$, respectively.
   }
    \label{fig:criticality}
\end{figure}

The spontaneous breaking of the strong $Z_2$ symmetry to a weak one in model (\ref{ising}) happens at the critical coupling $J_c=2$, where the order parameter correlation (\ref{wightman correlator}) decays in a universal power law:
\bea
\lim_{|i-j|\rightarrow\infty}C_0(i,j)=\lim_{|i-j|\rightarrow\infty}\expval{\psi|Z_i^AZ_i^BZ_j^AZ_j^B|\psi}_{J=2}\sim|i-j|^{-1/4},
\eea
which is the same to the behavior observed at the critical point of the transverse Ising model and is numerically verified in Fig.~\ref{fig:criticality} (a). In this specific model, the fidelity correlator (\ref{fidelity correlator}) is exactly the same as the Wightman correlator $C_0(i,j)$:
\bea
F_O(i,j)=\sum_{\{\sigma^x_k=\pm1\}}|C_{\{\sigma_k^x\}}C_{\sigma_1^x,\cdots,-\sigma_i^x,\cdots,-\sigma_j^x,\cdots}|=C_{0}(i,j).
\eea

Meanwhile, 
the continuous family (\ref{family of correlation}) of correlation functions can also be straightforwardly calculated using $C_{\{\sigma_{i}^{x}\}}$:
\bea
C_t(i,j)=\frac{\sum_{\{\sigma^x_k=\pm1\}}|C_{\{\sigma_k^x\}}C_{\sigma_1^x,\cdots,-\sigma_i^x,\cdots,-\sigma_j^x,\cdots}|^{t+1}}{\sum_{\{\sigma^x_i=\pm1\}}|C_{\{\sigma_i^x\}}|^{2(t+1)}},
\eea 
In Fig.~\ref{fig:criticality}(a), we numerically show that $C_{t}(i,j)$ also decays algebraically with the distance $|i-j|$ at the critical point $J_c=2$, and
\bea
C_{t}(i,j)\sim |i-j|^{-\lambda(t)}
\eea
with $\lambda(t)$ varies with $t$ (see Fig. \ref{fig:criticality}). We have numerically confirmed that the power law scaling of $C_{t}(i,j)$ at the critical point $J_{c}=2$ is robust against perturbation terms $H_{\text{pert}}=-\sum_{i}(J_{1}Z^{A}_{i}Z^{A}_{i+1}+J_{2}Z^{B}_{i}Z^{B}_{i+1})$ that break the integrability of Hamiltonian \eqref{ising}. Specifically, we observed that for various sets of parameters $(J_1,J_2)$ with $0\leq J_{1,2}\leq 1$, the scaling components obtained from linear fitting remain invariant, up to numerical errors in the calculations and fittings.



\subsection{Breaking of continuous strong symmetry $G=U(1)$ by interlay superconductivity in a bilayer superconductor}\label{sec:example:U(1)}

In this section, we investigate a symmetry-breaking state in a bilayer fermionic system with independent fermion number conservation in each layer. We demonstrate that the corresponding mixed state exhibits spontaneous (in the exact ground state) or explicit (in the mean-field ground state) breaking of its strong symmetry down to weak symmetry, exemplifying the scenario illustrated in the 2nd and 3rd rows of Table~\ref{tab:swssb}. Specifically, we study interlayer
superconductivity involving Cooper pairs formed between fermions from different layers, which spontaneously breaks the $U(1)$ symmetry associated with the net fermion number conservation. 

We focus on the situation where the superconductivity is induced by attractive interactions similar to the case of undoped graphene~\cite{uchoa_superconducting_2007, roy_unconventional_2010, marino_quantum_2006, black-schaffer_chiral_2014, castro_neto_charge_2001, zhao_bcs-bec_2006, kotov_electron-electron_2012}. Specifically, we examine an A-A stacked bilayer graphene model of spinless fermions, with an attractive interlayer interactions:

\bea\notag
\hat H &= \sum_{\langle m, n\rangle}\left( t \xi_{1,m}^\dagger \xi_{1,n} + t \xi_{2,m}^\dagger \xi_{2,n} + h.c.\right) 
\\
&- \sum_{m} V \xi_{2,m}^\dagger \xi_{2,m} \xi_{1,m}^\dagger \xi_{1,m}\label{sc:interacting}
\eea
where $\xi_{l,m} $ annihilates a fermion at site $m$ in layer $l\in\{1,2\}$ and $\langle m, n\rangle$ denotes the nearest neighbors. For the honeycomb lattice, $m$ can be further represented by $(\alpha, i)$ where $i$ labels the unit cell and $\alpha\in\{1,2\}$ labels the sublattice in each unit cell.  

Clearly the above model has a strong $U(1)_c\times U(1)_d=\{e^{\imth(\theta_1\hat N_c-\theta_2\hat N_d)}|0\leq\theta_{1,2}<2\pi\}$ symmetry of particle number conservation in each layer:
\bea
e^{\imth\theta\hat N_c}\xi_{1,m}e^{-\imth\theta\hat N_c}=e^{-\imth\theta}\xi_{1,m},~~~\hat N_c\equiv\sum_m\xi_{1,m}^\dagger\xi_{1,m},\\
e^{-\imth\theta\hat N_d}\xi_{2,m}e^{\imth\theta\hat N_d}=e^{\imth\theta}\xi_{2,m},~~~\hat N_d\equiv\sum_m\xi_{2,m}^\dagger\xi_{2,m}.
\eea

In the absence of interactions, the free fermions form two Dirac points at high-symmetry momenta $K$ and $K^\prime$. When the interaction strength $V$ is smaller than the critical value $V_{c}$, the ground state remains a semimetal which preserves the fermion number conservation in each layer. However, for interaction strengths $V>V_{c}$, interlayer Cooper pairs condense, leading to a superconducting ground state that spontaneously breaks the $U(1)_\text{charge}=\{e^{\imth\theta(\hat N_c+\hat N_d)}\}$ symmetry corresponding to the net fermion number conservation. Meanwhile, the diagonal subgroup of weak $U(1)_\text{diag}$ symmetry:
\bea\label{SC:weak symmetry}
U(1)_\text{diag}=\{e^{\imth\theta(\hat N_c-\hat N_d)}|0\leq\theta\leq2\pi\}
\eea
is still preserved by interlayer superconductivity. The mean-field Hamiltonian for the superconducting phase is the following:
\bea\notag
\hat H_\text{MF} &= \sum_{a, \mathbf{k}} \epsilon_{a, \mathbf{k}} \left(c_{a, \mathbf{k}}^\dagger c_{a,\mathbf{k}} + d_{a,\mathbf{k}}^\dagger d_{a, \mathbf{k}}\right) 
\\
&-\left( \Delta  c_{a, {-\mathbf{k}}} d_{a,\mathbf{k}} + H.c\right) + |\Delta|^2 \label{sc:mean field}
\eea
where the $U(1)_\text{charge}$ symmetry is explicitly broken, and $a = 1, 2$ is the band index. $c^{\dag}_{a,\mathbf{k}}=\sum_{\alpha}U_{\mathbf{k},a\alpha}\xi^{\dag}_{1,\alpha,\mathbf{k}}$ and $d^{\dag}_{a,\mathbf{k}}=\sum_{\alpha}U_{\mathbf{k},a\alpha}\xi^{\dag}_{2,\alpha,\mathbf{k}}$, where $\xi^{\dag}_{l,\alpha,\mathbf{k}}\equiv\frac{1}{\sqrt{N_{uc}}}\sum_{i}e^{i\mathbf{k}\cdot\mathbf{x}_{i}}\xi^{\dag}_{l,\alpha,i}$ with $N_{uc}$ the number of unit cell. If we denote the vectors pointing from a lattice to its three nearest neighbors as $\boldsymbol{\delta}_{i}, \ i\in\{1,2,3\}$, then $\epsilon_{1,\mathbf{k}} = t|\sum_{i=1}^3 e^{i\mathbf{k}\cdot \boldsymbol{\delta}_i}|$ and $\epsilon_{2,\mathbf{k}} = -\epsilon_{1,\mathbf{k}}$.

In the superconducting phase, $\Delta$ acquires a finite value. After diagonalizing the Bogoliubov-de Gennes Hamiltonian, the ground state of the mean-field Hamiltonian is:
\bea
|\psi_\text{MF} \rangle = \prod_{(a, \mathbf{k})} \left( u_{a, \mathbf{k}}-v_{a,\mathbf{k}} d_{a,\mathbf{k}}^\dagger c_{a, -\mathbf{k}}^\dagger \right) |0\rangle
\eea
where the coefficients $u_{a, \mathbf{k}}, v_{a, \mathbf{k}}$ satisfy $v_{a, \mathbf{k}} / u_{a, \mathbf{k}} = (\sqrt{\epsilon_{a, \mathbf{k}}^2 + |\Delta|^2} - \epsilon_{a, \mathbf{k}} ) / \Delta$ and $u_{a, \mathbf{k}} ^ 2 + v_{a, \mathbf{k}} ^ 2 = 1$.
By tracing out the $d$-fermions, we obtain the density matrix of layer 1 for the $c$-fermions:
\bea\notag
\rho_{c} &= \prod_{(a, \mathbf{k})} \left( |u_{a, \mathbf{k}}|^2  + (|v_{a, \mathbf{k}}|^2 -|u_{a, \mathbf{k}}|^2) 
c_{a,\mathbf{k}}^\dagger c_{a, \mathbf{k}}\right)
\\
&=\sum_{\{(a, \mathbf{k})\}} p_{\{(a, \mathbf{k})\}}\dket{\{(a, \mathbf{k})\}}\dbra{\{(a, \mathbf{k})\}},
\eea
where $p_{\{(a,\mathbf{k})\}}=\prod_{(a,\mathbf{k})\in\{(a,\mathbf{k})\}}|v_{a,\mathbf{k}}|^2\prod_{(a,\mathbf{k}^{\prime})\not\in \{(a,\mathbf{k})\}}|u_{a,\mathbf{k}^{\prime}}|^2$, and 
$\dket{\{(a,\mathbf{k})\}}$ represents a Fock state such that all $c$-fermions with momentum and band index (not) in  $\{(a,\mathbf{k})\}$ is occupied (unoccupied). 
Since the mean-field ground state $\dket{\psi_\text{MF}}$ preserves the diagonal $U(1)$ subgroup (\ref{SC:weak symmetry}):
\bea
e^{\imth\theta(\hat N_c-\hat N_d)}\dket{\psi_\text{MF}}=\dket{\psi_\text{MF}},~~~\forall~\theta.
\eea
the density matrix $\rho_c$ preserves weak $U(1)$ symmetry:
\bea
\rho_{c} = e^{i \hat{N}_{c} \theta} \rho_{c} e^{-i \hat{N}_{c} \theta},
\eea
while the strong $U(1)_c\times U(1)_d$ symmetry is explicitly broken. Here, the order parameter corresponding to the superconductivity in the bilayer system is simply the inter-layer Cooper pair $\sum_{\alpha}\xi_{1,\alpha,i}\xi_{2,\alpha,i}=O_{A,i}\cdot O^{\ast}_{B,i}$, so $O^{\alpha}_{A,i}=\xi_{1,\alpha,i}$ and $O^{\alpha \ast}_{B,i}=\xi_{2,\alpha,i}$. 

In the following, we will calculate the four physical quantities listed in Table~\ref{tab:swssb} to verify that $\rho_{c}$ belongs to the scenario where the strong symmetry is explicitly broken while the weak symmetry is preserved. First, the superconductivity in the bilayer system is a long range order,  and thus $\lim_{|i-j|\rightarrow \infty}C_{0}(i,j)\neq 0$, which can be explicitly verfied as following:
\begin{eqnarray}
&&\sum_{\alpha\beta}\operatorname{Tr}(\sqrt{\rho_{c}}\xi_{1,\alpha,i}\xi_{1,\beta,j}^{\dag}\sqrt{\rho_{c}}\xi_{1,\beta,j}\xi_{1,\alpha,i}^{\dag})\nonumber
\\
&&=\sum_{\alpha\beta}\sum_{\{a,\mathbf{k})\},\{(a',\mathbf{k}^{\prime})\}}\sqrt{p_{\{(a,\mathbf{k})\}}p_{\{(a',\mathbf{k}^{\prime})\}}}|\dbra{\{(a,\mathbf{k})\}}\xi_{1,\alpha,i}\xi^{\dag}_{1,\beta,j}\dket{\{(a',\mathbf{k}^{\prime})\}}|^2 \nonumber
\\
&&=\frac{1}{N^2_{uc}}\sum_{\alpha\beta}\sum_{(a_1,\mathbf{k}_{1})}\sum_{(a_2,\mathbf{k}_{2})} \left( |v_{a_1,\mathbf{k}_1}u_{a_1,\mathbf{k}_1}v_{a_2,\mathbf{k}_2}u_{a_2,\mathbf{k}_2}||U_{\mathbf{k}_1, a_1\alpha}U^{\ast}_{\mathbf{k}_2, a_2\beta}|^2 \right.\nonumber
\\
&& \left. + |u_{a_1,\mathbf{k}_1}u_{a_2,\mathbf{k}_2} |^2 e^{i(\mathbf{k}_{1}-\mathbf{k}_{2})\cdot(\mathbf{x}_{i}- \mathbf{x}_{j})} U_{\mathbf{k}_1, a_1\alpha}U^{\ast}_{\mathbf{k}_1, a_1\beta} U_{\mathbf{k}_2, a_2\beta}U^{\ast}_{\mathbf{k}_2, a_2\alpha} \right).
\end{eqnarray}
Taking the limit $|\mathbf{x}_{i}-\mathbf{x}_{j}|\rightarrow \infty$ leading to
\bea
&\lim_{|\mathbf{x}_{i}-\mathbf{x}_{j}|\rightarrow \infty} \frac{1}{N_{uc}^2}\sum_{\alpha\beta}\operatorname{Tr}(\sqrt{\rho_{c}}\xi_{1,\alpha,i}\xi_{1,\beta,j}^{\dag}\sqrt{\rho_{c}}\xi_{1,\beta,j}\xi_{1,\alpha,i}^{\dag})\nonumber
\\
&=\frac{4}{N^{2}_{uc}}\sum_{\mathbf{k}_{1}}|v_{1,\mathbf{k}_1}u_{1,\mathbf{k}_1}|\sum_{\mathbf{k}_{2}}|v_{1,\mathbf{k}_2}u_{1,\mathbf{k}_2}|\sim O(1),
\eea
where we have used the fact that $|v_{1,\mathbf{k}}||u_{1,\mathbf{k}}|=|v_{2,\mathbf{k}}||u_{2,\mathbf{k}}|$.

Next, we calculate $\sum_{\alpha}\operatorname{Tr}(\sqrt{\rho_{c}}\xi_{1,\alpha,i}\sqrt{\rho_{c}}\xi_{1,\alpha,i}^{\dag})$:
\bea
&\sum_{\alpha}\operatorname{Tr}(\sqrt{\rho_{c}}\xi_{1,\alpha,i}\sqrt{\rho_{c}}\xi_{1,\alpha,i}^{\dag})\nonumber
\\
&=\sum_{\alpha}\sum_{\{a,\mathbf{k})\},\{(a',\mathbf{k}^{\prime})\}}\sqrt{p_{\{(a,\mathbf{k})\}}p_{\{(a',\mathbf{k}^{\prime})\}}}|\dbra{\{(a,\mathbf{k})\}}\xi_{1,\alpha,i}\dket{\{(a',\mathbf{k}^{\prime})\}}|^2 \nonumber
\\
&=\frac{1}{N_{uc}}\sum_{\alpha}\sum_{(a_1,\mathbf{k}_{1})}|v_{a_1,\mathbf{k}_1}u_{a_1,\mathbf{k}_1}||U_{a_{1}\alpha}^{\ast}|^2\nonumber
\\
&=\frac{2}{N_{uc}}\sum_{\mathbf{k}_{1}}|v_{1,\mathbf{k}_1}u_{1,\mathbf{k}_1}|\sim O(1)
\eea
Similarly, we can prove that $\lim_{|i-j|\rightarrow \infty}\langle \sum_{\alpha}\xi_{1,\alpha,i}\xi_{1,\alpha,j}^{\dag}\rangle$ is vanishing
through
\bea
&\sum_{\alpha}\operatorname{Tr}[\rho_{c}\xi_{1,\alpha,i}\xi_{1,\alpha,j}^{\dag}]\nonumber
\\
&=\frac{1}{N_{uc}}\sum_{(a_1,\mathbf{k}_1)}e^{i\mathbf{k}_{1}\cdot(\mathbf{x}_{i}-\mathbf{x}_{j})}|u_{a_1,\mathbf{k}_1}|^2=0,
\eea
The last step is because that $|u_{a_1,\mathbf{k}_1}|^2$ is a smooth function of $\mathbf{k}_1$, and its Fourier transformation vanishes exponentially when $|\mathbf{x}_{i}-\mathbf{x}_{j}|$ goes to infinity. Lastly, $\langle \xi_{1,\alpha,i}\rangle=0$ due to the weak symmetry of $\rho_{c}$. These calculations match the third row in Table~\ref{tab:swssb}, which demonstrates that mixed state $\rho_{c}$ corresponds to a bilayer superconducting mean-field ground state, which explicitly breaks the strong $U(1)$ symmetry down to a weak one.

Finally, we notice that the interlayer superconductivity in model (\ref{sc:interacting}) and (\ref{sc:mean field}) can be mapped to a bilayer exciton condensate\cite{Eisenstein2014}, by implementing a particle-hole transformation on $d$ fermions in layer $2$. Note that symmetry $\bst$ requires that the kinetic energy of electrons in the two layers must have opposite signs in the bilayer exciton condensate.

\section{Bilayer construction for mixed state phenomena with strong and weak symmetries}\label{sec:realizations}

\subsection{Realizing decoherence of $\rho_A$ as quench dynamics of $\dket{\psi}$}\label{sec:decoherence}

Recently, lots of efforts have been devoted to study the effects of decoherence on a pure state, and mixed state phase transitions induced by decoherence\cite{Dennis2002,Fan2024,Bao2023,Lee2023,Sala2024a,Sala2024b,Chirame2024,Chen2024,Eckstein2024,Chen2024a,Ma2024a,Lessa2024,Lavasani2024,Hwang2024,Wang2023,Ellison2024}. Most of previous efforts focused on decoherence in terms of local errors, described by the following quantum channel:
\bea\label{decoherence:local}
&\rho\rightarrow\mathcal{E}[\rho]=\prod_j\mathcal{E}_j[\rho],\\
\notag&\mathcal{E}_j[\rho]\equiv(1-p)\rho+p O_j\rho O_j^\dagger.
\eea
where $O_j$ is a local unitary operator. Increasing the error rate $p\in[0,1]$ can induce phase transitions on mixed state $\rho$. A natural question is, how to use a bilayer pure state $\dket\psi$ to understand decoherence on a mixed state $\rho$ and decoherence induced transition in the bilayer construction? 

Note that the bilayer pure state $\dket\psi$ in (\ref{schmidt:psi}) is nothing but a specific purification for mixed state $\rho_A$ in (\ref{schmidt:rho}). It is well known\cite{Nielsen_Chuang_2010} that a quantum channel for mixed state $\rho_A$ is mapped to a unitary evolution on pure state $\dket\psi$ through the purification. Therefore quite generally, the above quantum channel (\ref{decoherence:local}) on mixed state $\rho_A$ can be mapped to unitary time evolution on state $\dket\psi$ under Hamiltonian $\sum_j h_j$:
\bea\label{unitary evolution}
\dket\psi\rightarrow U(t)\dket\psi=e^{-\imth t\sum_jh_j}\dket\psi
\eea
Specifically in the state-sum representation\cite{Nielsen_Chuang_2010}, if we consider a generic quantum channel acting on $\rho_A=\dket{\psi_A}\dbra{\psi_A}$ in system layer $A$, 
\bea
\mathcal{{E}}[\rho_A]=\sum_kE_k\rho_AE_k^\dagger,~~~\sum_kE^\dagger_kE_k=\hat 1.
\eea
it can be purified in the bilayer by unitary operator $U$ acting on the bilayer pure state $\dket{\psi}=\dket{\psi_A}\otimes\dket{\psi_B^\ast}$:
\bea
E_k=\expval{e_k,B|U|\psi_B^\ast}
\eea
where $\{\dket{e_k,B}\}$ is a set of complete orthonormal basis for environment layer $B$. Therefore decohering a pure state $\dket{\psi_A}$ by a local quantum channel (\ref{decoherence:local}) is equivalent to unitary evolution (\ref{unitary evolution}) on bilayer pure state $\dket{\psi}=\dket{\psi_A}\otimes\dket{\psi_B^\ast}$. 

To demonstrate this fact, we examine a specific example of local errors (\ref{decoherence:local}), where $O_j=X_j$ and $X_{j}$ is the $x$-Pauli matrix for the $j$-th qubit. This example has been extensively studied in a system of qubits\cite{Dennis2002,Fan2024,Bao2023,Chen2024a,Sala2024b} where the initial pure state $\dket{\psi_A}$ (and thus $\dket{\psi_B^\ast}$) is a two-dimensional topological order, such as a toric code\cite{Dennis2002,Fan2024,Bao2023,Chen2024a} or a $D_4$ topological order\cite{Sala2024b}.  For concreteness, consider $\dket{\psi_A}$ to be the ground state of the toric code on a disc with $e$-type open boundaries\cite{Bravyi1998}. Straightforward calculations reveal that the product of $X_{j}$ over the \textit{horizontal} links, denoted as $\prod^{\prime}_{j}$, satisfies
\bea
\expval{\psi_A|\textstyle\prod^{\prime}_{j\in s}X_j|\psi_A}=0,
\eea
for all possible sets $s$. Consequently,
the unitary time evolution of the bilayer state $\dket{\psi}=\dket{\psi_A}\otimes\dket{\psi_B^\ast}$ under the following bilayer quench Hamiltonian
\bea
\label{quench ham}
\hat H_f= \textstyle \sum^{\prime}_jX_{A,j}\otimes X_{B,j}
\eea
produces the same layer-$A$ density matrix as acting local quantum channel on $\rho_A=\dket{\psi_A}\dbra{\psi_A}$, i.e.,
\bea
&\trace{_B\big[e^{-\imth\hat H_ft}\dket{\psi}\dbra{\psi}e^{\imth\hat H_ft}\big]}=\prod^{\prime}_j\mathcal{E}_j(\rho_A),\\
&\mathcal{E}_j(\rho)=\cos^2t \cdot\rho+\sin^2t~X_j\rho X_j\notag,
\eea
where $\sum^{\prime}_j$ and $\prod^{\prime}_j$ indicate the summation and product over qubits sitting on the \textit{horizontal} bonds. Note that Hamiltonian $\hat H_f$ preserves the anti-unitary layer-exchange symmetry $\bst$. In contrast to previous studies where quantum channels act on all qubits, here we restrict the channels to qubits residing on horizontal bonds. Despite this restriction, the toric code ground state $\dket{\psi_A}$ still decoheres into the mixed classical toric code state under $\prod^{\prime}_j\mathcal{E}_j$.

In this example, since the initial pure state $\rho_A=\dket{\psi_A}\dbra{\psi_A}$ is the ground state of a Pauli stablizer model\cite{KITAEV2003} $\hat H_i$, time evolution under a different Hamiltonian $\hat H_f$ is nothing but a quantum quench\cite{Mitra2018} applied to the bilayer. Therefore, it is clear that the dynamics of $\rho_A$ generated by decoherence exactly corresponds to dynamics of the bilayer pure state $\dket\psi$ generated by a quantum quench. The decoherence induced phase transitions on mixed state $\rho_A$\cite{Dennis2002,Fan2024,Lee2023,Chen2024a,Chirame2024,Lessa2024} can therefore be realized by certain dynamical phase transitions\cite{Heyl2018} in bilayer pure state $\dket\psi$ induced by the quantum quench. 

We have established above that by tracing out layer $B$, the reduced density matrix of layer $A$ in the bilayer pure state $U(t)\dket\psi$ in (\ref{unitary evolution}) is the same as the decohered density matrix $\mathcal{E}[\rho_A]$ in (\ref{decoherence:local}). It is important to note the following: (i) Whether an arbitrary decoherence through local quantum channels can be represented as quench dynamics under \textit{local} bilayer Hamiltonians remains an open question. (ii) Even though the initial state $\dket{\psi}=\dket{\psi_A}\otimes\dket{\psi_B^\ast}$ purifies $\rho_A=\dket{\psi_A}\dbra{\psi_A}$ canonically, the final state $U(t)\dket\psi$ is not the canonical purification of $\mathcal{E}[\rho_A]$, for it violates both $\bst$ symmetry and the non-negativity constraint. As a result, the mixed state phase transition in $\mathcal{E}[\rho_A]$ may not correspond to a usual thermodynamic phase transition in bilayer state $U(t)\dket\psi$. How to diagnose the decoherence-induced mixed state transition in bilayer pure state $U(t)\dket\psi$ remains an open question for future works.

\subsection{Topology of mixed states with strong and weak symmetries}\label{sec:mSPT+mTO}

Due to the one-to-one correspondence between mixed states and non-negative $\bst$-symmetric bilayer pure states, the topology of mixed states can be studied by examining the topology of their corresponding bilayer pure states. Since the quantum phases of pure states have been thoroughly studied in the past few decades\cite{KITAEV2006,Wen2017,Zeng2019}, the above correspondence naturally defines the equivalence relation between mixed state phases: {\bf two mixed states belong to the same mixed state phase if and only if their corresponding $\bst$-symmetric non-negative bilayer pure states belong to the same phase.} Below we follow this approach to analyze the classification and properties of symmetry protected topological (SPT) mixed states in section \ref{sec:mSPT}, and mixed state topological orders in section \ref{sec:mixed TO}. 

It is important to note that the topology of mixed states is not the same as the topology of $\bst$-symmetric bilayer pure states without the non-negativity constraint. Specifically, the non-negativity constraint $\psi\geq0$ of wavefunction (\ref{bilayer pure state}) will change the topology of $\bst$-symmetric pure states in two important ways: (1) Certain $\bst$-symmetric bilayer pure states do not correspond to any mixed state, if they violate the constraint of a non-negative wavefunction. One example is the absence of weak-symmetry protected topological mixed states as shown in Ref.\cite{Ma2024a}. (2) One phase of $\bst$-symmetric bilayer pure states can correspond to multiple phases of mixed states, in the presence of non-negativity constraint. This is demonstrated by an example in section \ref{sec:mixed TO}, where two distinct mixed state topological orders obtained from decohering toric code\cite{Dennis2002,Fan2024,Bao2023,Lee2023,Wang2023} belong to the same $\bst$-enriched toric code phase in their bilayer counterparts. Nevertheless, the bilayer construction provides many insights for the classification, construction and realization of mixed state phases as we will discuss below. 


Before proceeding, we briefly discuss the important issue of how to define phases of mixed states\cite{Lin2021b,deGroot2022,Ma2023,Ma2023a,Sang2024,Sang2024a}. In the bilayer construction, it is natural to define mixed state phases in terms of their bilayer counterparts of pure state phases, under the extra constraints of $\bst$-symmetry and non-negativity. For example, in the bilayer construction, for the symmetry protected topological mixed states and mixed state topological orders discussed later in this section, they are defined as distinct gapped phases of $\bst$-symmetric non-negative bilayer pure states. They are classified by $\pi_0(\mathcal{S}^\bst_\text{NN})$, where $\mathcal{S}^\bst_\text{NN}$ is the space of short-range-correlated $\bst$-symmetric non-negative bilayer pure states, which preserve certain symmetries in the case of classifying symmetry protected and symmetry enriched topological states. Two mixed states belong to the same ``gapped'' phase if and only if their bilayer counterparts are continuously connected to each other by a smooth path in space $\mathcal{S}^\bst_\text{NN}$. The above definition adopted in this work, is \emph{not the same} as the definitions used in many previous works: e.g. Ref.\cite{Ma2023a} defines two mixed states to belong to the same phase if and only if they are two-way connected by finite-depth local quantum channels, and Ref.\cite{Sang2024a} proposes a finite Markov length as the definition for ``gapped'' mixed states. The difference in the definitions can lead to different classification of mixed states: e.g. the decohered toric code\cite{Dennis2002,Fan2024} discussed later in Eq.(\ref{decohered toric code:strong m}) belong to the same trivial phase as the product state by the definition of finite-depth local quantum channels\cite{Ma2023a}, however in our bilayer definition they are clearly distinct phases of mixed states. There are indeed important physical differences between the trivial product and the decohered toric code, such as the existence of classical memory in decohered toric code\cite{Dennis2002,Fan2024,Wang2023}. Compared to other proposals, we believe that the bilayer construction defines a natural equivalence relation between mixed states, and it naturally defines the topology of mixed states via $\pi_0(\mathcal{S}^\bst_\text{NN})$. Nonetheless, how to reconcile the different definitions of mixed state phases remains an interesting open question for future studies.

\subsubsection{Symmetry protected topological mixed states}\label{sec:mSPT}

The bilayer construction naturally provides a framework to construct and to realize mixed states $\rho_A$ with various strong and weak symmetries, in terms of the purified bilayer pure state $\dket\psi$. Consider a general symmetry group $G$ including both strong and weak symmetries in the bilayer (not including $\bst$), where each weak symmetry operation $g$ has the diagonal form of $g_Ag_B$, and each strong symmetry operation $h$ implies both $h_A$ and $h_B$ are preserved. This is a consequence of anti-unitary layer-exchange symmetry $\bst$ due to property described in Eq.~\eqref{T exchange}. Symmetry group $G$ can be understood as a subgroup of a larger product group $\mathcal{G}_A\times \mathcal{G}_B$, and can be obtained by breaking some strong symmetries down to weak symmetries in $\mathcal{G}_A\times \mathcal{G}_B$. Clearly all strong symmetries in $G$ form a normal subgroup $S=S_A\times S_B$ of $G$, since 
\bea\notag
(g_Ag_B)h_A(g_Ag_B)^{-1}\in S_A,~\forall~h_A\in S_A.
\eea
All the weak symmetries also form a subgroup $W=\{g_Ag_B\in G\}$ because $W$ is nothing but the diagonal subgroup of $\mathcal{G}_A\times \mathcal{G}_B$. 
Meanwhile, those weak symmetry elements $\{g_Ag_B|g_Ag_B\in G,g_A\notin S_A\}$ in $G$ that do not belong to strong symmetry subgroup $S$, together with the identity element, form a set isomorphic to quotient group $G/S$, giving rise to the following short exact sequence 
\bea\label{short exact sequence}
1\rightarrow S\rightarrow G\rightarrow G/S\rightarrow 1
\eea
which describes the interplay of strong symmetries in $S$ and weak symmetries in $G/S$. In other words, symmetry group $G$ is an extension of weak symmetry $G/S$ by strong symmetry $S$. So far we have not taken the anti-unitary layer-exchange symmetry $\bst$ into account. After considering $\bst$ symmetry, the full symmetry group of the bilayer construction is given by
\bea\label{sym:bilayer}
G_\text{bilayer}=G\rtimes Z_2^\bst
\eea
where the $\bst$-action on strong and weak symmetries in $G$ are defined by (\ref{T exchange}). 

After elucidating the full symmetry group $G_\text{bilayer}$ of the bilayer system, one can classify and construct symmetry protected topological (SPT) states in the bilayer. Since mixed states with symmetry $G$ are in one-to-one correspondence with non-negative bilayer pure state with symmetry $G_\text{bilayer}$, the topology of mixed state $\rho_A$ is completely determined by the topology of bilayer pure state $\dket\psi$, under the important constraint that pure state $\dket\psi$ must be non-negative. Therefore studying the topology of mixed states becomes a problem of constrained topology: the topology of $G_\text{bilayer}$-symmetric bilayer pure states fully determine the topology of $G$-symmetric mixed states, in the sense that topologically distinct bilayer pure states must correspond to topologically distinct mixed states. However, not all bilayer pure states have a mixed counterpart due to the non-negativity requirement, and those pure states with negative wavefunctions must be discarded. Since the fixed-point wavefunctions for pure $G_\text{bilayer}$-SPT states can be written down explicitly\cite{Chen2013}, it is straightforward to check which pure SPT phases have non-negative wavefunctions, and practices following this strategy can lead to a partial classification of mixed SPT states. The next step is to understand whether multiple mixed $G$-SPT phases can emerge out of the same pure $G_\text{bilayer}$-SPT phase due to the non-negativity constraint. 

While a complete classification of mixed SPT states in this bilayer construction is beyond the scope of this work, here we make some brief comments on the classification problem, in connection to previous works on this subject\cite{Ma2023,Ma2023a,Ma2024a,You2024,Sun2024}. For the same mixed state $\rho$ in (\ref{general mixed state}), the bilayer purification $\dket\psi$ in (\ref{bilayer purification}) used in this work has the same symmetry $G_\text{bilayer}$ as the Choi state $\dket{\psi_\text{Choi}}$ in (\ref{choi state}). Therefore the bilayer construction and the Choi-Jamiołkowski isomorphism leads to the same classification for mixed SPTs. Therefore, in terms of classification, our analysis is in parallel with the Choi states discussed in Ref.\cite{Ma2024a}. The anti-unitary layer-exchange symmetry $\bst$ is called ``modular conjugation'' in Ref.\cite{Ma2024a}.

First, a large class of bilayer pure $G_\text{bilayer}$-SPTs in $d$-spatial dimension are classified by the group cohomology\cite{Chen2013} $H^{d+1}(G_\text{bilayer},U(1))$. As pointed out by Ref.\cite{Ma2024a}, the non-negativity requirement rules out pure bilayer SPTs protected by anti-unitary symmetry $\bst$. Therefore we can focus on $G$-SPTs in the bilayer construction, classified by $H^{d+1}(G,U(1))$. Due to the group extension structure (\ref{short exact sequence}) for group $G$, the cohomology group $H^{d+1}(G,U(1))$ can be computed using spectral sequences\cite{Atiyah1972,Brown1982,Adem1994}. The full classification for mixed $G$-SPTs has been considered in Ref.\cite{Ma2024a} by excluding all bilayer $G_\text{bilayer}$-SPTs that violate the non-negative constraint, and the obtained spectral sequence is consistent with those proposed earlier in Ref.\cite{Ma2023a}. 

One obvious observation is that without weak symmetries, mixed SPTs have the same structure as pure (monolayer) SPTs. In the absence of weak symmetries, $G=S=S_A\times S_B$ and $G_\text{bilayer}=(S_A\times S_B)\rtimes Z_2^\bst$. Note that with strong symmetry only, all states $\dket{k,A}$ in (\ref{schmidt:rho}) share the same quantum numbers of strong symmetry, and hence a generic mixed state $\rho_A$ in (\ref{schmidt:rho}) can be continuously deformed into a pure state $\rho_A=\dket{0,A}\dbra{0,A}$, corresponding to a $\bst$-symmetric bilayer product state $\dket\psi=\dket{0,A}\otimes\dket{0,B}^\ast$. In the presence of strong symmetry $S_A$, short-range-entangled pure state $\dket{0,A}$ are classified by a topological index $\nu$, and can be continuously deformed into a fixed-point wavefunction $\dket{\psi_\nu,A}$. As a result, any mixed state with only strong symmetry $S=S_A\times S_B$ is continuosly connected to the following $\bst$-symmetric non-negative bilayer SPTs:
\bea
\dket\psi_{G=S_A\times S_B}=\dket{\psi_{\nu},A}\otimes\dket{\psi_{\nu}^\ast\simeq\psi_{-\nu},B}
\eea
where $\dket{\psi_\nu,A}$ is a pure $S_A$-SPT with topological index $\nu$. They are hence fully classified by the (monolayer) pure $S_A$-SPT $\dket{\psi_\nu}$. Therefore, {\bf the presence of both weak and strong symmetries are necessary in order for mixed SPTs to have a richer structure than pure SPTs}.

It is instructive to consider one example, where the bilayer system initially have a $Z_4$ strong symmetry generated by operation $g_A$ and $g_B$ satisfying $g_A^4=g_B^4=1$. After breaking $g_A$ and $g_B$ but preserving the weak symmetry $g_Ag_B$ and strong symmetry $g_A^2$ and $g_B^2$, similar to a nematic order that breaks 4-fold rotation down to 2-fold, we arrive at the following bilayer group structure 
\bea
G=\{(g_Ag_B)^ng_A^{2m}|0\leq n\leq 3,0\leq m\leq 1\}\simeq Z_4\times Z_2
\eea
with a normal subgroup of strong symmetries
\bea
S=\{g_A^{2a}g_B^{2b}|0\leq a,b\leq 1\}\simeq Z^A_2\times Z^B_2
\eea
and $G/S=Z_2$. This is a simple example where $G$ is not a direct product of strong symmetry $S$ and weak symmetry $G/S$. Instead, the short exact sequence (\ref{short exact sequence}) is not splitting, featuring a nontrivial 2-cocycle $\omega\in H^2(Z_2,Z_2\times Z_2)$. According to Kunneth formula, the cohomology group $H^2(G,U(1))$ in this case is given by
\bea
H^2(Z_4\times Z_2,U(1))=H^1(Z_4,H^1(Z_2,U(1)))=\mathbb{Z}_2
\eea
The physical meaning of this bilayer pure SPT is to decorate the domain wall\cite{Chen2014,Ma2023a} of weak symmetry $g_Ag_B$ with the charge of strong symmetry $g_A^2$ (or $g_B^2$). In previous studies of mixed SPTs\cite{Ma2023a}, this state was known as an intrinsic mixed SPT, for it does not have a monolayer gapped pure SPT counterpart. Instead, its counterpart as a monolayer pure state is an intrinsically gapless SPT\cite{Thorngren2021} with $Z_4$ symmetry. Nevertheless, in our bilayer construction, this intrinsic mixed SPT corresponds to nothing but a bilayer pure $G$-SPT state. 

It is straightforward to write down an exactly solvable $\bst$-symmetric bilayer Hamiltonian for this bilayer SPT state. We consider $Z_4$ generalization of Pauli matrices:
\bea
X_j^4=Z_j^4=1,~~~X_jZ_j=\imth Z_jX_j,~~~\forall~j.
\eea
The solvable bilayer Hamiltonian is given by\cite{Geraedts2014,Sun2024}
\bea
&\notag\hat H_{Z_4\times Z_2}=-J_\text{SB}\sum_j(Z_j^AZ_j^B)^2-J\sum_j(X_j^AX_j^B)^2\\
\notag&-1/2\sum_j\big[(Z_j^AX_j^{A}Z_{j+1}^A)^2+A\leftrightarrow B\big]\\
&-1/2\sum_j\big[Z_{j-1}^A(Z_{j-1}^B)^{-1}X_j^AX_j^BZ_j^A(Z_j^B)^{-1}+h.c.\big].
\eea
Note that the $J_\text{SB}$ term explicitly breaks the $Z_4^A\times Z_4^B$ symmetry of other terms down to group $G\simeq Z_4\times Z_2$. Its strong symmetry group $S=Z_2^A\times Z_2^B$ is generated by 
\bea
g_A^2=\prod_j(X_j^A)^2,~~g_B^2=\prod_j(X_j^B)^2.
\eea
while the weak symmetry is preserved as
\bea
g_Ag_B=\prod_j(X_j^AX_j^B).
\eea
Details regarding the construction of this model and the demonstration of the non-negativity of its ground-state wavefunction can be found in Appendix~\ref{app:SPT}.

\subsubsection{Mixed state topological orders}\label{sec:mixed TO}

Now that the topology of mixed states is fully determined by the topology of bilayer non-negative pure states with an extra $\bst$ symmetry, all mixed state topological orders\cite{Wang2023,Soha2024,Zhang2024,Ellison2024} can be realized as intrinsic topological orders in bilayer pure state $\dket\psi$. Specifically in two spatial dimensions, the presence of anti-unitary symmetry $\bst$ indicates that bilayer state $\dket\psi$ can only exhibit non-chiral topological orders, which are fully classified by Levin-Wen string-net models\cite{Levin2005,Lin2021a}. By examining the bilayer fixed-point wavefunctions of $\bst$-symmetric Levin-Wen models\cite{Levin2005,Lin2021a}, one can check the non-negativity of $\bst$-symmetric string-net ground states and rule out those with negative wavefunctions. For example, while the toric code topological order in the bilayer pure state has a positive ``loop soup'' wavefunction in the string-net picture, the wavefunction for the double semion topological order does have negative components\cite{Levin2005}. Meanwhile, the non-negativity constraint can also give rise to multiple mixed state topological orders belonging to the same bilayer pure state topological order, as we will demonstrate later in bilayer Hamiltonians (\ref{decohered toric code:strong m}) and (\ref{decohered toric code:strong f}). Progress towards a complete classification of mixed state topological orders can be achieved following the aforementioned strategy, which is beyond the scope of this work. 

Here, we take a different point of view to examine two-dimensional mixed state topological orders, as spontaneous breaking of one-form symmetries\cite{Gaiotto2015,Zhang2024} (in the case of Abelian anyon strings) or categorical symmetries\cite{Ji2020,Kong2020} (in the case of non-Abelian anyon strings). This picture can also be generalized to higher spatial dimensions in terms of spontaneously breaking higher form symmetries\cite{Gaiotto2015,Ji2020,Kong2020}. Similar to long-range correlation of local order parameters in Landau-type 0-form symmetry breaking phenomena, topological order can be diagnosed as long-range correlation of extended non-local order parameters, such as one-dimensional string operators (associated with 1-form symmetry breaking) along non-contractible loops in two-dimensional topological orders\cite{Gaiotto2015}. This analogy naturally motivates the concept of strong and weak higher-form\cite{Wang2023,Zhang2024} (and categorical) symmetries, via simply replacing the local order parameters $O_j$ in Table \ref{tab:swssb} by string and other extended operators in various correlators, as criteria for strong and weak higher-form (and category) symmetry breakings.  

Below we focus on two spatial dimensions. Recall that in order for mixed SPTs to have new structures beyond pure SPTs, it is necessary for the system to preserve both strong and weak symmetries. In the bilayer construction, a strong 0-form (global) symmetry is generated by $g_A$ and $g_B$, while a weak 0-form symmetry is generated by a diagonal element $g_Ag_B$. In analogy to 0-form symmetry, in the case of intrinsic topological orders in two dimensions, a strong 1-form symmetry is generated by intra-layer string operators $\mathcal{L}_{A,g}$ and $\mathcal{L}_{B,\bar{g}}$, while a weak 1-form symmetry is generated by inter-layer string operator $\mathcal{L}_{A,g}\otimes\mathcal{L}_{B,\bar g}$. Similar to the interlayer order parameter $O_{A,i}\cdot (O_{B,i})^{\ast}$ for spontaneous breaking of strong 0-form symmetry down to weak 0-form symmetry, the string order parameter for a strong 1-form symmetry breaking (down to weak 1-form symmetry) should be the product $\mathcal{W}_{A,a}\cdot \mathcal{W}_{B,\bar a}$ of an anyon-$a$ string $\mathcal{W}_{A,a}$ in layer $A$ and an anyon-$\bar a$ string $\mathcal{W}_{B,\bar a}$ in layer $B$\cite{Wang2023,Zhang2024}, where anyon $a$ has nontrivial braiding statistics with anyon $g$ in strong 1-form symmetry generator $\mathcal{L}_{A,g}$. The spontaneous breaking of a strong 1-form symmetry is therefore characterized by long-range correlation of interlayer-product string operator $\mathcal{W}_{A,a}\cdot \mathcal{W}_{B,\bar a}$, but not of $\mathcal{W}_{A,a}$ or $\mathcal{W}_{B,\bar a}$, implying the condensation of composite anyon $b_A\otimes\bar{b}_B$ where anyon $b$ has nontrivial braiding statistics with aforementioned anyon $a$. Condensation of $b_A\otimes\bar{b}_B$ confines anyon $a_A$ and $\bar{a}_B$, but leaves their composite $a_A\otimes\bar{a}_B$ as deconfined excitations in the bilayer. On the other hand, spontaneous breaking of a weak 1-form symmetry $\mathcal{L}_{A,g}\otimes\mathcal{L}_{B,\bar g}$ is characterized by long-range correlation of both $\mathcal{W}_{A,a}$ and $\mathcal{W}_{B,\bar a}$\cite{Wang2023,Zhang2024}, suggesting deconfined $a$ anyons in layer $A$ and $\bar{a}$ anyons in layer $B$. 

In the bilayer construction, it is straightforward to understand the mixed state topological orders obtained by decohering a pure state topological order\cite{Dennis2002,Fan2024,Bao2023,Lee2023,Wang2023,Soha2024,Ellison2024,Sala2024b,Hwang2024}, and the interplay of weak and strong 1-form symmetries\cite{Wang2023,Zhang2024} therein. Based on previous discussions in section \ref{sec:decoherence}, decohering a pure state topological order $\dket{\psi_A}$ is equivalent to applying a quantum quench to the ``doubled'' pure state $\dket\psi=\dket{\psi_A}\otimes\dket{\psi_{B}^\ast}$ in the bilayer construction. If the initial (monolayer) topological order is captured by a unitary modular tensor category (UMTC) $\mathcal{C}$\cite{KITAEV2006}, the bilayer state corresponds to a doubled topological order described by UMTC $\mathcal{C}_A\otimes\mathcal{\bar C}_B$. The quench Hamiltonian (\ref{quench ham}) generally drives an anyon-condensation transition\cite{KONG2014,Burnell2018} in the bilayer pure state $\dket\psi$, by condensing a composite anyon $\lambda_A\otimes\bar{\lambda}_B$. After the anyon condensation transition, the bilayer pure state $\dket{\psi^\prime}$ exhibit a new $\bst$-symmetric topological order $\mathcal{D}$, which have both intra-layer anyon strings related to strong 1-form symmetry and inter-layer anyon strings related to weak 1-form symmetry. The mixed state decohered topological order is nothing but density matrix $\rho_A^\prime=\trace{_B(\dket{\psi^\prime}\dbra{\psi^\prime})}$, and can be diagnosed by the correlators in Table \ref{tab:swssb} for intra- and inter-layer string order parameters. 

A peculiar feature of mixed state topological order $\rho_A^\prime$ is non-modularity\cite{Wang2023,Soha2024,Ellison2024}, meaning the existence of ``invisible'' anyons that cannot be detected by other anyons via braiding\cite{KITAEV2006,Kong2014a}. This point is also straightforward to understand from the bilayer construction. Since the bilayer topological order $\dket{\psi^\prime}$ has both inter-layer anyon strings and intra-layer anyon strings, viewing $\rho_A^\prime$ as a mixed state of layer $A$, only the anyons with intra-layer strings are accessible, while the coherence of inter-layer anyons are lost after tracing out layer $B$. In order for $\rho_A^\prime$ to be an intrinsic mixed state topological order beyond pure state counterparts, the intra-layer-$A$ anyons must have nontrivial braiding with some interlayer anyons, otherwise they can factor out as a pure state topological order within layer $A$. This means some layer-$A$ anyon may have a trivial braiding with other layer-$A$ anyons, hence non-modularity in mixed state $\rho^\prime_A$. 

Below we briefly discuss the mixed state topological orders in two examples. The first example is the well-studied toric code\cite{KITAEV2003} decohered by local error\cite{Dennis2002,Fan2024,Bao2023,Lee2023} $O_j=X_j$ in (\ref{decoherence:local}). In the bilayer construction, the initial state $\dket{\psi}$ as a doubled toric code has the following anyon content
\bea
\{1,e_A,m_A,\epsilon_A=e_A\times m_A\}\otimes\{1,e_B,m_B,\epsilon_B\}
\eea
The quantum quench (\ref{quench ham}) induces condensation of composite anyon $m_A\otimes m_B$, leaving the following deconfined anyons in the condensed phase $\dket{\psi^\prime}$:
\bea
\{1,m_A\sim m_B,e_A\otimes e_B,\epsilon_A\otimes e_B\sim e_A\otimes\epsilon_B\}
\eea
From the viewpoint of mixed state $\rho_A^\prime$, there are only anyons $\{1,m_A\}$ associated with strong 1-form $Z_2$ symmetry, corresponding to a non-modular anyon theory\cite{KITAEV2006}.  By stating a density matrix $\rho_{A}^{\prime}$ has an anyon $m_{A}$, we mean $\rho_{A}^{\prime}$ preserves the 1-form strong symmetry generated by $m_A$-loop, $\mathcal{L}_{A,m}$. 

In the second example, we consider decohering a chiral topological order described by $U(1)_4$ Chern-Simons theory\cite{Wen1995}, also known as $\nu=2$ state in Kitaev's 16-fold way\cite{KITAEV2006}. The corresponding bilayer state has the following anyon content:
\bea
U_A(1)_4\otimes U_B(1)_{-4}=\{1,a_A,a_A^2,a_A^3\}\otimes\{1,\bar{a}_B,\bar{a}_B^2,\bar{a}_B^3\}
\eea
satisfying the $Z_4$ fusion rule $a^4\sim\bar{a}^4\sim1$, with self-braiding statistics $\theta_a=\theta_{\bar{a}}^\ast=e^{\imth\pi/4}$. We consider a local-error-type decoherence that condenses interlayer anyon $a_A^2\otimes\bar{a}_B^2$, which drives a transition into a bilayer toric code state $\dket{\psi^\prime}$ with anyon content
\bea
\{1,a_A\otimes\bar{a}_B\sim e,~a_A\otimes\bar{b}^3_B\sim m,a_A^2\sim\epsilon\}
\eea
From the viewpoint of mixed state $\rho_A^\prime$, the accessible anyon content becomes $\{1,a_A^2\sim\epsilon\}$, leading to another non-modular anyon theory.

It is straightforward to construct bilayer Hamiltonians for mixed state topological orders in the bilayer construction. For example, the bilayer Hamiltonian for the decohered toric code discussed earlier\cite{Dennis2002,Fan2024,Lee2023} is:
\bea\notag
&\hat H^{(m)}_\text{dTC}=-\sum_\text{star}\prod_{j\in\text{star}}(X^A_{j}+X^B_j)\\
&-J\sum_jX_j^AX_j^B-\sum_\text{plaquet}\prod_{j\in\text{plaquet}}Z_j^AZ_j^B.\label{decohered toric code:strong m}
\eea
In the $J\gg1$ limit, we work in the low-energy subspace with $X_j^AX_j^B\equiv1,~\forall~j$, and the above Hamiltonian is nothing but a toric code for the effective qubit defined as:
\bea
\sigma_j^x=X_j^A=X_j^B,~~\sigma_j^z=Z_j^AZ_j^B
\eea
similar to the one-dimensional 2-leg Ising ladder studied in section \ref{sec:example:1d Ising}. Its bilayer ground state $\dket{\psi^{(m)}_\text{dTC}}$ corresponds to mixed state $\rho_{A,\text{dTC}}^{(m)}=\trace{_B(\dket{\psi^{(m)}_{\text{dTC}}}\dbra{\psi^{(m)}_{\text{dTC}}}}$ in layer $A$, which is the decohered toric code. It is straightforward to check that the bilayer ground state is non-negative in the Pauli-$X$ basis:
\bea
\dket{\psi^{(m)}_\text{dTC}}=\prod_\text{plaquet}\frac{1+\prod_{j\in\text{plaquet}}Z_j^A Z_j^B}2\otimes_j\dket{X_j^{A/B}\equiv 1}.
\eea
and $\rho_{A,\text{dTC}}^{(m)}$ is simply a classical summation of all loop configurations (in Pauli-$X$ basis) with equal probability. Clearly there is a strong 1-form $Z_2$ symmetry generated by an intra-layer string operator of $m$ particle:
\bea
\mathcal{L}_{A,m}=\prod_{l\in\text{string}}X_l^A
\eea
and a weak 1-form $Z_2$ symmetry generated by an interlayer string operator of $e$ particle
\bea
\mathcal{L}_{A,e}\otimes\mathcal{L}_{B,e}=\prod_{l\in\text{string}}Z_l^AZ_l^B
\eea
consistent with previous discussions. The  strong to weak spontaneous symmetry breaking of the 1-form $Z_2$ symmetry generated by $\mathcal{L}_{A/B,m}$ corresponds to the condensation of $m_{A}\otimes m_{B}$ anyons.

Notice that we can also write down another bilayer Hamiltonian
\bea\notag
&\hat H^{(\epsilon)}_\text{dTC}=-\sum_\text{star}\big(\prod_{j\in\text{star}}X^A_{j}Z^A_{j+\delta}+\prod_{j\in\text{star}}X^B_{j}Z^B_{j+\delta}\big)\\
&-J\sum_jZ_j^AZ_j^BX^A_{j+\delta}X^B_{j+\delta}-\sum_\text{plaquet}\prod_{j\in\text{plaquet}}Z_j^AZ_j^B.\label{decohered toric code:strong f}
\eea
where $\delta=(\frac12,\frac12)$ on the square lattice with a lattice constant set as $1$.  
Its bilayer ground state describes the toric code decohered by local errors (e.g. $O_j=Z_jX_{j+\delta}$) that condenses the composite anyon $\epsilon_A\otimes\epsilon_B$\cite{Wang2023}. In parallel with above discussions of bilayer Hamiltonian $H^{(m)}_\text{dTC}$ (\ref{decohered toric code:strong m}), here in the $J\gg1$ limit, we can work in the $Z_j^AZ_j^B\equiv X^A_{j+\delta}X^B_{j+\delta},~\forall~j$ subspace. Since all terms in Hamiltonian (\ref{decohered toric code:strong f}) commute with each other, the bilayer ground state is given by
\bea\notag
&\dket{\psi_{\text{dTC}}^{(\epsilon)}}=\prod_j\frac{1+Z_j^AZ_j^BX^A_{j+\delta}X^B_{j+\delta}}2\cdot\\
\notag&\prod_\text{star}\frac{1+\prod_{j\in\text{star}}X^A_{j}Z^A_{j+\delta}}2\frac{1+\prod_{j\in\text{star}}X^B_{j}Z^B_{j+\delta}}2\otimes_j\dket{Z_j^{A/B}\equiv1}\\
&=\prod_j\frac{1+Z_j^AX^A_{j+\delta}Z_j^BX^B_{j+\delta}}2\dket{\text{Toric Code}}_A\otimes\dket{\text{Toric Code}}_B
\eea
Since $Z_j^AX^A_{j+\delta}$ simply creates a pair of fermions near site $j$ on top of the toric code ground state $\dket{\text{Toric Code}}_A$, $\dket{\psi_{\text{dTC}}^{(\epsilon)}}$ is nothing but a superposition of all (open and closed) fermion string configurations on top of the toric code ground state, where $A$ and $B$ layers share exactly the same configuration. Therefore $\dket{\psi_{\text{dTC}}^{(\epsilon)}}$ is also non-negative.  

The ground state $\dket{\psi_{\text{dTC}}^{(\epsilon)}}$ of $H^{(\epsilon)}_\text{dTC}$ features a strong 1-form $Z_2$ symmetry generated by an intralayer fermion string
\bea
\mathcal{L}_{A,\epsilon}=\prod_{l\in\text{string}}X_l^AZ_{l+\delta}^A
\eea
and a weak 1-form $Z_2$ symmetry generated by an interlayer string operator of $e$ particle
\bea
\mathcal{L}_{A,e}\otimes \mathcal{L}_{B,e}=\prod_{l\in\text{string}}Z_l^AZ_l^B.
\eea
The  strong to weak spontaneous symmetry breaking of the 1-form $Z_2$ symmetry generated by $\mathcal{L}_{A/B,\epsilon}$ corresponds to the condensation of $\epsilon_{A}\otimes \epsilon_{B}$ anyons.

Clearly the two bilayer ground states of $\hat H^{(m)}_\text{dTC}$ and $\hat H^{(\epsilon)}_\text{dTC}$ belong to two distinct mixed state topological orders in $\rho_A$, for they feature different deconfined intralayer anyons (and hence different strong 1-form symmetries) in layer $A$: boson $m_A$ for $\hat H^{(m)}_\text{dTC}$ versus fermion $\epsilon_A$ for $\hat H^{(\epsilon)}_\text{dTC}$. However, they correspond to the same $\bst$-enriched toric code topological order in terms of their bilayer ground states $\dket\psi$. Therefore, $\hat H^{(m)}_\text{dTC}$ in (\ref{decohered toric code:strong m}) and $\hat H^{(\epsilon)}_\text{dTC}$ in (\ref{decohered toric code:strong f}) provide an example where the non-negativity constraint splits one phase of $\bst$-symmetric bilayer pure state into more than one mixed state phases.

Finally, we notice that a similar construction for bilayer Hamiltonians can be carried out for a general mixed state topological order of the quantum double type, using the stablizer model for quantum doubles\cite{Ellison2022}.

\section{Summary and outlook}\label{sec:summary}

In this work we introduce the bilayer construction, consisting of layer $A$ and layer $B$ related by an anti-unitary layer-exchange symmetry $\bst$, as a realization of all mixed state phenomena in terms of monolayer (layer $A$) properties of a bilayer pure state. In the presence of $\bst$ symmetry, all non-negative bilayer pure states are in one-to-one correspondence with all mixed states, where the corresponding mixed state is nothing but the monolayer (layer $A$) reduced density matrix $\rho_A=\trace{_B(\dket\psi\dbra\psi)}$ of the bilayer pure state $\dket\psi$. This framework allows us to understand strong and weak symmetries in mixed states and a variety of mixed state phenomena, and to realize them in bilayer 2D materials with a layer-exchange mirror symmetry. For example, strong-to-weak symmetry breaking in a mixed state corresponds to breaking independent symmetries in each layer down to a diagonal subgroup in the bilayer pure state (section \ref{sec:symmetry}); decohering a pure state by local errors corresponds to quantum quench dynamics of the bilayer pure state (\ref{sec:decoherence}); both mixed symmetry protected topological (SPT) states and mixed state topological orders can be classified, characterized and realized as usual SPTs and topological orders in the corresponding bilayer pure states (section \ref{sec:mSPT+mTO}).

Thanks to the bilayer construction, physical properties of mixed states can now be studied based on well-established knowledge of their counterparts in corresponding bilayer pure states. This framework leads to numerous future directions, few of which has been lightly touched in the current work. Below we conclude this paper by listing a few directions to be studied in the future:
\begin{itemize}
    \item Mixed state phase transitions. While some mixed state phase transitions correspond to quantum phase transitions of bilayer ground states by tuning parameters of the bilayer Hamiltonian, such as those SWSSB transitions discussed in section \ref{sec:Ising:criticality}, this is not the full story. For example, decoherence induced mixed state phase transitions discussed in section \ref{sec:decoherence} corresponds to some sort of dynamical phase transitions\cite{Heyl2018} in the bilayer pure state after by a quantum quench\cite{Mitra2018}. How to diagnose the mixed state transition in the quantum-quenched bilayer pure state?    
    \item Characterization of topology in mixed states. While many diagnosis for nontrivial topology in pure states have been proposed understood, such as entanglement spectrum\cite{Li2008,Fidkowski2010,Pollmann2010} and topological entanglement entropy\cite{Kitaev2006a,Levin2006}, characterization of mixed state topology remains largely unexplored. Using the one-to-one correspondence between mixed state and bilayer pure state, the topology indicators in pure states can be translated to diagnose topology in mixed states. 
    \item Is Renyi-2 correlator a good diagnosis for SWSSB? In section \ref{sec:renyi-2} we used two examples introduced in Ref.\cite{Lessa2024} to show that the long-range correlations of the Renyi-2 correlator can fail to diagnose SWSSB. However these two mixed states are somewhat abnormal and pathological, in the sense that their bilayer pure state correspondents may not be the ground state of any local bilayer Hamiltonian. Indeed in the ``normal'' examples studied in section \ref{sec:example:swssb}, Renyi-2 correlators seem to be able to diagnose SWSSB realized by tuning a parameter in the bilayer Hamiltonian. One natural question is, if we restrict ourselves to those mixed states that correspond to the ground states of local bilayer Hamiltonians, can SWSSB in these mixed state be diagnosed by the Renyi-2 correlator? 
    \item Symmetry enriched mixed state topological orders. Building on the classification and characterization of symmetry enriched topological orders in pure states\cite{Essin2013,Mesaros2013,Lu2016,Tarantino2016,Barkeshli2019}, symmetry enriched mixed state topological orders can be understood using their bilayer counterparts, where the interplay of 0-form and 1-form strong/weak symmetries plays a crucial role. 
    \item ``Gapless'' and ``fracton'' mixed states. Gapless quantum liquids\cite{Baym1991,Armitage2018} and fracton phases\cite{Nandkishore2019,Pretko2020} are two well-established examples of pure state quantum matters beyond gapped topological orders. What are their counterparts in mixed states? The bilayer construction provides a route to explore possible ``gapless'' and ``fracton'' phases of mixed states. 

\end{itemize}

\acknowledgements{We thank Ruben Verresen and Yizhi You for their seminars at OSU and for useful discussions, Chong Wang, Meng Cheng, Tim Hsieh and Yizhi You for feedbacks on the manuscript and helpful explanations of their works. This work is supported by Center for Emergent Materials at The Ohio State University, a National Science Foundation (NSF) MRSEC through NSF Award No. NSF DMR-2011876.}




\appendix

\section{Solvable model for bilayer $Z_4\times Z_2$-SPT in one dimension}
\label{app:SPT}

Consider a one-dimensional (1d) chain with $Z_4\times Z_2$ symmetry, consisting of qubits on even sites
\bea
X_{2j}Z_{2j}=-Z_{2j}X_{2j},~~~X_{2j}^2=Z_{2j}^2=1.
\eea
where $\{X_{2j},Y_{2j},Z_{2j}\}$ are Pauli matrices, and 4-dimensional qudits on odd sites
\bea
X_{2j-1}Z_{2j-1}=\imth Z_{2j-1}X_{2j-1},~~~X_{2j-1}^4=Z_{2j-1}^4=1.
\eea
The $Z_4$ symmetry is generated by
\bea
g=\prod_jX_{2j-1}
\eea
and the $Z_2$ symmetry generated by
\bea
h=\prod_jX_{2j}
\eea
The solvable model for $Z_4\times Z_2$ SPT is given by
\bea
\hat{H}=-\sum_j\big[Z_{2j-1}^2X_{2j}Z_{2j+1}^2+Z_{2j}X_{2j+1}Z_{2j+2}+h.c.\big]
\eea
where the 1st term decorates each $Z_4$ domain wall by unit $Z_2$ charge, and the 2nd term decorates each $Z_2$ domain wall by two $Z_4$ charges. Notice that in the low-energy subspace of $X_{2j-1}=\pm1$, one can define an effective qubit on odd sites:
\bea
\sigma_{2j-1}^x\equiv X_{2j-1},~~~\sigma_{2j-1}^z\equiv Z_{2j-1}^2
\eea
In this language, model $\hat H$ is nothing but the well-known cluster state Hamiltonian for $Z_2\times Z_2$-SPT in 1d\cite{Son2012}. 

We follow the same strategy to write down the bilayer model for $Z_4\times Z_2$ SPT phase. Now we consider a pair of 4-dimensional qudits on each site:
\bea
X^{A(B)}_jZ^{A(B)}_j=\imth Z^{A(B)}_jX^{A(B)}_j,~~~\left(X^{A(B)}_j\right)^4=\left(Z^{A(B)}_j\right)^4=1.
\eea
with $Z_4^{A/B}$ symmetry generated by 
\bea
g_{A/B}=\prod_jX_j^{A/B}
\eea
Next we explicitly break the $Z_4^A\times Z_4^B$ symmetry down to a $Z_4\times Z_2$ subgroup, by adding a $J_\text{SB}$ term:
\bea
\hat H_0=-J_\text{SB}\sum_j(Z_j^AZ_j^B)^2-J\sum_j(X_j^AX_j^B)^2
\eea
In addition to the symmetry-breaking term, the purpose of the 2nd $J$ term is to achieve a low-energy Hilbert space on each site $j$, satisfying
\bea
(Z_j^A)^2\equiv(Z_j^B)^2=\pm1,~~~(X_j^A)^2\equiv(X_j^B)^2=\pm1,~~~\forall~j.
\eea
We can define two qubits on each site in this 4-dimensional low-energy subspace:
\bea
&\sigma_j^x\equiv X_j^AX_j^B,~~~\sigma_j^z\equiv(Z_j^A)^2=(Z_j^B)^2;\\
&\tau_{j+\frac12}^x\equiv (X_j^A)^2=(X_j^B)^2,~~~\tau_{j+\frac12}^z\equiv Z_j^A(Z_j^B)^{-1}.
\eea
All these Pauli operators are invariant under layer-exchange anti-unitary symmetry $\bst$ in this $\bst$-invariant low-energy subspace. Clearly in this subspace, the $Z_4$ weak symmetry is generated by
\bea
g_Ag_B=\prod_j\sigma_j^x
\eea
while the strong $Z_2$ symmetry is generated by
\bea
g_A^2=g_B^2=\prod_j\tau_{j+\frac12}^x
\eea
Therefore we can write down the solvable bilayer Hamiltonian
\bea
\hat H_{Z_4\times Z_2}=\hat H_0-\sum_j\big(\sigma_j^z\tau_{j+\frac12}^x\sigma^z_{j+1}+\tau_{j-\frac12}^z\sigma_j^x\tau_{j+\frac12}^z\big)
\eea
which translates into
\bea
&\notag\hat H_{Z_4\times Z_2}=-J_\text{SB}\sum_j(Z_j^AZ_j^B)^2-J\sum_j(X_j^AX_j^B)^2\\
\notag&-1/2\sum_j\big[(Z_j^AX_j^{A}Z_{j+1}^A)^2+A\leftrightarrow B\big]\\
&-1/2\sum_j\big[Z_{j-1}^A(Z_{j-1}^B)^{-1}X_j^AX_j^BZ_j^A(Z_j^B)^{-1}+h.c.\big]
\eea
Since all terms in the Hamiltonian are commuting with each other, the ground state takes the form
\begin{widetext}
\bea
\dket{\Psi_{G}} = \prod_{j}\left[1+Z_{j-1}^A(Z_{j-1}^B)^{-1}X_j^AX_j^BZ_j^A(Z_j^B)^{-1}\right]\otimes_{j} \left[ \dket{(Z_{j}^{A}) ^ 2=+1, (X_{j}^{A}) ^ 2= +1} \otimes \dket{(Z_{j}^{B}) ^ 2=+1, (X_{j}^{B}) ^ 2= +1} \right]
\eea
where the conditions $(Z_{j}^{A}) ^ 2=+1, (X_{j}^{A}) ^ 2= +1$ fix the state as $(\dket{Z_j^A = 1} + \dket{Z_j^A = -1}) / \sqrt{2}$.
Since $\prod_{j}\left[1+Z_{j-1}^A(Z_{j-1}^B)^{-1}X_j^AX_j^BZ_j^A(Z_j^B)^{-1}\right]$
acts on $A$ and $B$ layer $\mathcal{T}$-symmetrically and positively, we can conclude that $\dket{\Psi_{G}}$ is non-negative.
\end{widetext}

\bibliography{refs}

\begin{thebibliography}{91}%
\makeatletter
\providecommand \@ifxundefined [1]{%
 \@ifx{#1\undefined}
}%
\providecommand \@ifnum [1]{%
 \ifnum #1\expandafter \@firstoftwo
 \else \expandafter \@secondoftwo
 \fi
}%
\providecommand \@ifx [1]{%
 \ifx #1\expandafter \@firstoftwo
 \else \expandafter \@secondoftwo
 \fi
}%
\providecommand \natexlab [1]{#1}%
\providecommand \enquote  [1]{``#1''}%
\providecommand \bibnamefont  [1]{#1}%
\providecommand \bibfnamefont [1]{#1}%
\providecommand \citenamefont [1]{#1}%
\providecommand \href@noop [0]{\@secondoftwo}%
\providecommand \href [0]{\begingroup \@sanitize@url \@href}%
\providecommand \@href[1]{\@@startlink{#1}\@@href}%
\providecommand \@@href[1]{\endgroup#1\@@endlink}%
\providecommand \@sanitize@url [0]{\catcode `\\12\catcode `\$12\catcode
  `\&12\catcode `\#12\catcode `\^12\catcode `\_12\catcode `\%12\relax}%
\providecommand \@@startlink[1]{}%
\providecommand \@@endlink[0]{}%
\providecommand \url  [0]{\begingroup\@sanitize@url \@url }%
\providecommand \@url [1]{\endgroup\@href {#1}{\urlprefix }}%
\providecommand \urlprefix  [0]{URL }%
\providecommand \Eprint [0]{\href }%
\providecommand \doibase [0]{https://doi.org/}%
\providecommand \selectlanguage [0]{\@gobble}%
\providecommand \bibinfo  [0]{\@secondoftwo}%
\providecommand \bibfield  [0]{\@secondoftwo}%
\providecommand \translation [1]{[#1]}%
\providecommand \BibitemOpen [0]{}%
\providecommand \bibitemStop [0]{}%
\providecommand \bibitemNoStop [0]{.\EOS\space}%
\providecommand \EOS [0]{\spacefactor3000\relax}%
\providecommand \BibitemShut  [1]{\csname bibitem#1\endcsname}%
\let\auto@bib@innerbib\@empty
\bibitem [{\citenamefont {Prange}(1990)}]{Prange1990}%
  \BibitemOpen
  \bibfield  {author} {\bibinfo {author} {\bibfnamefont {R.~E.}\ \bibnamefont
  {Prange}},\ }\bibinfo {title} {Introduction},\ in\ \href
  {https://doi.org/10.1007/978-1-4612-3350-3_1} {\emph {\bibinfo {booktitle}
  {The Quantum Hall Effect}}},\ \bibinfo {editor} {edited by\ \bibinfo {editor}
  {\bibfnamefont {R.~E.}\ \bibnamefont {Prange}}\ and\ \bibinfo {editor}
  {\bibfnamefont {S.~M.}\ \bibnamefont {Girvin}}}\ (\bibinfo  {publisher}
  {Springer New York},\ \bibinfo {address} {New York, NY},\ \bibinfo {year}
  {1990})\ pp.\ \bibinfo {pages} {1--35}\BibitemShut {NoStop}%
\bibitem [{\citenamefont {Wen}(2007)}]{Wen2007B}%
  \BibitemOpen
  \bibfield  {author} {\bibinfo {author} {\bibfnamefont {X.-G.}\ \bibnamefont
  {Wen}},\ }\bibinfo {title} {{Quantum Field Theory of Many-Body Systems: From
  the Origin of Sound to an Origin of Light and Electrons}}\ (\bibinfo
  {publisher} {Oxford University Press},\ \bibinfo {year} {2007})\BibitemShut
  {NoStop}%
\bibitem [{\citenamefont {Nayak}\ \emph {et~al.}(2008)\citenamefont {Nayak},
  \citenamefont {Simon}, \citenamefont {Stern}, \citenamefont {Freedman},\ and\
  \citenamefont {Das~Sarma}}]{Nayak2008}%
  \BibitemOpen
  \bibfield  {author} {\bibinfo {author} {\bibfnamefont {C.}~\bibnamefont
  {Nayak}}, \bibinfo {author} {\bibfnamefont {S.~H.}\ \bibnamefont {Simon}},
  \bibinfo {author} {\bibfnamefont {A.}~\bibnamefont {Stern}}, \bibinfo
  {author} {\bibfnamefont {M.}~\bibnamefont {Freedman}},\ and\ \bibinfo
  {author} {\bibfnamefont {S.}~\bibnamefont {Das~Sarma}},\ }\bibfield  {title}
  {\bibinfo {title} {Non-abelian anyons and topological quantum computation},\
  }\href {https://doi.org/10.1103/RevModPhys.80.1083} {\bibfield  {journal}
  {\bibinfo  {journal} {Rev. Mod. Phys.}\ }\textbf {\bibinfo {volume} {80}},\
  \bibinfo {pages} {1083} (\bibinfo {year} {2008})}\BibitemShut {NoStop}%
\bibitem [{\citenamefont {Hasan}\ and\ \citenamefont {Kane}(2010)}]{Hasan2010}%
  \BibitemOpen
  \bibfield  {author} {\bibinfo {author} {\bibfnamefont {M.~Z.}\ \bibnamefont
  {Hasan}}\ and\ \bibinfo {author} {\bibfnamefont {C.~L.}\ \bibnamefont
  {Kane}},\ }\bibfield  {title} {\bibinfo {title} {Colloquium: Topological
  insulators},\ }\href {https://doi.org/10.1103/RevModPhys.82.3045} {\bibfield
  {journal} {\bibinfo  {journal} {Rev. Mod. Phys.}\ }\textbf {\bibinfo {volume}
  {82}},\ \bibinfo {pages} {3045} (\bibinfo {year} {2010})}\BibitemShut
  {NoStop}%
\bibitem [{\citenamefont {Qi}\ and\ \citenamefont {Zhang}(2011)}]{Qi2011}%
  \BibitemOpen
  \bibfield  {author} {\bibinfo {author} {\bibfnamefont {X.-L.}\ \bibnamefont
  {Qi}}\ and\ \bibinfo {author} {\bibfnamefont {S.-C.}\ \bibnamefont {Zhang}},\
  }\bibfield  {title} {\bibinfo {title} {Topological insulators and
  superconductors},\ }\href {https://doi.org/10.1103/RevModPhys.83.1057}
  {\bibfield  {journal} {\bibinfo  {journal} {Rev. Mod. Phys.}\ }\textbf
  {\bibinfo {volume} {83}},\ \bibinfo {pages} {1057} (\bibinfo {year}
  {2011})}\BibitemShut {NoStop}%
\bibitem [{\citenamefont {Balents}(2010)}]{Balents2010}%
  \BibitemOpen
  \bibfield  {author} {\bibinfo {author} {\bibfnamefont {L.}~\bibnamefont
  {Balents}},\ }\bibfield  {title} {\bibinfo {title} {Spin liquids in
  frustrated magnets},\ }\href {https://doi.org/10.1038/nature08917} {\bibfield
   {journal} {\bibinfo  {journal} {Nature}\ }\textbf {\bibinfo {volume}
  {464}},\ \bibinfo {pages} {199} (\bibinfo {year} {2010})}\BibitemShut
  {NoStop}%
\bibitem [{\citenamefont {Senthil}(2015)}]{Senthil2015}%
  \BibitemOpen
  \bibfield  {author} {\bibinfo {author} {\bibfnamefont {T.}~\bibnamefont
  {Senthil}},\ }\bibfield  {title} {\bibinfo {title} {Symmetry-protected
  topological phases of quantum matter},\ }\href
  {https://doi.org/https://doi.org/10.1146/annurev-conmatphys-031214-014740}
  {\bibfield  {journal} {\bibinfo  {journal} {Annual Review of Condensed Matter
  Physics}\ }\textbf {\bibinfo {volume} {6}},\ \bibinfo {pages} {299} (\bibinfo
  {year} {2015})}\BibitemShut {NoStop}%
\bibitem [{\citenamefont {Ando}\ and\ \citenamefont {Fu}(2015)}]{Ando2015}%
  \BibitemOpen
  \bibfield  {author} {\bibinfo {author} {\bibfnamefont {Y.}~\bibnamefont
  {Ando}}\ and\ \bibinfo {author} {\bibfnamefont {L.}~\bibnamefont {Fu}},\
  }\bibfield  {title} {\bibinfo {title} {Topological crystalline insulators and
  topological superconductors: From concepts to materials},\ }\href
  {https://doi.org/https://doi.org/10.1146/annurev-conmatphys-031214-014501}
  {\bibfield  {journal} {\bibinfo  {journal} {Annual Review of Condensed Matter
  Physics}\ }\textbf {\bibinfo {volume} {6}},\ \bibinfo {pages} {361} (\bibinfo
  {year} {2015})}\BibitemShut {NoStop}%
\bibitem [{\citenamefont {Chiu}\ \emph {et~al.}(2016)\citenamefont {Chiu},
  \citenamefont {Teo}, \citenamefont {Schnyder},\ and\ \citenamefont
  {Ryu}}]{Chiu2016}%
  \BibitemOpen
  \bibfield  {author} {\bibinfo {author} {\bibfnamefont {C.-K.}\ \bibnamefont
  {Chiu}}, \bibinfo {author} {\bibfnamefont {J.~C.~Y.}\ \bibnamefont {Teo}},
  \bibinfo {author} {\bibfnamefont {A.~P.}\ \bibnamefont {Schnyder}},\ and\
  \bibinfo {author} {\bibfnamefont {S.}~\bibnamefont {Ryu}},\ }\bibfield
  {title} {\bibinfo {title} {Classification of topological quantum matter with
  symmetries},\ }\href {https://doi.org/10.1103/RevModPhys.88.035005}
  {\bibfield  {journal} {\bibinfo  {journal} {Rev. Mod. Phys.}\ }\textbf
  {\bibinfo {volume} {88}},\ \bibinfo {pages} {035005} (\bibinfo {year}
  {2016})}\BibitemShut {NoStop}%
\bibitem [{\citenamefont {Wen}(2017)}]{Wen2017}%
  \BibitemOpen
  \bibfield  {author} {\bibinfo {author} {\bibfnamefont {X.-G.}\ \bibnamefont
  {Wen}},\ }\bibfield  {title} {\bibinfo {title} {Colloquium: Zoo of
  quantum-topological phases of matter},\ }\href
  {https://doi.org/10.1103/RevModPhys.89.041004} {\bibfield  {journal}
  {\bibinfo  {journal} {Rev. Mod. Phys.}\ }\textbf {\bibinfo {volume} {89}},\
  \bibinfo {pages} {041004} (\bibinfo {year} {2017})}\BibitemShut {NoStop}%
\bibitem [{\citenamefont {Armitage}\ \emph {et~al.}(2018)\citenamefont
  {Armitage}, \citenamefont {Mele},\ and\ \citenamefont
  {Vishwanath}}]{Armitage2018}%
  \BibitemOpen
  \bibfield  {author} {\bibinfo {author} {\bibfnamefont {N.~P.}\ \bibnamefont
  {Armitage}}, \bibinfo {author} {\bibfnamefont {E.~J.}\ \bibnamefont {Mele}},\
  and\ \bibinfo {author} {\bibfnamefont {A.}~\bibnamefont {Vishwanath}},\
  }\bibfield  {title} {\bibinfo {title} {Weyl and dirac semimetals in
  three-dimensional solids},\ }\href
  {https://doi.org/10.1103/RevModPhys.90.015001} {\bibfield  {journal}
  {\bibinfo  {journal} {Rev. Mod. Phys.}\ }\textbf {\bibinfo {volume} {90}},\
  \bibinfo {pages} {015001} (\bibinfo {year} {2018})}\BibitemShut {NoStop}%
\bibitem [{\citenamefont {Zeng}\ \emph {et~al.}(2019)\citenamefont {Zeng},
  \citenamefont {Chen}, \citenamefont {Zhou},\ and\ \citenamefont
  {Wen}}]{Zeng2019}%
  \BibitemOpen
  \bibfield  {author} {\bibinfo {author} {\bibfnamefont {B.}~\bibnamefont
  {Zeng}}, \bibinfo {author} {\bibfnamefont {X.}~\bibnamefont {Chen}}, \bibinfo
  {author} {\bibfnamefont {D.-L.}\ \bibnamefont {Zhou}},\ and\ \bibinfo
  {author} {\bibfnamefont {X.-G.}\ \bibnamefont {Wen}},\ }\href
  {https://doi.org/10.1007/978-1-4939-9084-9_11} {\emph {\bibinfo {title}
  {Quantum Information Meets Quantum Matter: From Quantum Entanglement to
  Topological Phases of Many-Body Systems}}}\ (\bibinfo  {publisher} {Springer
  New York},\ \bibinfo {address} {New York, NY},\ \bibinfo {year} {2019})\ pp.\
  \bibinfo {pages} {335--364}\BibitemShut {NoStop}%
\bibitem [{\citenamefont {de~Groot}\ \emph {et~al.}(2022)\citenamefont
  {de~Groot}, \citenamefont {Turzillo},\ and\ \citenamefont
  {Schuch}}]{deGroot2022}%
  \BibitemOpen
  \bibfield  {author} {\bibinfo {author} {\bibfnamefont {C.}~\bibnamefont
  {de~Groot}}, \bibinfo {author} {\bibfnamefont {A.}~\bibnamefont {Turzillo}},\
  and\ \bibinfo {author} {\bibfnamefont {N.}~\bibnamefont {Schuch}},\
  }\bibfield  {title} {\bibinfo {title} {Symmetry {P}rotected {T}opological
  {O}rder in {O}pen {Q}uantum {S}ystems},\ }\href
  {https://doi.org/10.22331/q-2022-11-10-856} {\bibfield  {journal} {\bibinfo
  {journal} {{Quantum}}\ }\textbf {\bibinfo {volume} {6}},\ \bibinfo {pages}
  {856} (\bibinfo {year} {2022})}\BibitemShut {NoStop}%
\bibitem [{\citenamefont {{Lee}}\ \emph {et~al.}(2022)\citenamefont {{Lee}},
  \citenamefont {{You}},\ and\ \citenamefont {{Xu}}}]{Lee2022}%
  \BibitemOpen
  \bibfield  {author} {\bibinfo {author} {\bibfnamefont {J.~Y.}\ \bibnamefont
  {{Lee}}}, \bibinfo {author} {\bibfnamefont {Y.-Z.}\ \bibnamefont {{You}}},\
  and\ \bibinfo {author} {\bibfnamefont {C.}~\bibnamefont {{Xu}}},\ }\bibfield
  {title} {\bibinfo {title} {{Symmetry protected topological phases under
  decoherence}},\ }\href {https://doi.org/10.48550/arXiv.2210.16323} {\bibfield
   {journal} {\bibinfo  {journal} {arXiv e-prints}\ ,\ \bibinfo {eid}
  {arXiv:2210.16323}} (\bibinfo {year} {2022})},\ \Eprint
  {https://arxiv.org/abs/2210.16323} {arXiv:2210.16323 [cond-mat.str-el]}
  \BibitemShut {NoStop}%
\bibitem [{\citenamefont {Ma}\ and\ \citenamefont {Wang}(2023)}]{Ma2023}%
  \BibitemOpen
  \bibfield  {author} {\bibinfo {author} {\bibfnamefont {R.}~\bibnamefont
  {Ma}}\ and\ \bibinfo {author} {\bibfnamefont {C.}~\bibnamefont {Wang}},\
  }\bibfield  {title} {\bibinfo {title} {Average symmetry-protected topological
  phases},\ }\href {https://doi.org/10.1103/PhysRevX.13.031016} {\bibfield
  {journal} {\bibinfo  {journal} {Phys. Rev. X}\ }\textbf {\bibinfo {volume}
  {13}},\ \bibinfo {pages} {031016} (\bibinfo {year} {2023})}\BibitemShut
  {NoStop}%
\bibitem [{\citenamefont {{Wang}}\ \emph {et~al.}(2023)\citenamefont {{Wang}},
  \citenamefont {{Wu}},\ and\ \citenamefont {{Wang}}}]{Wang2023}%
  \BibitemOpen
  \bibfield  {author} {\bibinfo {author} {\bibfnamefont {Z.}~\bibnamefont
  {{Wang}}}, \bibinfo {author} {\bibfnamefont {Z.}~\bibnamefont {{Wu}}},\ and\
  \bibinfo {author} {\bibfnamefont {Z.}~\bibnamefont {{Wang}}},\ }\bibfield
  {title} {\bibinfo {title} {{Intrinsic Mixed-state Quantum Topological
  Order}},\ }\href {https://doi.org/10.48550/arXiv.2307.13758} {\bibfield
  {journal} {\bibinfo  {journal} {arXiv e-prints}\ ,\ \bibinfo {eid}
  {arXiv:2307.13758}} (\bibinfo {year} {2023})},\ \Eprint
  {https://arxiv.org/abs/2307.13758} {arXiv:2307.13758 [quant-ph]} \BibitemShut
  {NoStop}%
\bibitem [{\citenamefont {{Ma}}\ and\ \citenamefont
  {{Turzillo}}(2024)}]{Ma2024a}%
  \BibitemOpen
  \bibfield  {author} {\bibinfo {author} {\bibfnamefont {R.}~\bibnamefont
  {{Ma}}}\ and\ \bibinfo {author} {\bibfnamefont {A.}~\bibnamefont
  {{Turzillo}}},\ }\bibfield  {title} {\bibinfo {title} {{Symmetry Protected
  Topological Phases of Mixed States in the Doubled Space}},\ }\href
  {https://doi.org/10.48550/arXiv.2403.13280} {\bibfield  {journal} {\bibinfo
  {journal} {arXiv e-prints}\ ,\ \bibinfo {eid} {arXiv:2403.13280}} (\bibinfo
  {year} {2024})},\ \Eprint {https://arxiv.org/abs/2403.13280}
  {arXiv:2403.13280 [quant-ph]} \BibitemShut {NoStop}%
\bibitem [{\citenamefont {Fan}\ \emph {et~al.}(2024)\citenamefont {Fan},
  \citenamefont {Bao}, \citenamefont {Altman},\ and\ \citenamefont
  {Vishwanath}}]{Fan2024}%
  \BibitemOpen
  \bibfield  {author} {\bibinfo {author} {\bibfnamefont {R.}~\bibnamefont
  {Fan}}, \bibinfo {author} {\bibfnamefont {Y.}~\bibnamefont {Bao}}, \bibinfo
  {author} {\bibfnamefont {E.}~\bibnamefont {Altman}},\ and\ \bibinfo {author}
  {\bibfnamefont {A.}~\bibnamefont {Vishwanath}},\ }\bibfield  {title}
  {\bibinfo {title} {Diagnostics of mixed-state topological order and breakdown
  of quantum memory},\ }\href {https://doi.org/10.1103/PRXQuantum.5.020343}
  {\bibfield  {journal} {\bibinfo  {journal} {PRX Quantum}\ }\textbf {\bibinfo
  {volume} {5}},\ \bibinfo {pages} {020343} (\bibinfo {year}
  {2024})}\BibitemShut {NoStop}%
\bibitem [{\citenamefont {{Bao}}\ \emph {et~al.}(2023)\citenamefont {{Bao}},
  \citenamefont {{Fan}}, \citenamefont {{Vishwanath}},\ and\ \citenamefont
  {{Altman}}}]{Bao2023}%
  \BibitemOpen
  \bibfield  {author} {\bibinfo {author} {\bibfnamefont {Y.}~\bibnamefont
  {{Bao}}}, \bibinfo {author} {\bibfnamefont {R.}~\bibnamefont {{Fan}}},
  \bibinfo {author} {\bibfnamefont {A.}~\bibnamefont {{Vishwanath}}},\ and\
  \bibinfo {author} {\bibfnamefont {E.}~\bibnamefont {{Altman}}},\ }\bibfield
  {title} {\bibinfo {title} {{Mixed-state topological order and the errorfield
  double formulation of decoherence-induced transitions}},\ }\href
  {https://doi.org/10.48550/arXiv.2301.05687} {\bibfield  {journal} {\bibinfo
  {journal} {arXiv e-prints}\ ,\ \bibinfo {eid} {arXiv:2301.05687}} (\bibinfo
  {year} {2023})},\ \Eprint {https://arxiv.org/abs/2301.05687}
  {arXiv:2301.05687 [quant-ph]} \BibitemShut {NoStop}%
\bibitem [{\citenamefont {Lee}\ \emph {et~al.}(2023)\citenamefont {Lee},
  \citenamefont {Jian},\ and\ \citenamefont {Xu}}]{Lee2023}%
  \BibitemOpen
  \bibfield  {author} {\bibinfo {author} {\bibfnamefont {J.~Y.}\ \bibnamefont
  {Lee}}, \bibinfo {author} {\bibfnamefont {C.-M.}\ \bibnamefont {Jian}},\ and\
  \bibinfo {author} {\bibfnamefont {C.}~\bibnamefont {Xu}},\ }\bibfield
  {title} {\bibinfo {title} {Quantum criticality under decoherence or weak
  measurement},\ }\href {https://doi.org/10.1103/PRXQuantum.4.030317}
  {\bibfield  {journal} {\bibinfo  {journal} {PRX Quantum}\ }\textbf {\bibinfo
  {volume} {4}},\ \bibinfo {pages} {030317} (\bibinfo {year}
  {2023})}\BibitemShut {NoStop}%
\bibitem [{\citenamefont {Sala}\ \emph {et~al.}(2024)\citenamefont {Sala},
  \citenamefont {Gopalakrishnan}, \citenamefont {Oshikawa},\ and\ \citenamefont
  {You}}]{Sala2024}%
  \BibitemOpen
  \bibfield  {author} {\bibinfo {author} {\bibfnamefont {P.}~\bibnamefont
  {Sala}}, \bibinfo {author} {\bibfnamefont {S.}~\bibnamefont
  {Gopalakrishnan}}, \bibinfo {author} {\bibfnamefont {M.}~\bibnamefont
  {Oshikawa}},\ and\ \bibinfo {author} {\bibfnamefont {Y.}~\bibnamefont
  {You}},\ }\bibfield  {title} {\bibinfo {title} {Spontaneous strong symmetry
  breaking in open systems: Purification perspective},\ }\href
  {https://doi.org/10.1103/PhysRevB.110.155150} {\bibfield  {journal} {\bibinfo
   {journal} {Phys. Rev. B}\ }\textbf {\bibinfo {volume} {110}},\ \bibinfo
  {pages} {155150} (\bibinfo {year} {2024})}\BibitemShut {NoStop}%
\bibitem [{\citenamefont {{You}}\ and\ \citenamefont
  {{Oshikawa}}(2024)}]{You2024}%
  \BibitemOpen
  \bibfield  {author} {\bibinfo {author} {\bibfnamefont {Y.}~\bibnamefont
  {{You}}}\ and\ \bibinfo {author} {\bibfnamefont {M.}~\bibnamefont
  {{Oshikawa}}},\ }\bibfield  {title} {\bibinfo {title} {{Intrinsic
  symmetry-protected topological mixed state from modulated symmetries and
  hierarchical structure of boundary anomaly}},\ }\href
  {https://doi.org/10.1103/PhysRevB.110.165160} {\bibfield  {journal} {\bibinfo
   {journal} {\prb}\ }\textbf {\bibinfo {volume} {110}},\ \bibinfo {eid}
  {165160} (\bibinfo {year} {2024})},\ \Eprint
  {https://arxiv.org/abs/2407.08786} {arXiv:2407.08786 [quant-ph]} \BibitemShut
  {NoStop}%
\bibitem [{\citenamefont {{Lessa}}\ \emph {et~al.}(2024)\citenamefont
  {{Lessa}}, \citenamefont {{Ma}}, \citenamefont {{Zhang}}, \citenamefont
  {{Bi}}, \citenamefont {{Cheng}},\ and\ \citenamefont {{Wang}}}]{Lessa2024}%
  \BibitemOpen
  \bibfield  {author} {\bibinfo {author} {\bibfnamefont {L.~A.}\ \bibnamefont
  {{Lessa}}}, \bibinfo {author} {\bibfnamefont {R.}~\bibnamefont {{Ma}}},
  \bibinfo {author} {\bibfnamefont {J.-H.}\ \bibnamefont {{Zhang}}}, \bibinfo
  {author} {\bibfnamefont {Z.}~\bibnamefont {{Bi}}}, \bibinfo {author}
  {\bibfnamefont {M.}~\bibnamefont {{Cheng}}},\ and\ \bibinfo {author}
  {\bibfnamefont {C.}~\bibnamefont {{Wang}}},\ }\bibfield  {title} {\bibinfo
  {title} {{Strong-to-Weak Spontaneous Symmetry Breaking in Mixed Quantum
  States}},\ }\href {https://doi.org/10.48550/arXiv.2405.03639} {\bibfield
  {journal} {\bibinfo  {journal} {arXiv e-prints}\ ,\ \bibinfo {eid}
  {arXiv:2405.03639}} (\bibinfo {year} {2024})},\ \Eprint
  {https://arxiv.org/abs/2405.03639} {arXiv:2405.03639 [quant-ph]} \BibitemShut
  {NoStop}%
\bibitem [{\citenamefont {{Zhang}}\ \emph {et~al.}(2024)\citenamefont
  {{Zhang}}, \citenamefont {{Xu}}, \citenamefont {{Zhang}}, \citenamefont
  {{Xu}}, \citenamefont {{Bi}},\ and\ \citenamefont {{Luo}}}]{Zhang2024}%
  \BibitemOpen
  \bibfield  {author} {\bibinfo {author} {\bibfnamefont {C.}~\bibnamefont
  {{Zhang}}}, \bibinfo {author} {\bibfnamefont {Y.}~\bibnamefont {{Xu}}},
  \bibinfo {author} {\bibfnamefont {J.-H.}\ \bibnamefont {{Zhang}}}, \bibinfo
  {author} {\bibfnamefont {C.}~\bibnamefont {{Xu}}}, \bibinfo {author}
  {\bibfnamefont {Z.}~\bibnamefont {{Bi}}},\ and\ \bibinfo {author}
  {\bibfnamefont {Z.-X.}\ \bibnamefont {{Luo}}},\ }\bibfield  {title} {\bibinfo
  {title} {{Strong-to-weak spontaneous breaking of 1-form symmetry and
  intrinsically mixed topological order}},\ }\href
  {https://doi.org/10.48550/arXiv.2409.17530} {\bibfield  {journal} {\bibinfo
  {journal} {arXiv e-prints}\ ,\ \bibinfo {eid} {arXiv:2409.17530}} (\bibinfo
  {year} {2024})},\ \Eprint {https://arxiv.org/abs/2409.17530}
  {arXiv:2409.17530 [quant-ph]} \BibitemShut {NoStop}%
\bibitem [{\citenamefont {{Sohal}}\ and\ \citenamefont
  {{Prem}}(2024)}]{Soha2024}%
  \BibitemOpen
  \bibfield  {author} {\bibinfo {author} {\bibfnamefont {R.}~\bibnamefont
  {{Sohal}}}\ and\ \bibinfo {author} {\bibfnamefont {A.}~\bibnamefont
  {{Prem}}},\ }\bibfield  {title} {\bibinfo {title} {{A Noisy Approach to
  Intrinsically Mixed-State Topological Order}},\ }\href
  {https://doi.org/10.48550/arXiv.2403.13879} {\bibfield  {journal} {\bibinfo
  {journal} {arXiv e-prints}\ ,\ \bibinfo {eid} {arXiv:2403.13879}} (\bibinfo
  {year} {2024})},\ \Eprint {https://arxiv.org/abs/2403.13879}
  {arXiv:2403.13879 [cond-mat.str-el]} \BibitemShut {NoStop}%
\bibitem [{\citenamefont {{Ellison}}\ and\ \citenamefont
  {{Cheng}}(2024)}]{Ellison2024}%
  \BibitemOpen
  \bibfield  {author} {\bibinfo {author} {\bibfnamefont {T.}~\bibnamefont
  {{Ellison}}}\ and\ \bibinfo {author} {\bibfnamefont {M.}~\bibnamefont
  {{Cheng}}},\ }\bibfield  {title} {\bibinfo {title} {{Towards a classification
  of mixed-state topological orders in two dimensions}},\ }\href
  {https://doi.org/10.48550/arXiv.2405.02390} {\bibfield  {journal} {\bibinfo
  {journal} {arXiv e-prints}\ ,\ \bibinfo {eid} {arXiv:2405.02390}} (\bibinfo
  {year} {2024})},\ \Eprint {https://arxiv.org/abs/2405.02390}
  {arXiv:2405.02390 [cond-mat.str-el]} \BibitemShut {NoStop}%
\bibitem [{\citenamefont {{Guo}}\ \emph {et~al.}(2024)\citenamefont {{Guo}},
  \citenamefont {{Ding}},\ and\ \citenamefont {{Yang}}}]{Guo2024}%
  \BibitemOpen
  \bibfield  {author} {\bibinfo {author} {\bibfnamefont {Y.}~\bibnamefont
  {{Guo}}}, \bibinfo {author} {\bibfnamefont {K.}~\bibnamefont {{Ding}}},\ and\
  \bibinfo {author} {\bibfnamefont {S.}~\bibnamefont {{Yang}}},\ }\bibfield
  {title} {\bibinfo {title} {{A New Framework for Quantum Phases in Open
  Systems: Steady State of Imaginary-Time Lindbladian Evolution}},\ }\href
  {https://doi.org/10.48550/arXiv.2408.03239} {\bibfield  {journal} {\bibinfo
  {journal} {arXiv e-prints}\ ,\ \bibinfo {eid} {arXiv:2408.03239}} (\bibinfo
  {year} {2024})},\ \Eprint {https://arxiv.org/abs/2408.03239}
  {arXiv:2408.03239 [quant-ph]} \BibitemShut {NoStop}%
\bibitem [{\citenamefont {Chen}\ \emph {et~al.}(2013)\citenamefont {Chen},
  \citenamefont {Gu}, \citenamefont {Liu},\ and\ \citenamefont
  {Wen}}]{Chen2013}%
  \BibitemOpen
  \bibfield  {author} {\bibinfo {author} {\bibfnamefont {X.}~\bibnamefont
  {Chen}}, \bibinfo {author} {\bibfnamefont {Z.-C.}\ \bibnamefont {Gu}},
  \bibinfo {author} {\bibfnamefont {Z.-X.}\ \bibnamefont {Liu}},\ and\ \bibinfo
  {author} {\bibfnamefont {X.-G.}\ \bibnamefont {Wen}},\ }\bibfield  {title}
  {\bibinfo {title} {Symmetry protected topological orders and the group
  cohomology of their symmetry group},\ }\href
  {https://doi.org/10.1103/PhysRevB.87.155114} {\bibfield  {journal} {\bibinfo
  {journal} {Phys. Rev. B}\ }\textbf {\bibinfo {volume} {87}},\ \bibinfo
  {pages} {155114} (\bibinfo {year} {2013})}\BibitemShut {NoStop}%
\bibitem [{\citenamefont {Levin}\ and\ \citenamefont {Gu}(2012)}]{Levin2012}%
  \BibitemOpen
  \bibfield  {author} {\bibinfo {author} {\bibfnamefont {M.}~\bibnamefont
  {Levin}}\ and\ \bibinfo {author} {\bibfnamefont {Z.-C.}\ \bibnamefont {Gu}},\
  }\bibfield  {title} {\bibinfo {title} {Braiding statistics approach to
  symmetry-protected topological phases},\ }\href
  {https://doi.org/10.1103/PhysRevB.86.115109} {\bibfield  {journal} {\bibinfo
  {journal} {Phys. Rev. B}\ }\textbf {\bibinfo {volume} {86}},\ \bibinfo
  {pages} {115109} (\bibinfo {year} {2012})}\BibitemShut {NoStop}%
\bibitem [{\citenamefont {Lu}\ and\ \citenamefont {Vishwanath}(2012)}]{Lu2012}%
  \BibitemOpen
  \bibfield  {author} {\bibinfo {author} {\bibfnamefont {Y.-M.}\ \bibnamefont
  {Lu}}\ and\ \bibinfo {author} {\bibfnamefont {A.}~\bibnamefont
  {Vishwanath}},\ }\bibfield  {title} {\bibinfo {title} {Theory and
  classification of interacting integer topological phases in two dimensions: A
  chern-simons approach},\ }\href {https://doi.org/10.1103/PhysRevB.86.125119}
  {\bibfield  {journal} {\bibinfo  {journal} {Phys. Rev. B}\ }\textbf {\bibinfo
  {volume} {86}},\ \bibinfo {pages} {125119} (\bibinfo {year}
  {2012})}\BibitemShut {NoStop}%
\bibitem [{\citenamefont {Essin}\ and\ \citenamefont
  {Hermele}(2013)}]{Essin2013}%
  \BibitemOpen
  \bibfield  {author} {\bibinfo {author} {\bibfnamefont {A.~M.}\ \bibnamefont
  {Essin}}\ and\ \bibinfo {author} {\bibfnamefont {M.}~\bibnamefont
  {Hermele}},\ }\bibfield  {title} {\bibinfo {title} {Classifying
  fractionalization: Symmetry classification of gapped ${\mathbb{z}}_{2}$ spin
  liquids in two dimensions},\ }\href
  {https://doi.org/10.1103/PhysRevB.87.104406} {\bibfield  {journal} {\bibinfo
  {journal} {Phys. Rev. B}\ }\textbf {\bibinfo {volume} {87}},\ \bibinfo
  {pages} {104406} (\bibinfo {year} {2013})}\BibitemShut {NoStop}%
\bibitem [{\citenamefont {Mesaros}\ and\ \citenamefont
  {Ran}(2013)}]{Mesaros2013}%
  \BibitemOpen
  \bibfield  {author} {\bibinfo {author} {\bibfnamefont {A.}~\bibnamefont
  {Mesaros}}\ and\ \bibinfo {author} {\bibfnamefont {Y.}~\bibnamefont {Ran}},\
  }\bibfield  {title} {\bibinfo {title} {Classification of symmetry enriched
  topological phases with exactly solvable models},\ }\href
  {https://doi.org/10.1103/PhysRevB.87.155115} {\bibfield  {journal} {\bibinfo
  {journal} {Phys. Rev. B}\ }\textbf {\bibinfo {volume} {87}},\ \bibinfo
  {pages} {155115} (\bibinfo {year} {2013})}\BibitemShut {NoStop}%
\bibitem [{\citenamefont {Lu}\ and\ \citenamefont {Vishwanath}(2016)}]{Lu2016}%
  \BibitemOpen
  \bibfield  {author} {\bibinfo {author} {\bibfnamefont {Y.-M.}\ \bibnamefont
  {Lu}}\ and\ \bibinfo {author} {\bibfnamefont {A.}~\bibnamefont
  {Vishwanath}},\ }\bibfield  {title} {\bibinfo {title} {Classification and
  properties of symmetry-enriched topological phases: Chern-simons approach
  with applications to ${Z}_{2}$ spin liquids},\ }\href
  {https://doi.org/10.1103/PhysRevB.93.155121} {\bibfield  {journal} {\bibinfo
  {journal} {Phys. Rev. B}\ }\textbf {\bibinfo {volume} {93}},\ \bibinfo
  {pages} {155121} (\bibinfo {year} {2016})}\BibitemShut {NoStop}%
\bibitem [{\citenamefont {Tarantino}\ \emph {et~al.}(2016)\citenamefont
  {Tarantino}, \citenamefont {Lindner},\ and\ \citenamefont
  {Fidkowski}}]{Tarantino2016}%
  \BibitemOpen
  \bibfield  {author} {\bibinfo {author} {\bibfnamefont {N.}~\bibnamefont
  {Tarantino}}, \bibinfo {author} {\bibfnamefont {N.~H.}\ \bibnamefont
  {Lindner}},\ and\ \bibinfo {author} {\bibfnamefont {L.}~\bibnamefont
  {Fidkowski}},\ }\bibfield  {title} {\bibinfo {title} {Symmetry
  fractionalization and twist defects},\ }\href
  {https://doi.org/10.1088/1367-2630/18/3/035006} {\bibfield  {journal}
  {\bibinfo  {journal} {New Journal of Physics}\ }\textbf {\bibinfo {volume}
  {18}},\ \bibinfo {pages} {035006} (\bibinfo {year} {2016})}\BibitemShut
  {NoStop}%
\bibitem [{\citenamefont {Barkeshli}\ \emph {et~al.}(2019)\citenamefont
  {Barkeshli}, \citenamefont {Bonderson}, \citenamefont {Cheng},\ and\
  \citenamefont {Wang}}]{Barkeshli2019}%
  \BibitemOpen
  \bibfield  {author} {\bibinfo {author} {\bibfnamefont {M.}~\bibnamefont
  {Barkeshli}}, \bibinfo {author} {\bibfnamefont {P.}~\bibnamefont
  {Bonderson}}, \bibinfo {author} {\bibfnamefont {M.}~\bibnamefont {Cheng}},\
  and\ \bibinfo {author} {\bibfnamefont {Z.}~\bibnamefont {Wang}},\ }\bibfield
  {title} {\bibinfo {title} {Symmetry fractionalization, defects, and gauging
  of topological phases},\ }\href {https://doi.org/10.1103/PhysRevB.100.115147}
  {\bibfield  {journal} {\bibinfo  {journal} {Phys. Rev. B}\ }\textbf {\bibinfo
  {volume} {100}},\ \bibinfo {pages} {115147} (\bibinfo {year}
  {2019})}\BibitemShut {NoStop}%
\bibitem [{\citenamefont {{Ma}}\ \emph {et~al.}(2023)\citenamefont {{Ma}},
  \citenamefont {{Zhang}}, \citenamefont {{Bi}}, \citenamefont {{Cheng}},\ and\
  \citenamefont {{Wang}}}]{Ma2023a}%
  \BibitemOpen
  \bibfield  {author} {\bibinfo {author} {\bibfnamefont {R.}~\bibnamefont
  {{Ma}}}, \bibinfo {author} {\bibfnamefont {J.-H.}\ \bibnamefont {{Zhang}}},
  \bibinfo {author} {\bibfnamefont {Z.}~\bibnamefont {{Bi}}}, \bibinfo {author}
  {\bibfnamefont {M.}~\bibnamefont {{Cheng}}},\ and\ \bibinfo {author}
  {\bibfnamefont {C.}~\bibnamefont {{Wang}}},\ }\bibfield  {title} {\bibinfo
  {title} {{Topological Phases with Average Symmetries: the Decohered, the
  Disordered, and the Intrinsic}},\ }\href
  {https://doi.org/10.48550/arXiv.2305.16399} {\bibfield  {journal} {\bibinfo
  {journal} {arXiv e-prints}\ ,\ \bibinfo {eid} {arXiv:2305.16399}} (\bibinfo
  {year} {2023})},\ \Eprint {https://arxiv.org/abs/2305.16399}
  {arXiv:2305.16399 [cond-mat.str-el]} \BibitemShut {NoStop}%
\bibitem [{\citenamefont {{Weinstein}}(2024)}]{Weinstein2024}%
  \BibitemOpen
  \bibfield  {author} {\bibinfo {author} {\bibfnamefont {Z.}~\bibnamefont
  {{Weinstein}}},\ }\bibfield  {title} {\bibinfo {title} {{Efficient Detection
  of Strong-To-Weak Spontaneous Symmetry Breaking via the
  R\textbackslash'enyi-1 Correlator}},\ }\href@noop {} {\bibfield  {journal}
  {\bibinfo  {journal} {arXiv e-prints}\ ,\ \bibinfo {eid} {arXiv:2410.23512}}
  (\bibinfo {year} {2024})},\ \Eprint {https://arxiv.org/abs/2410.23512}
  {arXiv:2410.23512 [quant-ph]} \BibitemShut {NoStop}%
\bibitem [{\citenamefont {{Liu}}\ \emph {et~al.}(2024)\citenamefont {{Liu}},
  \citenamefont {{Chen}}, \citenamefont {{Zhang}}, \citenamefont {{Zhou}},\
  and\ \citenamefont {{Zhang}}}]{Liu2024}%
  \BibitemOpen
  \bibfield  {author} {\bibinfo {author} {\bibfnamefont {Z.}~\bibnamefont
  {{Liu}}}, \bibinfo {author} {\bibfnamefont {L.}~\bibnamefont {{Chen}}},
  \bibinfo {author} {\bibfnamefont {Y.}~\bibnamefont {{Zhang}}}, \bibinfo
  {author} {\bibfnamefont {S.}~\bibnamefont {{Zhou}}},\ and\ \bibinfo {author}
  {\bibfnamefont {P.}~\bibnamefont {{Zhang}}},\ }\bibfield  {title} {\bibinfo
  {title} {{Diagnosing Strong-to-Weak Symmetry Breaking via Wightman
  Correlators}},\ }\href {https://doi.org/10.48550/arXiv.2410.09327} {\bibfield
   {journal} {\bibinfo  {journal} {arXiv e-prints}\ ,\ \bibinfo {eid}
  {arXiv:2410.09327}} (\bibinfo {year} {2024})},\ \Eprint
  {https://arxiv.org/abs/2410.09327} {arXiv:2410.09327 [quant-ph]} \BibitemShut
  {NoStop}%
\bibitem [{\citenamefont {{Gu}}\ \emph {et~al.}(2024)\citenamefont {{Gu}},
  \citenamefont {{Wang}},\ and\ \citenamefont {{Wang}}}]{Ding2024}%
  \BibitemOpen
  \bibfield  {author} {\bibinfo {author} {\bibfnamefont {D.}~\bibnamefont
  {{Gu}}}, \bibinfo {author} {\bibfnamefont {Z.}~\bibnamefont {{Wang}}},\ and\
  \bibinfo {author} {\bibfnamefont {Z.}~\bibnamefont {{Wang}}},\ }\bibfield
  {title} {\bibinfo {title} {{Spontaneous symmetry breaking in open quantum
  systems: strong, weak, and strong-to-weak}},\ }\href
  {https://doi.org/10.48550/arXiv.2406.19381} {\bibfield  {journal} {\bibinfo
  {journal} {arXiv e-prints}\ ,\ \bibinfo {eid} {arXiv:2406.19381}} (\bibinfo
  {year} {2024})},\ \Eprint {https://arxiv.org/abs/2406.19381}
  {arXiv:2406.19381 [quant-ph]} \BibitemShut {NoStop}%
\bibitem [{\citenamefont {Dennis}\ \emph {et~al.}(2002)\citenamefont {Dennis},
  \citenamefont {Kitaev}, \citenamefont {Landahl},\ and\ \citenamefont
  {Preskill}}]{Dennis2002}%
  \BibitemOpen
  \bibfield  {author} {\bibinfo {author} {\bibfnamefont {E.}~\bibnamefont
  {Dennis}}, \bibinfo {author} {\bibfnamefont {A.}~\bibnamefont {Kitaev}},
  \bibinfo {author} {\bibfnamefont {A.}~\bibnamefont {Landahl}},\ and\ \bibinfo
  {author} {\bibfnamefont {J.}~\bibnamefont {Preskill}},\ }\bibfield  {title}
  {\bibinfo {title} {{Topological quantum memory}},\ }\href
  {https://doi.org/10.1063/1.1499754} {\bibfield  {journal} {\bibinfo
  {journal} {Journal of Mathematical Physics}\ }\textbf {\bibinfo {volume}
  {43}},\ \bibinfo {pages} {4452} (\bibinfo {year} {2002})},\ \Eprint
  {https://arxiv.org/abs/https://pubs.aip.org/aip/jmp/article-pdf/43/9/4452/19183135/4452\_1\_online.pdf}
  {https://pubs.aip.org/aip/jmp/article-pdf/43/9/4452/19183135/4452\_1\_online.pdf}
  \BibitemShut {NoStop}%
\bibitem [{\citenamefont {Chen}\ and\ \citenamefont
  {Grover}(2024{\natexlab{a}})}]{Chen2024a}%
  \BibitemOpen
  \bibfield  {author} {\bibinfo {author} {\bibfnamefont {Y.-H.}\ \bibnamefont
  {Chen}}\ and\ \bibinfo {author} {\bibfnamefont {T.}~\bibnamefont {Grover}},\
  }\bibfield  {title} {\bibinfo {title} {Unconventional topological mixed-state
  transition and critical phase induced by self-dual coherent errors},\ }\href
  {https://doi.org/10.1103/PhysRevB.110.125152} {\bibfield  {journal} {\bibinfo
   {journal} {Phys. Rev. B}\ }\textbf {\bibinfo {volume} {110}},\ \bibinfo
  {pages} {125152} (\bibinfo {year} {2024}{\natexlab{a}})}\BibitemShut
  {NoStop}%
\bibitem [{\citenamefont {{Sala}}\ \emph {et~al.}(2024)\citenamefont {{Sala}},
  \citenamefont {{Alicea}},\ and\ \citenamefont {{Verresen}}}]{Sala2024b}%
  \BibitemOpen
  \bibfield  {author} {\bibinfo {author} {\bibfnamefont {P.}~\bibnamefont
  {{Sala}}}, \bibinfo {author} {\bibfnamefont {J.}~\bibnamefont {{Alicea}}},\
  and\ \bibinfo {author} {\bibfnamefont {R.}~\bibnamefont {{Verresen}}},\
  }\bibfield  {title} {\bibinfo {title} {{Decoherence and wavefunction
  deformation of $D_4$ non-Abelian topological order}},\ }\href
  {https://doi.org/10.48550/arXiv.2409.12948} {\bibfield  {journal} {\bibinfo
  {journal} {arXiv e-prints}\ ,\ \bibinfo {eid} {arXiv:2409.12948}} (\bibinfo
  {year} {2024})},\ \Eprint {https://arxiv.org/abs/2409.12948}
  {arXiv:2409.12948 [cond-mat.str-el]} \BibitemShut {NoStop}%
\bibitem [{\citenamefont {{Chirame}}\ \emph {et~al.}(2024)\citenamefont
  {{Chirame}}, \citenamefont {{Prem}}, \citenamefont {{Gopalakrishnan}},\ and\
  \citenamefont {{Burnell}}}]{Chirame2024}%
  \BibitemOpen
  \bibfield  {author} {\bibinfo {author} {\bibfnamefont {S.}~\bibnamefont
  {{Chirame}}}, \bibinfo {author} {\bibfnamefont {A.}~\bibnamefont {{Prem}}},
  \bibinfo {author} {\bibfnamefont {S.}~\bibnamefont {{Gopalakrishnan}}},\ and\
  \bibinfo {author} {\bibfnamefont {F.~J.}\ \bibnamefont {{Burnell}}},\
  }\bibfield  {title} {\bibinfo {title} {{Stabilizing Non-Abelian Topological
  Order against Heralded Noise via Local Lindbladian Dynamics}},\ }\href
  {https://doi.org/10.48550/arXiv.2410.21402} {\bibfield  {journal} {\bibinfo
  {journal} {arXiv e-prints}\ ,\ \bibinfo {eid} {arXiv:2410.21402}} (\bibinfo
  {year} {2024})},\ \Eprint {https://arxiv.org/abs/2410.21402}
  {arXiv:2410.21402 [quant-ph]} \BibitemShut {NoStop}%
\bibitem [{\citenamefont {Mitra}(2018)}]{Mitra2018}%
  \BibitemOpen
  \bibfield  {author} {\bibinfo {author} {\bibfnamefont {A.}~\bibnamefont
  {Mitra}},\ }\bibfield  {title} {\bibinfo {title} {Quantum quench dynamics},\
  }\href
  {https://doi.org/https://doi.org/10.1146/annurev-conmatphys-031016-025451}
  {\bibfield  {journal} {\bibinfo  {journal} {Annual Review of Condensed Matter
  Physics}\ }\textbf {\bibinfo {volume} {9}},\ \bibinfo {pages} {245} (\bibinfo
  {year} {2018})}\BibitemShut {NoStop}%
\bibitem [{\citenamefont {{Moudgalya}}\ and\ \citenamefont
  {{Motrunich}}(2023)}]{Moudgalya2023}%
  \BibitemOpen
  \bibfield  {author} {\bibinfo {author} {\bibfnamefont {S.}~\bibnamefont
  {{Moudgalya}}}\ and\ \bibinfo {author} {\bibfnamefont {O.~I.}\ \bibnamefont
  {{Motrunich}}},\ }\bibfield  {title} {\bibinfo {title} {{Symmetries as Ground
  States of Local Superoperators}},\ }\href
  {https://doi.org/10.48550/arXiv.2309.15167} {\bibfield  {journal} {\bibinfo
  {journal} {arXiv e-prints}\ ,\ \bibinfo {eid} {arXiv:2309.15167}} (\bibinfo
  {year} {2023})},\ \Eprint {https://arxiv.org/abs/2309.15167}
  {arXiv:2309.15167 [cond-mat.stat-mech]} \BibitemShut {NoStop}%
\bibitem [{\citenamefont {Zeng}\ and\ \citenamefont {Wen}(2015)}]{zeng2015}%
  \BibitemOpen
  \bibfield  {author} {\bibinfo {author} {\bibfnamefont {B.}~\bibnamefont
  {Zeng}}\ and\ \bibinfo {author} {\bibfnamefont {X.-G.}\ \bibnamefont {Wen}},\
  }\bibfield  {title} {\bibinfo {title} {Gapped quantum liquids and topological
  order, stochastic local transformations and emergence of unitarity},\ }\href
  {https://doi.org/10.1103/PhysRevB.91.125121} {\bibfield  {journal} {\bibinfo
  {journal} {Phys. Rev. B}\ }\textbf {\bibinfo {volume} {91}},\ \bibinfo
  {pages} {125121} (\bibinfo {year} {2015})}\BibitemShut {NoStop}%
\bibitem [{\citenamefont {Uchoa}\ and\ \citenamefont
  {Castro~Neto}(2007)}]{uchoa_superconducting_2007}%
  \BibitemOpen
  \bibfield  {author} {\bibinfo {author} {\bibfnamefont {B.}~\bibnamefont
  {Uchoa}}\ and\ \bibinfo {author} {\bibfnamefont {A.~H.}\ \bibnamefont
  {Castro~Neto}},\ }\bibfield  {title} {\bibinfo {title} {Superconducting
  {States} of {Pure} and {Doped} {Graphene}},\ }\href
  {https://doi.org/10.1103/PhysRevLett.98.146801} {\bibfield  {journal}
  {\bibinfo  {journal} {Physical Review Letters}\ }\textbf {\bibinfo {volume}
  {98}},\ \bibinfo {pages} {146801} (\bibinfo {year} {2007})}\BibitemShut
  {NoStop}%
\bibitem [{\citenamefont {Roy}\ and\ \citenamefont
  {Herbut}(2010)}]{roy_unconventional_2010}%
  \BibitemOpen
  \bibfield  {author} {\bibinfo {author} {\bibfnamefont {B.}~\bibnamefont
  {Roy}}\ and\ \bibinfo {author} {\bibfnamefont {I.~F.}\ \bibnamefont
  {Herbut}},\ }\bibfield  {title} {\bibinfo {title} {Unconventional
  superconductivity on honeycomb lattice: {Theory} of {Kekule} order
  parameter},\ }\href {https://doi.org/10.1103/PhysRevB.82.035429} {\bibfield
  {journal} {\bibinfo  {journal} {Physical Review B}\ }\textbf {\bibinfo
  {volume} {82}},\ \bibinfo {pages} {035429} (\bibinfo {year}
  {2010})}\BibitemShut {NoStop}%
\bibitem [{\citenamefont {Marino}\ and\ \citenamefont
  {Nunes}(2006)}]{marino_quantum_2006}%
  \BibitemOpen
  \bibfield  {author} {\bibinfo {author} {\bibfnamefont {E.}~\bibnamefont
  {Marino}}\ and\ \bibinfo {author} {\bibfnamefont {L.~H.}\ \bibnamefont
  {Nunes}},\ }\bibfield  {title} {\bibinfo {title} {Quantum criticality and
  superconductivity in quasi-two-dimensional {Dirac} electronic systems},\
  }\href {https://doi.org/10.1016/j.nuclphysb.2006.02.025} {\bibfield
  {journal} {\bibinfo  {journal} {Nuclear Physics B}\ }\textbf {\bibinfo
  {volume} {741}},\ \bibinfo {pages} {404} (\bibinfo {year}
  {2006})}\BibitemShut {NoStop}%
\bibitem [{\citenamefont {Black-Schaffer}\ and\ \citenamefont
  {Honerkamp}(2014)}]{black-schaffer_chiral_2014}%
  \BibitemOpen
  \bibfield  {author} {\bibinfo {author} {\bibfnamefont {A.~M.}\ \bibnamefont
  {Black-Schaffer}}\ and\ \bibinfo {author} {\bibfnamefont {C.}~\bibnamefont
  {Honerkamp}},\ }\bibfield  {title} {\bibinfo {title} {Chiral \textit{d} -wave
  superconductivity in doped graphene},\ }\href
  {https://doi.org/10.1088/0953-8984/26/42/423201} {\bibfield  {journal}
  {\bibinfo  {journal} {Journal of Physics: Condensed Matter}\ }\textbf
  {\bibinfo {volume} {26}},\ \bibinfo {pages} {423201} (\bibinfo {year}
  {2014})}\BibitemShut {NoStop}%
\bibitem [{\citenamefont {Castro~Neto}(2001)}]{castro_neto_charge_2001}%
  \BibitemOpen
  \bibfield  {author} {\bibinfo {author} {\bibfnamefont {A.~H.}\ \bibnamefont
  {Castro~Neto}},\ }\bibfield  {title} {\bibinfo {title} {Charge {Density}
  {Wave}, {Superconductivity}, and {Anomalous} {Metallic} {Behavior} in {2D}
  {Transition} {Metal} {Dichalcogenides}},\ }\href
  {https://doi.org/10.1103/PhysRevLett.86.4382} {\bibfield  {journal} {\bibinfo
   {journal} {Physical Review Letters}\ }\textbf {\bibinfo {volume} {86}},\
  \bibinfo {pages} {4382} (\bibinfo {year} {2001})}\BibitemShut {NoStop}%
\bibitem [{\citenamefont {Zhao}\ and\ \citenamefont
  {Paramekanti}(2006)}]{zhao_bcs-bec_2006}%
  \BibitemOpen
  \bibfield  {author} {\bibinfo {author} {\bibfnamefont {E.}~\bibnamefont
  {Zhao}}\ and\ \bibinfo {author} {\bibfnamefont {A.}~\bibnamefont
  {Paramekanti}},\ }\bibfield  {title} {\bibinfo {title} {{BCS}-{BEC}
  {Crossover} on the {Two}-{Dimensional} {Honeycomb} {Lattice}},\ }\href
  {https://doi.org/10.1103/PhysRevLett.97.230404} {\bibfield  {journal}
  {\bibinfo  {journal} {Physical Review Letters}\ }\textbf {\bibinfo {volume}
  {97}},\ \bibinfo {pages} {230404} (\bibinfo {year} {2006})}\BibitemShut
  {NoStop}%
\bibitem [{\citenamefont {Kotov}\ \emph {et~al.}(2012)\citenamefont {Kotov},
  \citenamefont {Uchoa}, \citenamefont {Pereira}, \citenamefont {Guinea},\ and\
  \citenamefont {Castro~Neto}}]{kotov_electron-electron_2012}%
  \BibitemOpen
  \bibfield  {author} {\bibinfo {author} {\bibfnamefont {V.~N.}\ \bibnamefont
  {Kotov}}, \bibinfo {author} {\bibfnamefont {B.}~\bibnamefont {Uchoa}},
  \bibinfo {author} {\bibfnamefont {V.~M.}\ \bibnamefont {Pereira}}, \bibinfo
  {author} {\bibfnamefont {F.}~\bibnamefont {Guinea}},\ and\ \bibinfo {author}
  {\bibfnamefont {A.~H.}\ \bibnamefont {Castro~Neto}},\ }\bibfield  {title}
  {\bibinfo {title} {Electron-{Electron} {Interactions} in {Graphene}:
  {Current} {Status} and {Perspectives}},\ }\href
  {https://doi.org/10.1103/RevModPhys.84.1067} {\bibfield  {journal} {\bibinfo
  {journal} {Reviews of Modern Physics}\ }\textbf {\bibinfo {volume} {84}},\
  \bibinfo {pages} {1067} (\bibinfo {year} {2012})}\BibitemShut {NoStop}%
\bibitem [{\citenamefont {Eisenstein}(2014)}]{Eisenstein2014}%
  \BibitemOpen
  \bibfield  {author} {\bibinfo {author} {\bibfnamefont {J.}~\bibnamefont
  {Eisenstein}},\ }\bibfield  {title} {\bibinfo {title} {Exciton condensation
  in bilayer quantum hall systems},\ }\href
  {https://doi.org/https://doi.org/10.1146/annurev-conmatphys-031113-133832}
  {\bibfield  {journal} {\bibinfo  {journal} {Annual Review of Condensed Matter
  Physics}\ }\textbf {\bibinfo {volume} {5}},\ \bibinfo {pages} {159} (\bibinfo
  {year} {2014})}\BibitemShut {NoStop}%
\bibitem [{\citenamefont {{Sala}}\ and\ \citenamefont
  {{Verresen}}(2024)}]{Sala2024a}%
  \BibitemOpen
  \bibfield  {author} {\bibinfo {author} {\bibfnamefont {P.}~\bibnamefont
  {{Sala}}}\ and\ \bibinfo {author} {\bibfnamefont {R.}~\bibnamefont
  {{Verresen}}},\ }\bibfield  {title} {\bibinfo {title} {{Stability and Loop
  Models from Decohering Non-Abelian Topological Order}},\ }\href
  {https://doi.org/10.48550/arXiv.2409.12230} {\bibfield  {journal} {\bibinfo
  {journal} {arXiv e-prints}\ ,\ \bibinfo {eid} {arXiv:2409.12230}} (\bibinfo
  {year} {2024})},\ \Eprint {https://arxiv.org/abs/2409.12230}
  {arXiv:2409.12230 [quant-ph]} \BibitemShut {NoStop}%
\bibitem [{\citenamefont {Chen}\ and\ \citenamefont
  {Grover}(2024{\natexlab{b}})}]{Chen2024}%
  \BibitemOpen
  \bibfield  {author} {\bibinfo {author} {\bibfnamefont {Y.-H.}\ \bibnamefont
  {Chen}}\ and\ \bibinfo {author} {\bibfnamefont {T.}~\bibnamefont {Grover}},\
  }\bibfield  {title} {\bibinfo {title} {Symmetry-enforced many-body
  separability transitions},\ }\href
  {https://doi.org/10.1103/PRXQuantum.5.030310} {\bibfield  {journal} {\bibinfo
   {journal} {PRX Quantum}\ }\textbf {\bibinfo {volume} {5}},\ \bibinfo {pages}
  {030310} (\bibinfo {year} {2024}{\natexlab{b}})}\BibitemShut {NoStop}%
\bibitem [{\citenamefont {Eckstein}\ \emph {et~al.}(2024)\citenamefont
  {Eckstein}, \citenamefont {Han}, \citenamefont {Trebst},\ and\ \citenamefont
  {Zhu}}]{Eckstein2024}%
  \BibitemOpen
  \bibfield  {author} {\bibinfo {author} {\bibfnamefont {F.}~\bibnamefont
  {Eckstein}}, \bibinfo {author} {\bibfnamefont {B.}~\bibnamefont {Han}},
  \bibinfo {author} {\bibfnamefont {S.}~\bibnamefont {Trebst}},\ and\ \bibinfo
  {author} {\bibfnamefont {G.-Y.}\ \bibnamefont {Zhu}},\ }\bibfield  {title}
  {\bibinfo {title} {Robust teleportation of a surface code and cascade of
  topological quantum phase transitions},\ }\href
  {https://doi.org/10.1103/PRXQuantum.5.040313} {\bibfield  {journal} {\bibinfo
   {journal} {PRX Quantum}\ }\textbf {\bibinfo {volume} {5}},\ \bibinfo {pages}
  {040313} (\bibinfo {year} {2024})}\BibitemShut {NoStop}%
\bibitem [{\citenamefont {{Lavasani}}\ and\ \citenamefont
  {{Vijay}}(2024)}]{Lavasani2024}%
  \BibitemOpen
  \bibfield  {author} {\bibinfo {author} {\bibfnamefont {A.}~\bibnamefont
  {{Lavasani}}}\ and\ \bibinfo {author} {\bibfnamefont {S.}~\bibnamefont
  {{Vijay}}},\ }\bibfield  {title} {\bibinfo {title} {{The Stability of Gapped
  Quantum Matter and Error-Correction with Adiabatic Noise}},\ }\href
  {https://doi.org/10.48550/arXiv.2402.14906} {\bibfield  {journal} {\bibinfo
  {journal} {arXiv e-prints}\ ,\ \bibinfo {eid} {arXiv:2402.14906}} (\bibinfo
  {year} {2024})},\ \Eprint {https://arxiv.org/abs/2402.14906}
  {arXiv:2402.14906 [cond-mat.str-el]} \BibitemShut {NoStop}%
\bibitem [{\citenamefont {Hwang}(2024)}]{Hwang2024}%
  \BibitemOpen
  \bibfield  {author} {\bibinfo {author} {\bibfnamefont {K.}~\bibnamefont
  {Hwang}},\ }\bibfield  {title} {\bibinfo {title} {Mixed-{S}tate {Q}uantum
  {S}pin {L}iquids and {D}ynamical {A}nyon {C}ondensations in {K}itaev
  {L}indbladians},\ }\href {https://doi.org/10.22331/q-2024-07-17-1412}
  {\bibfield  {journal} {\bibinfo  {journal} {{Quantum}}\ }\textbf {\bibinfo
  {volume} {8}},\ \bibinfo {pages} {1412} (\bibinfo {year} {2024})}\BibitemShut
  {NoStop}%
\bibitem [{\citenamefont {Nielsen}\ and\ \citenamefont
  {Chuang}(2010)}]{Nielsen_Chuang_2010}%
  \BibitemOpen
  \bibfield  {author} {\bibinfo {author} {\bibfnamefont {M.~A.}\ \bibnamefont
  {Nielsen}}\ and\ \bibinfo {author} {\bibfnamefont {I.~L.}\ \bibnamefont
  {Chuang}},\ }\bibinfo {title} {Quantum computation and quantum information:
  10th anniversary edition}\ (\bibinfo  {publisher} {Cambridge University
  Press},\ \bibinfo {year} {2010})\BibitemShut {NoStop}%
\bibitem [{\citenamefont {Ellison}\ \emph {et~al.}(2022)\citenamefont
  {Ellison}, \citenamefont {Chen}, \citenamefont {Dua}, \citenamefont
  {Shirley}, \citenamefont {Tantivasadakarn},\ and\ \citenamefont
  {Williamson}}]{Ellison2022}%
  \BibitemOpen
  \bibfield  {author} {\bibinfo {author} {\bibfnamefont {T.~D.}\ \bibnamefont
  {Ellison}}, \bibinfo {author} {\bibfnamefont {Y.-A.}\ \bibnamefont {Chen}},
  \bibinfo {author} {\bibfnamefont {A.}~\bibnamefont {Dua}}, \bibinfo {author}
  {\bibfnamefont {W.}~\bibnamefont {Shirley}}, \bibinfo {author} {\bibfnamefont
  {N.}~\bibnamefont {Tantivasadakarn}},\ and\ \bibinfo {author} {\bibfnamefont
  {D.~J.}\ \bibnamefont {Williamson}},\ }\bibfield  {title} {\bibinfo {title}
  {Pauli stabilizer models of twisted quantum doubles},\ }\href
  {https://doi.org/10.1103/PRXQuantum.3.010353} {\bibfield  {journal} {\bibinfo
   {journal} {PRX Quantum}\ }\textbf {\bibinfo {volume} {3}},\ \bibinfo {pages}
  {010353} (\bibinfo {year} {2022})}\BibitemShut {NoStop}%
\bibitem [{\citenamefont {Kitaev}(2003)}]{KITAEV2003}%
  \BibitemOpen
  \bibfield  {author} {\bibinfo {author} {\bibfnamefont {A.}~\bibnamefont
  {Kitaev}},\ }\bibfield  {title} {\bibinfo {title} {Fault-tolerant quantum
  computation by anyons},\ }\href
  {https://doi.org/https://doi.org/10.1016/S0003-4916(02)00018-0} {\bibfield
  {journal} {\bibinfo  {journal} {Annals of Physics}\ }\textbf {\bibinfo
  {volume} {303}},\ \bibinfo {pages} {2} (\bibinfo {year} {2003})}\BibitemShut
  {NoStop}%
\bibitem [{\citenamefont {Heyl}(2018)}]{Heyl2018}%
  \BibitemOpen
  \bibfield  {author} {\bibinfo {author} {\bibfnamefont {M.}~\bibnamefont
  {Heyl}},\ }\bibfield  {title} {\bibinfo {title} {Dynamical quantum phase
  transitions: a review},\ }\href {https://doi.org/10.1088/1361-6633/aaaf9a}
  {\bibfield  {journal} {\bibinfo  {journal} {Reports on Progress in Physics}\
  }\textbf {\bibinfo {volume} {81}},\ \bibinfo {pages} {054001} (\bibinfo
  {year} {2018})}\BibitemShut {NoStop}%
\bibitem [{\citenamefont {Kitaev}(2006)}]{KITAEV2006}%
  \BibitemOpen
  \bibfield  {author} {\bibinfo {author} {\bibfnamefont {A.}~\bibnamefont
  {Kitaev}},\ }\bibfield  {title} {\bibinfo {title} {Anyons in an exactly
  solved model and beyond},\ }\href
  {https://doi.org/https://doi.org/10.1016/j.aop.2005.10.005} {\bibfield
  {journal} {\bibinfo  {journal} {Annals of Physics}\ }\textbf {\bibinfo
  {volume} {321}},\ \bibinfo {pages} {2} (\bibinfo {year} {2006})},\ \bibinfo
  {note} {january Special Issue}\BibitemShut {NoStop}%
\bibitem [{\citenamefont {Lin}\ \emph {et~al.}(2021)\citenamefont {Lin},
  \citenamefont {Li},\ and\ \citenamefont {Hsieh}}]{Lin2021}%
  \BibitemOpen
  \bibfield  {author} {\bibinfo {author} {\bibfnamefont {C.-J.}\ \bibnamefont
  {Lin}}, \bibinfo {author} {\bibfnamefont {Z.}~\bibnamefont {Li}},\ and\
  \bibinfo {author} {\bibfnamefont {T.~H.}\ \bibnamefont {Hsieh}},\ }\bibfield
  {title} {\bibinfo {title} {Entanglement renormalization of thermofield double
  states},\ }\href {https://doi.org/10.1103/PhysRevLett.127.080602} {\bibfield
  {journal} {\bibinfo  {journal} {Phys. Rev. Lett.}\ }\textbf {\bibinfo
  {volume} {127}},\ \bibinfo {pages} {080602} (\bibinfo {year}
  {2021})}\BibitemShut {NoStop}%
\bibitem [{\citenamefont {Sang}\ \emph {et~al.}(2024)\citenamefont {Sang},
  \citenamefont {Zou},\ and\ \citenamefont {Hsieh}}]{Sang2024}%
  \BibitemOpen
  \bibfield  {author} {\bibinfo {author} {\bibfnamefont {S.}~\bibnamefont
  {Sang}}, \bibinfo {author} {\bibfnamefont {Y.}~\bibnamefont {Zou}},\ and\
  \bibinfo {author} {\bibfnamefont {T.~H.}\ \bibnamefont {Hsieh}},\ }\bibfield
  {title} {\bibinfo {title} {Mixed-state quantum phases: Renormalization and
  quantum error correction},\ }\href
  {https://doi.org/10.1103/PhysRevX.14.031044} {\bibfield  {journal} {\bibinfo
  {journal} {Phys. Rev. X}\ }\textbf {\bibinfo {volume} {14}},\ \bibinfo
  {pages} {031044} (\bibinfo {year} {2024})}\BibitemShut {NoStop}%
\bibitem [{\citenamefont {{Sang}}\ and\ \citenamefont
  {{Hsieh}}(2024)}]{Sang2024a}%
  \BibitemOpen
  \bibfield  {author} {\bibinfo {author} {\bibfnamefont {S.}~\bibnamefont
  {{Sang}}}\ and\ \bibinfo {author} {\bibfnamefont {T.~H.}\ \bibnamefont
  {{Hsieh}}},\ }\bibfield  {title} {\bibinfo {title} {{Stability of mixed-state
  quantum phases via finite Markov length}},\ }\href
  {https://doi.org/10.48550/arXiv.2404.07251} {\bibfield  {journal} {\bibinfo
  {journal} {arXiv e-prints}\ ,\ \bibinfo {eid} {arXiv:2404.07251}} (\bibinfo
  {year} {2024})},\ \Eprint {https://arxiv.org/abs/2404.07251}
  {arXiv:2404.07251 [quant-ph]} \BibitemShut {NoStop}%
\bibitem [{\citenamefont {{Sun}}\ \emph {et~al.}(2024)\citenamefont {{Sun}},
  \citenamefont {{Zhang}}, \citenamefont {{Bi}},\ and\ \citenamefont
  {{You}}}]{Sun2024}%
  \BibitemOpen
  \bibfield  {author} {\bibinfo {author} {\bibfnamefont {S.}~\bibnamefont
  {{Sun}}}, \bibinfo {author} {\bibfnamefont {J.-H.}\ \bibnamefont {{Zhang}}},
  \bibinfo {author} {\bibfnamefont {Z.}~\bibnamefont {{Bi}}},\ and\ \bibinfo
  {author} {\bibfnamefont {Y.}~\bibnamefont {{You}}},\ }\bibfield  {title}
  {\bibinfo {title} {{Holographic View of Mixed-State Symmetry-Protected
  Topological Phases in Open Quantum Systems}},\ }\href
  {https://doi.org/10.48550/arXiv.2410.08205} {\bibfield  {journal} {\bibinfo
  {journal} {arXiv e-prints}\ ,\ \bibinfo {eid} {arXiv:2410.08205}} (\bibinfo
  {year} {2024})},\ \Eprint {https://arxiv.org/abs/2410.08205}
  {arXiv:2410.08205 [quant-ph]} \BibitemShut {NoStop}%
\bibitem [{\citenamefont {Atiyah}\ \emph {et~al.}(1972)\citenamefont {Atiyah},
  \citenamefont {Hirzebruch}, \citenamefont {Adams},\ and\ \citenamefont
  {Shepherd}}]{Atiyah1972}%
  \BibitemOpen
  \bibfield  {author} {\bibinfo {author} {\bibfnamefont {M.~F.}\ \bibnamefont
  {Atiyah}}, \bibinfo {author} {\bibfnamefont {F.}~\bibnamefont {Hirzebruch}},
  \bibinfo {author} {\bibfnamefont {J.~F.}\ \bibnamefont {Adams}},\ and\
  \bibinfo {author} {\bibfnamefont {G.~C.}\ \bibnamefont {Shepherd}},\
  }\bibinfo {title} {Vector bundles and homogeneous spaces},\ in\ \href@noop {}
  {\emph {\bibinfo {booktitle} {Algebraic Topology: A Student’s Guide}}},\
  \bibinfo {series and number} {London Mathematical Society Lecture Note
  Series}\ (\bibinfo  {publisher} {Cambridge University Press},\ \bibinfo
  {year} {1972})\ p.\ \bibinfo {pages} {196–222}\BibitemShut {NoStop}%
\bibitem [{\citenamefont {Brown}(1982)}]{Brown1982}%
  \BibitemOpen
  \bibfield  {author} {\bibinfo {author} {\bibfnamefont {K.~S.}\ \bibnamefont
  {Brown}},\ }\bibinfo {title} {Equivariant homology and spectral sequences},\
  in\ \href {https://doi.org/10.1007/978-1-4684-9327-6_8} {\emph {\bibinfo
  {booktitle} {Cohomology of Groups}}}\ (\bibinfo  {publisher} {Springer New
  York},\ \bibinfo {address} {New York, NY},\ \bibinfo {year} {1982})\ pp.\
  \bibinfo {pages} {161--182}\BibitemShut {NoStop}%
\bibitem [{\citenamefont {Adem}\ and\ \citenamefont
  {Milgram}(1994)}]{Adem1994}%
  \BibitemOpen
  \bibfield  {author} {\bibinfo {author} {\bibfnamefont {A.}~\bibnamefont
  {Adem}}\ and\ \bibinfo {author} {\bibfnamefont {R.~J.}\ \bibnamefont
  {Milgram}},\ }\bibinfo {title} {Spectral sequences and detection theorems},\
  in\ \href {https://doi.org/10.1007/978-3-662-06282-1_5} {\emph {\bibinfo
  {booktitle} {Cohomology of Finite Groups}}}\ (\bibinfo  {publisher} {Springer
  Berlin Heidelberg},\ \bibinfo {address} {Berlin, Heidelberg},\ \bibinfo
  {year} {1994})\ pp.\ \bibinfo {pages} {117--159}\BibitemShut {NoStop}%
\bibitem [{\citenamefont {Chen}\ \emph {et~al.}(2014)\citenamefont {Chen},
  \citenamefont {Lu},\ and\ \citenamefont {Vishwanath}}]{Chen2014}%
  \BibitemOpen
  \bibfield  {author} {\bibinfo {author} {\bibfnamefont {X.}~\bibnamefont
  {Chen}}, \bibinfo {author} {\bibfnamefont {Y.-M.}\ \bibnamefont {Lu}},\ and\
  \bibinfo {author} {\bibfnamefont {A.}~\bibnamefont {Vishwanath}},\ }\bibfield
   {title} {\bibinfo {title} {Symmetry-protected topological phases from
  decorated domain walls},\ }\href {https://doi.org/10.1038/ncomms4507}
  {\bibfield  {journal} {\bibinfo  {journal} {Nature Communications}\ }\textbf
  {\bibinfo {volume} {5}},\ \bibinfo {pages} {3507} (\bibinfo {year}
  {2014})}\BibitemShut {NoStop}%
\bibitem [{\citenamefont {Thorngren}\ \emph {et~al.}(2021)\citenamefont
  {Thorngren}, \citenamefont {Vishwanath},\ and\ \citenamefont
  {Verresen}}]{Thorngren2021}%
  \BibitemOpen
  \bibfield  {author} {\bibinfo {author} {\bibfnamefont {R.}~\bibnamefont
  {Thorngren}}, \bibinfo {author} {\bibfnamefont {A.}~\bibnamefont
  {Vishwanath}},\ and\ \bibinfo {author} {\bibfnamefont {R.}~\bibnamefont
  {Verresen}},\ }\bibfield  {title} {\bibinfo {title} {Intrinsically gapless
  topological phases},\ }\href {https://doi.org/10.1103/PhysRevB.104.075132}
  {\bibfield  {journal} {\bibinfo  {journal} {Phys. Rev. B}\ }\textbf {\bibinfo
  {volume} {104}},\ \bibinfo {pages} {075132} (\bibinfo {year}
  {2021})}\BibitemShut {NoStop}%
\bibitem [{\citenamefont {{Geraedts}}\ and\ \citenamefont
  {{Motrunich}}(2014)}]{Geraedts2014}%
  \BibitemOpen
  \bibfield  {author} {\bibinfo {author} {\bibfnamefont {S.~D.}\ \bibnamefont
  {{Geraedts}}}\ and\ \bibinfo {author} {\bibfnamefont {O.~I.}\ \bibnamefont
  {{Motrunich}}},\ }\bibfield  {title} {\bibinfo {title} {{Exact Models for
  Symmetry-Protected Topological Phases in One Dimension}},\ }\href
  {https://doi.org/10.48550/arXiv.1410.1580} {\bibfield  {journal} {\bibinfo
  {journal} {arXiv e-prints}\ ,\ \bibinfo {eid} {arXiv:1410.1580}} (\bibinfo
  {year} {2014})},\ \Eprint {https://arxiv.org/abs/1410.1580} {arXiv:1410.1580
  [cond-mat.stat-mech]} \BibitemShut {NoStop}%
\bibitem [{\citenamefont {Levin}\ and\ \citenamefont {Wen}(2005)}]{Levin2005}%
  \BibitemOpen
  \bibfield  {author} {\bibinfo {author} {\bibfnamefont {M.~A.}\ \bibnamefont
  {Levin}}\ and\ \bibinfo {author} {\bibfnamefont {X.-G.}\ \bibnamefont
  {Wen}},\ }\bibfield  {title} {\bibinfo {title} {String-net condensation: A
  physical mechanism for topological phases},\ }\href
  {https://doi.org/10.1103/PhysRevB.71.045110} {\bibfield  {journal} {\bibinfo
  {journal} {Phys. Rev. B}\ }\textbf {\bibinfo {volume} {71}},\ \bibinfo
  {pages} {045110} (\bibinfo {year} {2005})}\BibitemShut {NoStop}%
\bibitem [{\citenamefont {Gaiotto}\ \emph {et~al.}(2015)\citenamefont
  {Gaiotto}, \citenamefont {Kapustin}, \citenamefont {Seiberg},\ and\
  \citenamefont {Willett}}]{Gaiotto2015}%
  \BibitemOpen
  \bibfield  {author} {\bibinfo {author} {\bibfnamefont {D.}~\bibnamefont
  {Gaiotto}}, \bibinfo {author} {\bibfnamefont {A.}~\bibnamefont {Kapustin}},
  \bibinfo {author} {\bibfnamefont {N.}~\bibnamefont {Seiberg}},\ and\ \bibinfo
  {author} {\bibfnamefont {B.}~\bibnamefont {Willett}},\ }\bibfield  {title}
  {\bibinfo {title} {Generalized global symmetries},\ }\href
  {https://doi.org/10.1007/JHEP02(2015)172} {\bibfield  {journal} {\bibinfo
  {journal} {Journal of High Energy Physics}\ }\textbf {\bibinfo {volume}
  {2015}},\ \bibinfo {pages} {172} (\bibinfo {year} {2015})}\BibitemShut
  {NoStop}%
\bibitem [{\citenamefont {Ji}\ and\ \citenamefont {Wen}(2020)}]{Ji2020}%
  \BibitemOpen
  \bibfield  {author} {\bibinfo {author} {\bibfnamefont {W.}~\bibnamefont
  {Ji}}\ and\ \bibinfo {author} {\bibfnamefont {X.-G.}\ \bibnamefont {Wen}},\
  }\bibfield  {title} {\bibinfo {title} {Categorical symmetry and noninvertible
  anomaly in symmetry-breaking and topological phase transitions},\ }\href
  {https://doi.org/10.1103/PhysRevResearch.2.033417} {\bibfield  {journal}
  {\bibinfo  {journal} {Phys. Rev. Res.}\ }\textbf {\bibinfo {volume} {2}},\
  \bibinfo {pages} {033417} (\bibinfo {year} {2020})}\BibitemShut {NoStop}%
\bibitem [{\citenamefont {Kong}\ \emph {et~al.}(2020)\citenamefont {Kong},
  \citenamefont {Lan}, \citenamefont {Wen}, \citenamefont {Zhang},\ and\
  \citenamefont {Zheng}}]{Kong2020}%
  \BibitemOpen
  \bibfield  {author} {\bibinfo {author} {\bibfnamefont {L.}~\bibnamefont
  {Kong}}, \bibinfo {author} {\bibfnamefont {T.}~\bibnamefont {Lan}}, \bibinfo
  {author} {\bibfnamefont {X.-G.}\ \bibnamefont {Wen}}, \bibinfo {author}
  {\bibfnamefont {Z.-H.}\ \bibnamefont {Zhang}},\ and\ \bibinfo {author}
  {\bibfnamefont {H.}~\bibnamefont {Zheng}},\ }\bibfield  {title} {\bibinfo
  {title} {Algebraic higher symmetry and categorical symmetry: A holographic
  and entanglement view of symmetry},\ }\href
  {https://doi.org/10.1103/PhysRevResearch.2.043086} {\bibfield  {journal}
  {\bibinfo  {journal} {Phys. Rev. Res.}\ }\textbf {\bibinfo {volume} {2}},\
  \bibinfo {pages} {043086} (\bibinfo {year} {2020})}\BibitemShut {NoStop}%
\bibitem [{\citenamefont {Kong}(2014)}]{KONG2014}%
  \BibitemOpen
  \bibfield  {author} {\bibinfo {author} {\bibfnamefont {L.}~\bibnamefont
  {Kong}},\ }\bibfield  {title} {\bibinfo {title} {Anyon condensation and
  tensor categories},\ }\href
  {https://doi.org/https://doi.org/10.1016/j.nuclphysb.2014.07.003} {\bibfield
  {journal} {\bibinfo  {journal} {Nuclear Physics B}\ }\textbf {\bibinfo
  {volume} {886}},\ \bibinfo {pages} {436} (\bibinfo {year}
  {2014})}\BibitemShut {NoStop}%
\bibitem [{\citenamefont {Burnell}(2018)}]{Burnell2018}%
  \BibitemOpen
  \bibfield  {author} {\bibinfo {author} {\bibfnamefont {F.}~\bibnamefont
  {Burnell}},\ }\bibfield  {title} {\bibinfo {title} {Anyon condensation and
  its applications},\ }\href
  {https://doi.org/https://doi.org/10.1146/annurev-conmatphys-033117-054154}
  {\bibfield  {journal} {\bibinfo  {journal} {Annual Review of Condensed Matter
  Physics}\ }\textbf {\bibinfo {volume} {9}},\ \bibinfo {pages} {307} (\bibinfo
  {year} {2018})}\BibitemShut {NoStop}%
\bibitem [{\citenamefont {{Kong}}\ and\ \citenamefont
  {{Wen}}(2014)}]{Kong2014a}%
  \BibitemOpen
  \bibfield  {author} {\bibinfo {author} {\bibfnamefont {L.}~\bibnamefont
  {{Kong}}}\ and\ \bibinfo {author} {\bibfnamefont {X.-G.}\ \bibnamefont
  {{Wen}}},\ }\bibfield  {title} {\bibinfo {title} {{Braided fusion categories,
  gravitational anomalies, and the mathematical framework for topological
  orders in any dimensions}},\ }\href
  {https://doi.org/10.48550/arXiv.1405.5858} {\bibfield  {journal} {\bibinfo
  {journal} {arXiv e-prints}\ ,\ \bibinfo {eid} {arXiv:1405.5858}} (\bibinfo
  {year} {2014})},\ \Eprint {https://arxiv.org/abs/1405.5858} {arXiv:1405.5858
  [cond-mat.str-el]} \BibitemShut {NoStop}%
\bibitem [{\citenamefont {Wen}(1995)}]{Wen1995}%
  \BibitemOpen
  \bibfield  {author} {\bibinfo {author} {\bibfnamefont {X.-G.}\ \bibnamefont
  {Wen}},\ }\bibfield  {title} {\bibinfo {title} {Topological orders and edge
  excitations in fractional quantum hall states},\ }\href
  {https://doi.org/10.1080/00018739500101566} {\bibfield  {journal} {\bibinfo
  {journal} {Advances in Physics}\ }\textbf {\bibinfo {volume} {44}},\ \bibinfo
  {pages} {405} (\bibinfo {year} {1995})},\ \Eprint
  {https://arxiv.org/abs/https://doi.org/10.1080/00018739500101566}
  {https://doi.org/10.1080/00018739500101566} \BibitemShut {NoStop}%
\bibitem [{\citenamefont {Li}\ and\ \citenamefont {Haldane}(2008)}]{Li2008}%
  \BibitemOpen
  \bibfield  {author} {\bibinfo {author} {\bibfnamefont {H.}~\bibnamefont
  {Li}}\ and\ \bibinfo {author} {\bibfnamefont {F.~D.~M.}\ \bibnamefont
  {Haldane}},\ }\bibfield  {title} {\bibinfo {title} {Entanglement spectrum as
  a generalization of entanglement entropy: Identification of topological order
  in non-abelian fractional quantum hall effect states},\ }\href
  {https://doi.org/10.1103/PhysRevLett.101.010504} {\bibfield  {journal}
  {\bibinfo  {journal} {Phys. Rev. Lett.}\ }\textbf {\bibinfo {volume} {101}},\
  \bibinfo {pages} {010504} (\bibinfo {year} {2008})}\BibitemShut {NoStop}%
\bibitem [{\citenamefont {Fidkowski}(2010)}]{Fidkowski2010}%
  \BibitemOpen
  \bibfield  {author} {\bibinfo {author} {\bibfnamefont {L.}~\bibnamefont
  {Fidkowski}},\ }\bibfield  {title} {\bibinfo {title} {Entanglement spectrum
  of topological insulators and superconductors},\ }\href
  {https://doi.org/10.1103/PhysRevLett.104.130502} {\bibfield  {journal}
  {\bibinfo  {journal} {Phys. Rev. Lett.}\ }\textbf {\bibinfo {volume} {104}},\
  \bibinfo {pages} {130502} (\bibinfo {year} {2010})}\BibitemShut {NoStop}%
\bibitem [{\citenamefont {Pollmann}\ \emph {et~al.}(2010)\citenamefont
  {Pollmann}, \citenamefont {Turner}, \citenamefont {Berg},\ and\ \citenamefont
  {Oshikawa}}]{Pollmann2010}%
  \BibitemOpen
  \bibfield  {author} {\bibinfo {author} {\bibfnamefont {F.}~\bibnamefont
  {Pollmann}}, \bibinfo {author} {\bibfnamefont {A.~M.}\ \bibnamefont
  {Turner}}, \bibinfo {author} {\bibfnamefont {E.}~\bibnamefont {Berg}},\ and\
  \bibinfo {author} {\bibfnamefont {M.}~\bibnamefont {Oshikawa}},\ }\bibfield
  {title} {\bibinfo {title} {Entanglement spectrum of a topological phase in
  one dimension},\ }\href {https://doi.org/10.1103/PhysRevB.81.064439}
  {\bibfield  {journal} {\bibinfo  {journal} {Phys. Rev. B}\ }\textbf {\bibinfo
  {volume} {81}},\ \bibinfo {pages} {064439} (\bibinfo {year}
  {2010})}\BibitemShut {NoStop}%
\bibitem [{\citenamefont {Kitaev}\ and\ \citenamefont
  {Preskill}(2006)}]{Kitaev2006a}%
  \BibitemOpen
  \bibfield  {author} {\bibinfo {author} {\bibfnamefont {A.}~\bibnamefont
  {Kitaev}}\ and\ \bibinfo {author} {\bibfnamefont {J.}~\bibnamefont
  {Preskill}},\ }\bibfield  {title} {\bibinfo {title} {Topological entanglement
  entropy},\ }\href {https://doi.org/10.1103/PhysRevLett.96.110404} {\bibfield
  {journal} {\bibinfo  {journal} {Phys. Rev. Lett.}\ }\textbf {\bibinfo
  {volume} {96}},\ \bibinfo {pages} {110404} (\bibinfo {year}
  {2006})}\BibitemShut {NoStop}%
\bibitem [{\citenamefont {Levin}\ and\ \citenamefont {Wen}(2006)}]{Levin2006}%
  \BibitemOpen
  \bibfield  {author} {\bibinfo {author} {\bibfnamefont {M.}~\bibnamefont
  {Levin}}\ and\ \bibinfo {author} {\bibfnamefont {X.-G.}\ \bibnamefont
  {Wen}},\ }\bibfield  {title} {\bibinfo {title} {Detecting topological order
  in a ground state wave function},\ }\href
  {https://doi.org/10.1103/PhysRevLett.96.110405} {\bibfield  {journal}
  {\bibinfo  {journal} {Phys. Rev. Lett.}\ }\textbf {\bibinfo {volume} {96}},\
  \bibinfo {pages} {110405} (\bibinfo {year} {2006})}\BibitemShut {NoStop}%
\bibitem [{\citenamefont {Baym}\ and\ \citenamefont
  {Pethick}(1991)}]{Baym1991}%
  \BibitemOpen
  \bibfield  {author} {\bibinfo {author} {\bibfnamefont {G.}~\bibnamefont
  {Baym}}\ and\ \bibinfo {author} {\bibfnamefont {C.}~\bibnamefont {Pethick}},\
  }\bibinfo {title} {Landau fermi-liquid theory and low temperature properties
  of normal liquid 3he},\ in\ \href
  {https://doi.org/https://doi.org/10.1002/9783527617159.ch1} {\emph {\bibinfo
  {booktitle} {Landau Fermi‐Liquid Theory}}}\ (\bibinfo  {publisher} {John
  Wiley \& Sons, Ltd},\ \bibinfo {year} {1991})\ Chap.~\bibinfo {chapter} {1},
  pp.\ \bibinfo {pages} {1--121},\ \Eprint
  {https://arxiv.org/abs/https://onlinelibrary.wiley.com/doi/pdf/10.1002/9783527617159.ch1}
  {https://onlinelibrary.wiley.com/doi/pdf/10.1002/9783527617159.ch1}
  \BibitemShut {NoStop}%
\bibitem [{\citenamefont {Nandkishore}\ and\ \citenamefont
  {Hermele}(2019)}]{Nandkishore2019}%
  \BibitemOpen
  \bibfield  {author} {\bibinfo {author} {\bibfnamefont {R.~M.}\ \bibnamefont
  {Nandkishore}}\ and\ \bibinfo {author} {\bibfnamefont {M.}~\bibnamefont
  {Hermele}},\ }\bibfield  {title} {\bibinfo {title} {Fractons},\ }\href
  {https://doi.org/https://doi.org/10.1146/annurev-conmatphys-031218-013604}
  {\bibfield  {journal} {\bibinfo  {journal} {Annual Review of Condensed Matter
  Physics}\ }\textbf {\bibinfo {volume} {10}},\ \bibinfo {pages} {295}
  (\bibinfo {year} {2019})}\BibitemShut {NoStop}%
\bibitem [{\citenamefont {Pretko}\ \emph {et~al.}(2020)\citenamefont {Pretko},
  \citenamefont {Chen},\ and\ \citenamefont {You}}]{Pretko2020}%
  \BibitemOpen
  \bibfield  {author} {\bibinfo {author} {\bibfnamefont {M.}~\bibnamefont
  {Pretko}}, \bibinfo {author} {\bibfnamefont {X.}~\bibnamefont {Chen}},\ and\
  \bibinfo {author} {\bibfnamefont {Y.}~\bibnamefont {You}},\ }\bibfield
  {title} {\bibinfo {title} {Fracton phases of matter},\ }\href
  {https://doi.org/10.1142/S0217751X20300033} {\bibfield  {journal} {\bibinfo
  {journal} {International Journal of Modern Physics A}\ }\textbf {\bibinfo
  {volume} {35}},\ \bibinfo {pages} {2030003} (\bibinfo {year} {2020})},\
  \Eprint {https://arxiv.org/abs/https://doi.org/10.1142/S0217751X20300033}
  {https://doi.org/10.1142/S0217751X20300033} \BibitemShut {NoStop}%
\bibitem [{\citenamefont {Son}\ \emph {et~al.}(2012)\citenamefont {Son},
  \citenamefont {Amico},\ and\ \citenamefont {Vedral}}]{Son2012}%
  \BibitemOpen
  \bibfield  {author} {\bibinfo {author} {\bibfnamefont {W.}~\bibnamefont
  {Son}}, \bibinfo {author} {\bibfnamefont {L.}~\bibnamefont {Amico}},\ and\
  \bibinfo {author} {\bibfnamefont {V.}~\bibnamefont {Vedral}},\ }\bibfield
  {title} {\bibinfo {title} {Topological order in 1d cluster state protected by
  symmetry},\ }\href {https://doi.org/10.1007/s11128-011-0346-7} {\bibfield
  {journal} {\bibinfo  {journal} {Quantum Information Processing}\ }\textbf
  {\bibinfo {volume} {11}},\ \bibinfo {pages} {1961} (\bibinfo {year}
  {2012})}\BibitemShut {NoStop}%
\end{thebibliography}%

\end{document}